\documentclass[twocolumn]{aastex631}%linenumbers

\shorttitle{BWFields: Bubble Identification and Analysis}
\shortauthors{Jiang et al.}

\usepackage{mathrsfs}
\usepackage{amsmath}
\usepackage{amssymb}
\usepackage{appendix}
\usepackage{array}
\usepackage{url}
\usepackage{hyperref}
\usepackage{soul}
\usepackage{color, xcolor} 
\usepackage{makecell}
\usepackage{booktabs}
\usepackage{threeparttable}
\usepackage{tabularx}
\usepackage{graphicx}
\usepackage{enumitem}
\usepackage{rotating}
\usepackage{footmisc}

\begin{document}

% \title{Investigations of MWISP Bubbles: Identification and Analysis of Enclosed Molecular Bubbles}

\title{Investigations of MWISP Bubbles: Identification and Analysis of Enclosed Molecular Bubbles by Weight Fields}

% \title{Investigations of MWISP Bubbles: A PPV-based Framework for Identifying and Characterising Molecular-Bubble Candidates with Enclosed Cavities}

% \title{Investigations of MWISP Bubbles. I. Enclosed Bubble Identification and Analysis Algorithms}

% \title{Investigations of MWISP Bubbles. I. Enclosed Bubble Identification and Analysis Algorithms, and Source Catalog}

\author[0000-0002-3549-5029]{Yu Jiang}
\affiliation{Purple Mountain Observatory and Key Laboratory of Radio Astronomy, Chinese Academy of Sciences, 10 Yuanhua Road, Nanjing 210034, People’s Republic of China }
\affiliation{School of Astronomy and Space Science, University of Science and Technology of China, 96 Jinzhai Road, Hefei 230026, People’s Republic of China}
% \affiliation{Center for Astronomy and Space Sciences, China Three Gorges University, 8 University Road, Yichang 443002, People’s Republic of China}
%\affiliation{College of Science, China Three Gorges University, 8 University Road, Yichang, People’s Republic of China}

\author[0000-0003-3151-8964]{Xuepeng Chen}
\affiliation{Purple Mountain Observatory and Key Laboratory of Radio Astronomy, Chinese Academy of Sciences, 10 Yuanhua Road, Nanjing 210034, People’s Republic of China }
\affiliation{School of Astronomy and Space Science, University of Science and Technology of China, 96 Jinzhai Road, Hefei 230026, People’s Republic of China}

\author[0000-0001-7768-7320]{Ji Yang}
\affiliation{Purple Mountain Observatory and Key Laboratory of Radio Astronomy, Chinese Academy of Sciences, 10 Yuanhua Road, Nanjing 210034, People’s Republic of China }

\author[0000-0002-7237-3856]{Ke Wang}
\affiliation{Kavli Institute for Astronomy and Astrophysics, Peking University, Beijing 100871, People’s Republic of China}

\author{Zhibo Jiang}
\affiliation{Purple Mountain Observatory and Key Laboratory of Radio Astronomy, Chinese Academy of Sciences, 10 Yuanhua Road, Nanjing 210034, People’s Republic of China }

\author{Sheng Zheng}
\affiliation{Center for Astronomy and Space Sciences, China Three Gorges University, 8 University Road, Yichang 443002, People’s Republic of China}
% \affiliation{College of Science, China Three Gorges University, 8 University Road, Yichang 443002, People’s Republic of China}

\author[0000-0002-3904-1622]{Yan Sun}
\affiliation{Purple Mountain Observatory and Key Laboratory of Radio Astronomy, Chinese Academy of Sciences, 10 Yuanhua Road, Nanjing 210034, People’s Republic of China }

\author[0000-0002-0197-470X]{Yang Su}
\affiliation{Purple Mountain Observatory and Key Laboratory of Radio Astronomy, Chinese Academy of Sciences, 10 Yuanhua Road, Nanjing 210034, People’s Republic of China }

\author[0000-0001-8060-1321]{Min Fang}
\affiliation{Purple Mountain Observatory and Key Laboratory of Radio Astronomy, Chinese Academy of Sciences, 10 Yuanhua Road, Nanjing 210034, People’s Republic of China }

\author[0000-0003-4586-7751]{Qing-Zeng Yan}
\affiliation{Purple Mountain Observatory and Key Laboratory of Radio Astronomy, Chinese Academy of Sciences, 10 Yuanhua Road, Nanjing 210034, People’s Republic of China }

\correspondingauthor{Ji Yang}
\email{jiyang@pmo.ac.cn}

\correspondingauthor{Xuepeng Chen}
\email{xpchen@pmo.ac.cn}

% \correspondingauthor{Yu Jiang}
% \email{yujiang@pmo.ac.cn}

\begin{abstract}
Molecular bubbles are widely used as tracers of stellar feedback; yet, their identification in spectral-line surveys remains challenging because both cavity morphology and kinematic structure must be assessed consistently in position--position--velocity (PPV) space. 
We present the Bubble-Weight Fields (BWFields) framework, a PPV-based method that for the first time enables the automated and objective identification and analysis of enclosed molecular bubbles directly from spectral-line data cubes. 
BWFields constructs a bubble-weight field, $W_{l,b,v}$, which encodes cumulative evidence for cavity interiors by aggregating topological signatures across multiple signal-to-noise tiers and velocity-integration scales. 
Contiguous cavity interiors are segmented as weight-clumps and associated with surrounding molecular gas, linking candidate bubbles to the structure of their host clouds.
Shell morphology is characterized using radial intensity profiles and emission-defined intensity skeletons, which capture the shell geometry as traced by the observed emission.
Bubble kinematics are quantified using azimuthally sampled position-velocity (PV) diagnostics, along with a turbulence-normalized expansion significance, which serves as a direct measure of the expansion-like velocity organisation.
Applied to MWISP $^{13}$CO observations of the G17 region, BWFields identifies a population of bubble candidates with a broad range of morphologies and velocity structures in complex environments.
BWFields establishes a scalable and physically interpretable framework for molecular-bubble studies in large surveys, enabling systematic investigations of stellar feedback in the Galactic interstellar medium.
\end{abstract}


\section{Introduction}

Stellar feedback from massive stars shapes the interstellar medium by driving expanding bubbles that connect ionized regions, stellar populations, and surrounding molecular gas \citep[e.g.,][]{Review_5,Churchwell_1,Churchwell_2,Deharveng2010,Review_6,Review_2}. Large-area Galactic molecular-line surveys now enable the systematic identification of these structures and their connection to molecular cloud properties \citep[e.g.,][]{Review_13,Beuther2016,SOFIA_0,CO_Survey_MWISP_3}. 
Observations further reveal that such bubbles are intrinsically multiscale systems, exhibiting coherent expansion over tens of parsecs while simultaneously hosting subparsec substructures, including nested cavities and localized kinematic perturbations \citep{SOFIA_Orion_1,SOFIA_Orion_2,SOFIA_Orion_3,SOFIA_Orion_4,PHANGS_Bub_1,NGC628_1,SOFIA_Cygnus_1,M16_SOFIA_3,ALMA_Bub_1}. This multiscale nature motivates methods capable of capturing both global morphology and embedded substructure.

A key challenge is that molecular bubbles are intrinsically three-dimensional structures. They are expected to manifest as cavities surrounded by enhanced emission, persist over finite velocity ranges, and exhibit coherent velocity patterns reflecting multi-scale kinematics \citep[e.g.,][]{Arce2011,Li2015,Liu2024}. In practice, these signatures are often analysed separately using integrated intensity maps, channel maps, or position--velocity (PV) diagrams \citep{N74_1,Bubble_Catalog_1,Beaumont2014,Bubble_Catalog_2}. However, automated methods that jointly characterize cavity morphology and kinematic organization, particularly in crowded regions with overlapping shell-like structures, remain scarce \citep{Xu2017,Xu2020,Infrared_Bubble_Recognition,NGC628_2}.

To address this gap, we develop the Bubble-Weight Fields (BWFields) framework, the first PPV-native method for the automated identification and characterization of molecular bubbles in three-dimensional spectral-line data. BWFields constructs a bubble-weight field that aggregates evidence for cavity interiors across multiple velocity scales and signal-to-noise levels, enabling robust identification of cavity structures in PPV space. These cavities are segmented as weight-clumps using \texttt{FacetClumps} \citep{FacetClumps} and associated with surrounding molecular gas, thereby linking interior cavities to their parent clouds.

Each candidate is characterized using complementary morphological and kinematic diagnostics. Shell geometry is derived from radial intensity profiles and emission-defined ridge structures, providing estimates of radius, thickness, and symmetry that are directly tied to the observed emission. Kinematics are quantified using azimuthally sampled PV slices and a turbulence-normalised expansion significance, enabling model-independent assessment of expansion-like velocity structures. Together, these components provide an interpretable description of both the morphology and dynamics of molecular bubbles.

We apply BWFields to MWISP $^{13}$CO observations of the G17 region, which contains both isolated and confused structures spanning a wide range of environments, from relatively simple shells to complex star-forming regions, including the M16 complex \citep{M16_1,M16_2,M16_3,M16_4,M16_SOFIA_1,M16_SOFIA_2,M16_SOFIA_3}, the infrared bubble N19 \citep{Churchwell_1}, and additional cavity-like features. This diversity provides a comprehensive testbed for evaluating the robustness of the method under varying morphological and kinematic conditions.

The paper is organized as follows. Section~\ref{sec:data_mwisp} describes the MWISP survey and observational data. Section~\ref{sec:flow} presents BWFields, including the construction of the bubble-weight field, identification of weight-clumps, association with molecular gas, morphological and kinematic diagnostics, and the limitations of the method. Section~\ref{sec:application_g17} demonstrates the application to the G17 region. Section~\ref{sec:kinematic_interpretation} discusses the interpretation of kinematic diagnostics. Finally, Section~\ref{sec:conclusions} summarizes the results and outlines prospects for large-scale molecular-bubble population studies.

\begin{figure*}
\centering
\centerline{\includegraphics[width=6in]{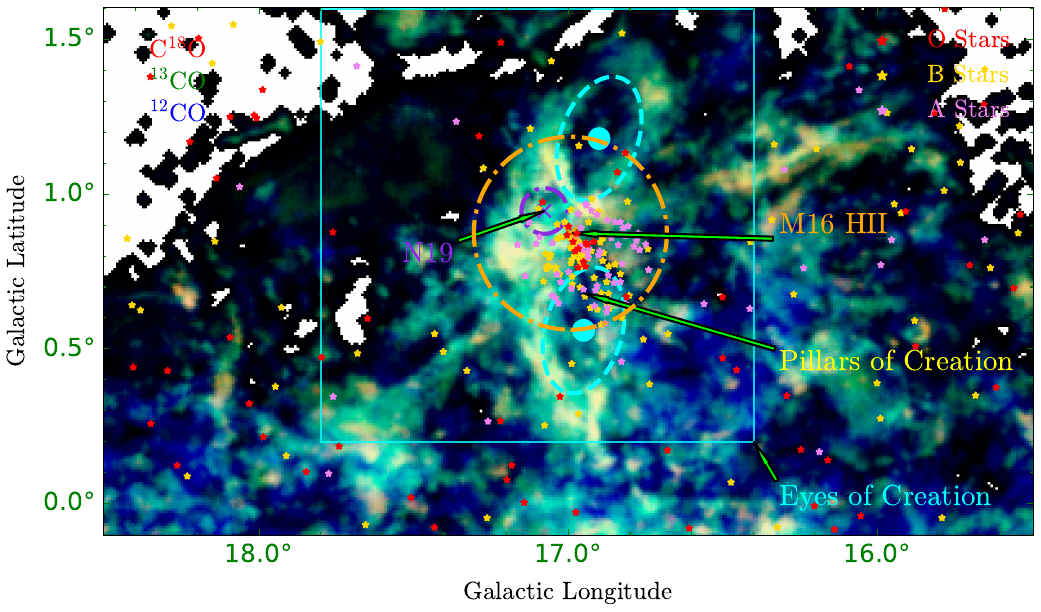}}
\caption{An illustrative view of the G17 region used to demonstrate BWFields. 
The background shows a three-color composite from the MWISP survey, with C$^{18}$O in red, $^{13}$CO in green, and $^{12}$CO in blue, integrated over velocities from $-10$ to $50~\mathrm{km\,s^{-1}}$. 
Prominent ionized regions and infrared bubbles are labeled, including M16 H\,\textsc{ii} and N19. 
O-, B-, and A-type stars are overlaid as red, gold, and violet star symbols, respectively. 
Two visually symmetric, eye-shaped cavity-like structures located on opposite sides of M16—referred to as EoC—are outlined by cyan contours; the southern component (EoC-S) contains the Pillars of Creation.
}
\label{Img_Origin_Data}
\end{figure*}

\section{MWISP Survey and Data}
\label{sec:data_mwisp}

We use molecular-line data from the Milky Way Imaging Scroll Painting (MWISP) project, an unbiased Galactic-plane survey of the $^{12}$CO, $^{13}$CO, and C$^{18}$O ($J$=1--0) transitions conducted with the Purple Mountain Observatory 13.7-m telescope \citep{CO_Survey_MWISP_2,CO_Survey_MWISP_1}. 
The three isotopologues are observed simultaneously, enabling a uniform view of molecular gas across a wide range of column densities and optical depths.

MWISP Phase~I covers Galactic longitudes $9\fdg75 \le l \le 229\fdg75$ and latitudes $|b| \le 5\fdg25$, with an angular resolution of $\sim50\arcsec$ and a velocity resolution of $\sim0.17~\mathrm{km~s^{-1}}$. 
Typical rms noise per channel is $\sim0.45~\mathrm{K}$ for $^{12}$CO and $\sim0.25~\mathrm{K}$ for $^{13}$CO and C$^{18}$O. 
Standard calibration and reduction—including baseline subtraction, bad-channel excision, and quality control—produce fully calibrated data cubes \citep{CO_Survey_MWISP_3}.

In this work, we adopt $^{13}$CO as the primary tracer of molecular shells. It effectively traces the cool, dense molecular gas associated with shell structures. Compared with $^{12}$CO and C$^{18}$O, $^{13}$CO offers a favorable balance between optical depth and signal-to-noise ratio (SNR), making it well suited for identifying cavity--shell morphologies and characterizing their velocity structure.

% In this work, we use the 13CO data to trace molecular shell structures and their kinematics. Compared with 12 CO, 13 CO suffers less from optical-depth effects and delineates shell rims more clearly, while retaining higher signal-to-noise than C18O over large areas. This makes 13CO well suited for identifying cavity–shell morpholo- gies and characterising velocity structure.

% Other tracers—such as [C~\textsc{ii}] 158~$\mu$m (e.g., SOFIA; \citealt{Pabst2021,Pabst2022}), HI 21~cm, and infrared continuum—probe different gas phases or dust emission, but their optical properties, sensitivity, or limited coverage make them less suitable for the systematic identification and characterisation of molecular cavities targeted here.

Figure~\ref{Img_Origin_Data} shows a three-color composite of the G17 region using MWISP $J$=1--0 data, illustrating the complex molecular environment around the prominent ionized regions M16 H\,\textsc{ii} \citep{Anderson2014} and N19. 
O-, B-, and A-type stars\footnote{Stellar data are retrieved from the SIMBAD database (\href{http://simbad.cds.unistra.fr/simbad/sim-fcoo}{http://simbad.cds.unistra.fr/simbad/sim-fcoo}) within the spatial extent of the MWISP data.} are overlaid on the molecular emission, highlighting the distribution of massive and intermediate-mass stellar populations relative to the surrounding molecular gas.
Two large, visually symmetric cavity-like features appear on opposite sides of the nebula. We refer to this pair as the \emph{Eyes of Creation} (EoC), consisting of a northern (EoC-N) and a southern (EoC-S) component, with EoC-S encompassing the projected location of the Pillars of Creation \citep{Poc}.

\begin{figure*}
\centering
\centerline{\includegraphics[width=7in]{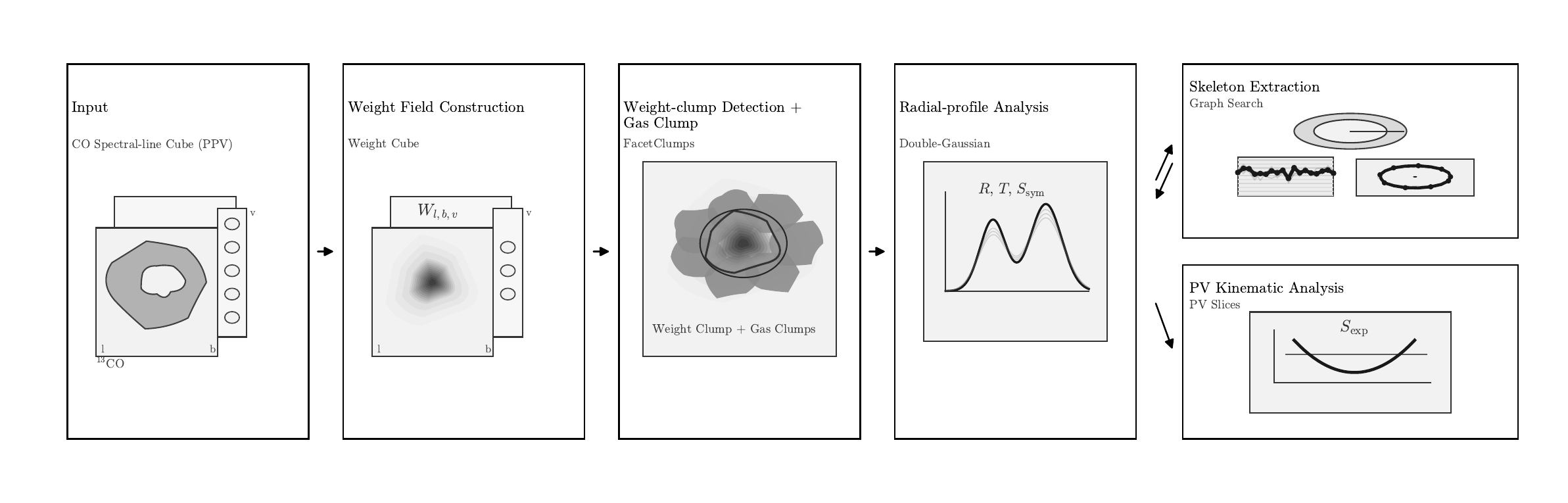}}
\caption{Overview of the BWFields pipeline for identifying molecular bubble candidates. 
A bubble-weight field $W_{l,b,v}$ is constructed from slab-integrated maps to highlight cavity interiors across multiple signal-to-noise tiers and velocity scales. 
Weight-clumps are segmented and linked to the surrounding molecular gas, with a fitted ellipse defining the initial geometric reference.
Based on this reference, shell morphology is first characterized using radial intensity profiles and double-Gaussian fitting, yielding estimates of the shell radius and thickness.
Using these morphological parameters as a guide, azimuth--radius unwrapping and graph-based ridge extraction are then applied to derive an emission-defined intensity skeleton. 
An ellipse fitted to the skeleton is subsequently used to perform a second radial-profile analysis, yielding refined estimates of the shell morphological parameters, which are then adopted in the kinematic analysis. 
Bubble kinematics are evaluated using azimuthally sampled PV slices and a turbulence-normalised expansion significance.
}
\label{Img_Pipeline}
\end{figure*}

\section{Bubble Identification and Analysis}
\label{sec:flow}
The identification of molecular bubbles in Galactic-plane CO surveys requires jointly analysing cavity morphology and velocity structure in PPV space, while remaining robust to projection effects and scalable to large data cubes. 

\begin{figure*}
\centering
\vspace{0cm}
\begin{minipage}[t]{0.3\textwidth}
    \centering
    \centerline{\includegraphics[width=2in]{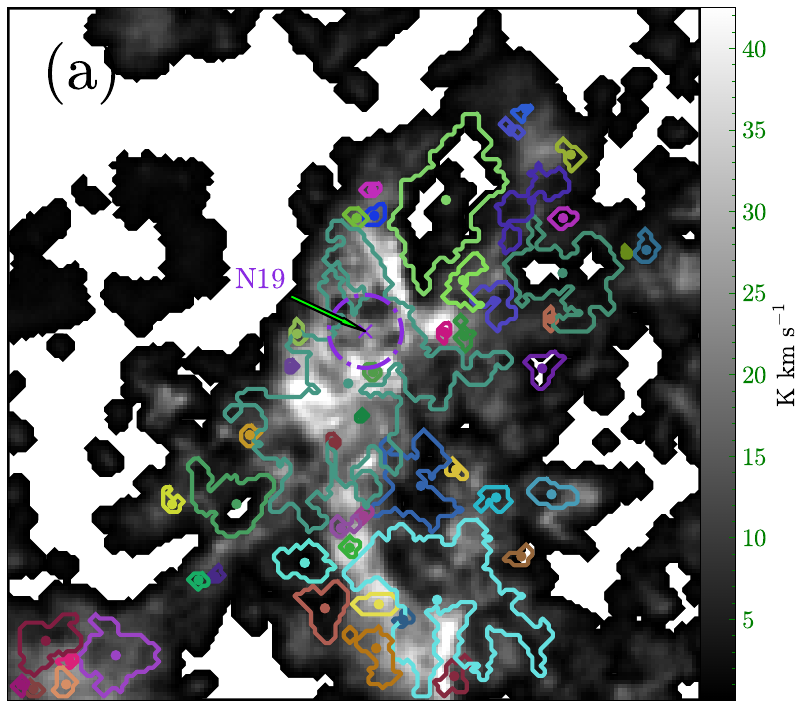}}
\end{minipage}
\begin{minipage}[t]{0.3\textwidth}
    \centering
    \centerline{\includegraphics[width=2in]{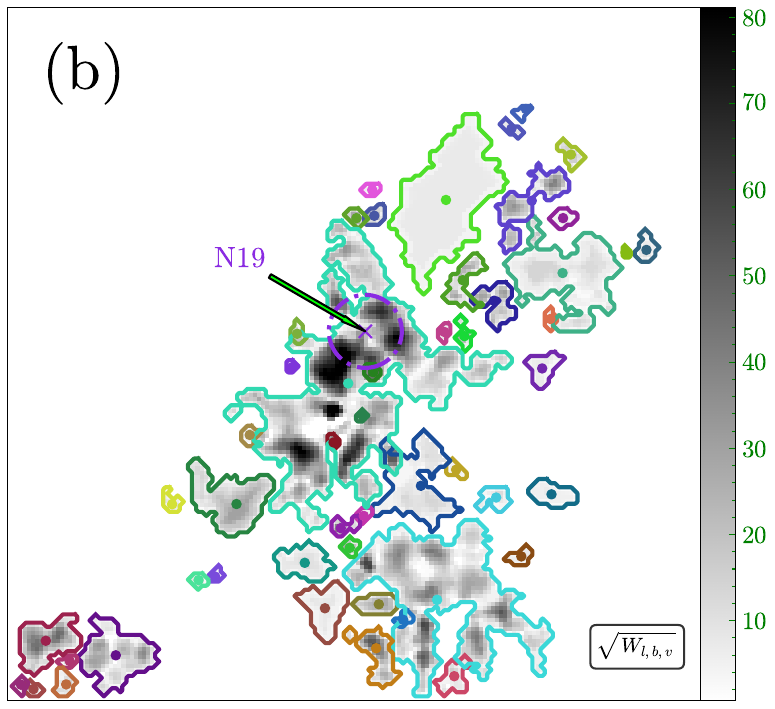}}
    %		\vspace{-4cm}
    %		\centerline{(a)}
\end{minipage}%
\begin{minipage}[t]{0.3\textwidth}
    \centering
    \centerline{\includegraphics[width=2in]{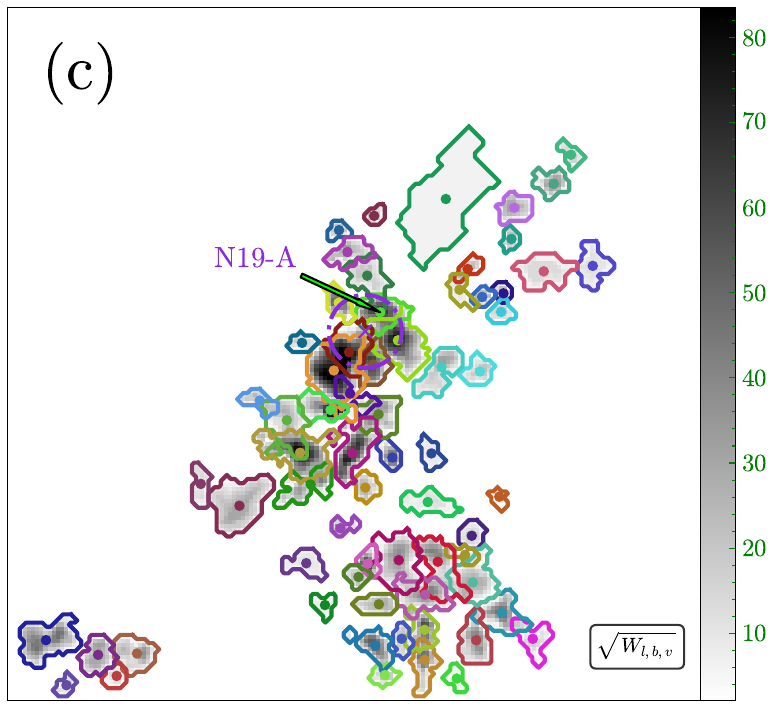}}
    %		\vspace{-4cm}
    %		\centerline{(a)}
\end{minipage}
\caption{Spatial and morphological distributions of molecular gas and identified bubble structures in the EoC region. 
(a) Integrated intensity map of the molecular emission. Colored points and contours show the positions and boundaries of bubble regions extracted from the bubble-weight cube $W_{l,b,v}$. The blue-violet dot and circle indicate the position and radius of bubble N19. 
(b) Velocity-integrated bubble-weight map, with contours showing the identified bubble regions. To reduce the contrast with background emission, the color scale and the grayscale background represent the square root of the bubble weight. 
(c) Weight-clumps identified from the bubble-weight cube, with different colors delineating individual clumps. 
N19-A marks a sub-bubble candidate identified by BWFields.}
\label{Img_Weighted_Clumps}
\end{figure*}

\begin{figure*}
\centering
\centerline{\includegraphics[width=6.5in]{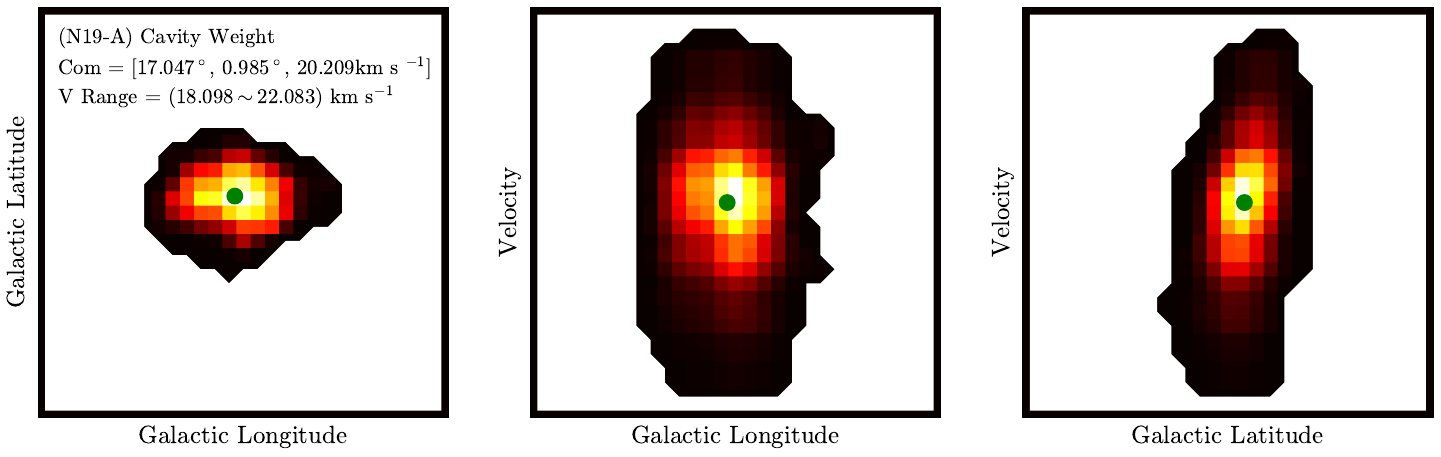}}
\caption{Three orthogonal projections of the N19--A weight-clump identified in the bubble-weight cube $W_{l,b,v}$ by \texttt{FacetClumps}. 
Left: integrated weight in the longitude--latitude plane, showing the projected cavity morphology. 
Middle and right: latitude- and longitude-integrated weight maps. 
The green marker indicates the weighted center of mass, with its coordinates and velocity range annotated. 
Colors encode the bubble weight.}
\label{Img_Weighted_Clump_N19A}
\end{figure*}

\begin{figure*}
\centering
\vspace{0cm}
\begin{minipage}[t]{0.4\textwidth}
    \centering
    \centerline{\includegraphics[width=2.5in]{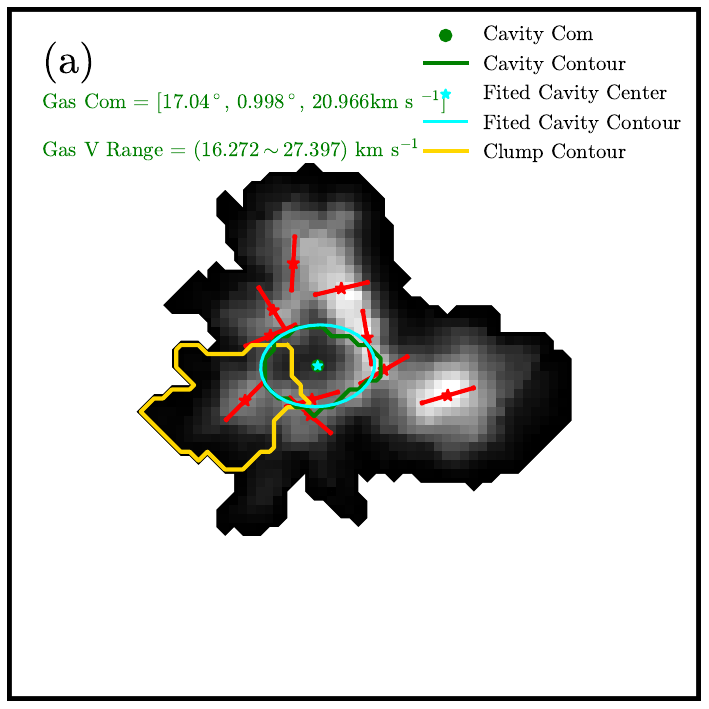}}
    %		\vspace{-4cm}
    %		\centerline{(a)}
\end{minipage}
\begin{minipage}[t]{0.4\textwidth}
    \centering
    \centerline{\includegraphics[width=2.5in]{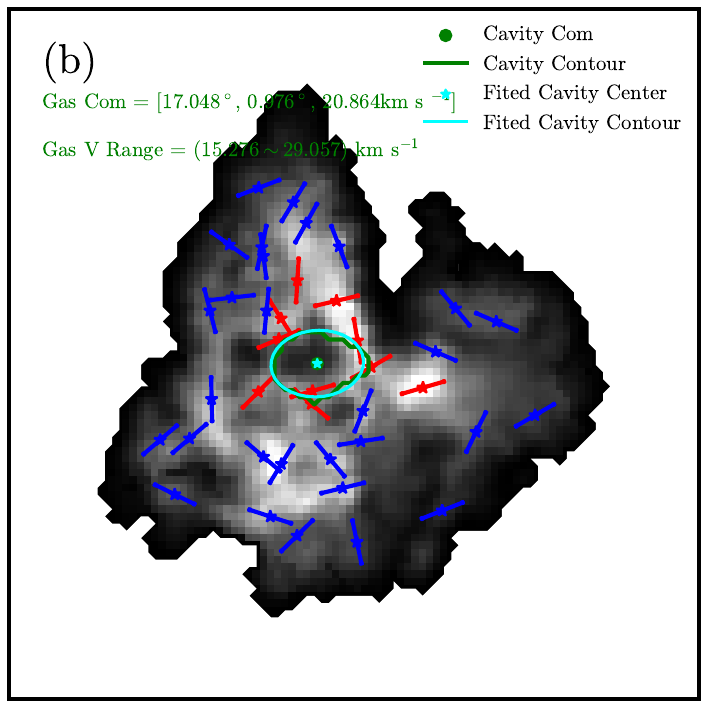}}
    %		\vspace{-4cm}
    %		\centerline{(a)}
\end{minipage}%
% \begin{minipage}[t]{0.3\textwidth}
%     \centering
%     \centerline{\includegraphics[width=2in]{Images/Imgs_Mor/Region_Slice_N19_1.pdf}}
%     %		\vspace{-4cm}
%     %		\centerline{(a)}
% \end{minipage}
\caption{Gas distribution and profile-extraction geometry for N19--A.
(a) Velocity-integrated molecular emission. The green contour and dot mark the boundary and centroid of the weight-clump, and the cyan ellipse shows the centroid-constrained fitted inner boundary. Gas clumps directly associated with the weight-clump are indicated by red stars and short line segments, while a representative clump boundary is outlined in gold; the gas centroid and velocity range are annotated.
(b) Hierarchical connections of the surrounding gas. Red markers denote clumps directly linked to the inner weight-clump, and blue markers indicate clumps connected to the red clumps at the next level. The intensity-weighted mean velocity of the red and blue clumps is adopted as the systemic velocity of the bubble.
}
\label{Img_Gas_Region_N19A}
\end{figure*}

We introduce the BWFields pipeline for molecular-bubble identification and characterization, schematically illustrated in Figure~\ref{Img_Pipeline}. 
First, a three-dimensional bubble-weight field $W_{l,b,v}$ (Figures~\ref{Img_Weighted_Clumps}(a) and (b)) is constructed to encode cumulative evidence for cavity-like morphology across multiple signal-to-noise tiers and velocity-integration scales. 
Cavity interiors are then segmented as \emph{weight-clumps} (Figure~\ref{Img_Weighted_Clumps}(c) and Figure~\ref{Img_Weighted_Clump_N19A}), from which a centroid-constrained geometric reference ellipse is derived for each candidate and associated with the surrounding molecular gas clumps (Figure~\ref{Img_Gas_Region_N19A}).

Based on this initial geometric reference, shell morphology is first characterized using ensembles of radial intensity profiles and constrained double-Gaussian fitting (Figure~\ref{Img_Intensity_Profile_N19A}(a) and (b)), yielding preliminary estimates of the shell radius and thickness. 
Using these parameters as a guide, an emission-defined intensity skeleton is then extracted via azimuth--radius unwrapping and graph-based ridge tracing (Figure~\ref{Img_Intensity_Skt_N19A}), yielding an emission-based representation of the shell structure. 

An ellipse fitted to the skeleton is subsequently used to perform a second radial-profile analysis, yielding refined estimates of the shell morphological parameters (Figure~\ref{Img_Intensity_Profile_N19A}(c) and (d)). These updated parameters are then adopted in the kinematic analysis. 
Bubble kinematics are quantified using azimuthally sampled PV-slice diagnostics, including a turbulence-normalised expansion significance (Figure~\ref{Img_PV_Analysis_N19A}). 

% Throughout this paper, the structures identified by BWFields are referred to as \textit{bubble candidates}, with \textit{cavity} and \textit{shell} denoting the interior depression and surrounding emission, respectively. Verified cases presented in the Appendix are referred to simply as \textit{bubbles}.

The present implementation of BWFields primarily targets enclosed cavities, for which the interior depression is topologically well defined. Nevertheless, the subsequent morphological and kinematic analyses can also be applied to partially open or deformed shells. The BWFields code is publicly available on GitHub\footnote{\href{https://github.com/JiangYuTS/BWFields}{https://github.com/JiangYuTS/BWFields}}, distributed as a Python package on PyPI\footnote{\href{https://pypi.org/project/BWFields}{https://pypi.org/project/BWFields}}, and archived on Zenodo \citep{BWFieldsZenodo}. Documentation and usage examples are available in the repository, and the package will be continuously maintained, with community contributions welcome.

\subsection{Bubble Weight Estimation in PPV Space}
\label{sec:method}

A cavity detected at higher SNRs and within narrower velocity intervals is generally more likely to represent a bubble. This motivates a weighted representation that accumulates cavity evidence across SNR levels and velocity-integration scales, such that the relative weight distribution, instead of its absolute magnitude, traces the most probable bubble interiors.

In BWFields, this concept is implemented through a three-dimensional bubble-weight field
\begin{equation}
\label{Weight_Cal}
W_{l,b,v} =
\sum_{i}\sum_{j}\sum_{k}
\frac{\mathrm{SNR}_i}{\Delta V_j}\,
\mathfrak{1}_{B_i^{(j,k)}(l,b,v)} ,
\end{equation}
where $W_{l,b,v}$ denotes the cumulative weight assigned to voxel $(l,b,v)$.
Here, $\mathrm{SNR}_i$ is the signal-to-noise ratio of the $i$th intensity tier, spanning from the \texttt{Threshold} to the maximum SNR, and $\Delta V_j$ is the $j$th velocity-integration width, ranging from the native channel spacing to the maximum scale \texttt{SliceDV}. The index $k$ labels the slab position at the current velocity-integration width. The indicator function $\mathfrak{1}_{B_i^{(j,k)}(l,b,v)}$ equals unity if the voxel $(l,b,v)$ lies within a cavity, and zero otherwise. The weighting factor $\mathrm{SNR}_i / \Delta V_j$ thus acts as a proxy for both contrast and kinematic coherence, preferentially enhancing high-contrast cavities detected at higher SNR and confined to narrower velocity intervals, as expected for kinematically coherent bubble interiors.

Let $N_v$ denote the total number of spectral channels along the velocity axis. 
For each velocity-integration width $\Delta V_j$, the spectral axis is partitioned into a sequence of centerd, nonoverlapping velocity slabs,
\begin{equation}
s_k^{(j)} = o_j + k\,\Delta V_j, \qquad k = 0,1,\dots,K_j ,
\end{equation}
with
\begin{equation}
o_j = \left\lfloor \frac{N_v \bmod \Delta V_j}{2} \right\rfloor,
\qquad
K_j = \left\lfloor \frac{N_v - o_j}{\Delta V_j} \right\rfloor .
\end{equation}
This construction partitions the usable velocity range into symmetrically distributed half-open intervals $\left[ s_k^{(j)}, s_{k+1}^{(j)} \right)$, ensuring balanced coverage along the velocity axis. 

For each combination of SNR tier $i$, velocity scale $j$, and slab position $k$, the emission within the corresponding velocity slab is integrated along the spectral axis to produce a two-dimensional projection. In this projection, cavity interiors are identified using morphological hole-filling \citep{scikit-image} and defined as $B_i^{(j,k)}$, which are then used to construct the indicator function $\mathfrak{1}_{B_i^{(j,k)}(l,b,v)}$. The corresponding weight contributions associated with these cavity regions are subsequently mapped back into PPV space over the same velocity interval, contributing to the weight field at the specific $(i,j,k)$ configuration. The final bubble-weight field is obtained by aggregating these contributions over all SNR tiers, velocity scales, and slab positions. 

Adopting an equivalent global detection scheme analogous to that of \cite{MWISP_Clumps}, all operations are restricted to signal-bearing subcubes to ensure computational tractability, improve efficiency, and suppress contamination from unrelated line-of-sight emission. The resulting local bubble-weight fields are subsequently merged into a global cube for further analysis.

Figure~\ref{Img_Weighted_Clumps}(a) shows the velocity-integrated molecular gas map, where regions of enhanced bubble weight closely trace the observed cavities. 
Figure~\ref{Img_Weighted_Clumps}(b) presents the integrated bubble-weight map, in which the bubble-weight field selectively enhances cavity interiors relative to their surroundings. 
The resulting weight distribution often exhibits locally clustered structures, which may arise when multiple distinct bubbles in PPV space overlap or become connected in projection.

\begin{figure*}
\centering
\centerline{\includegraphics[width=7in]{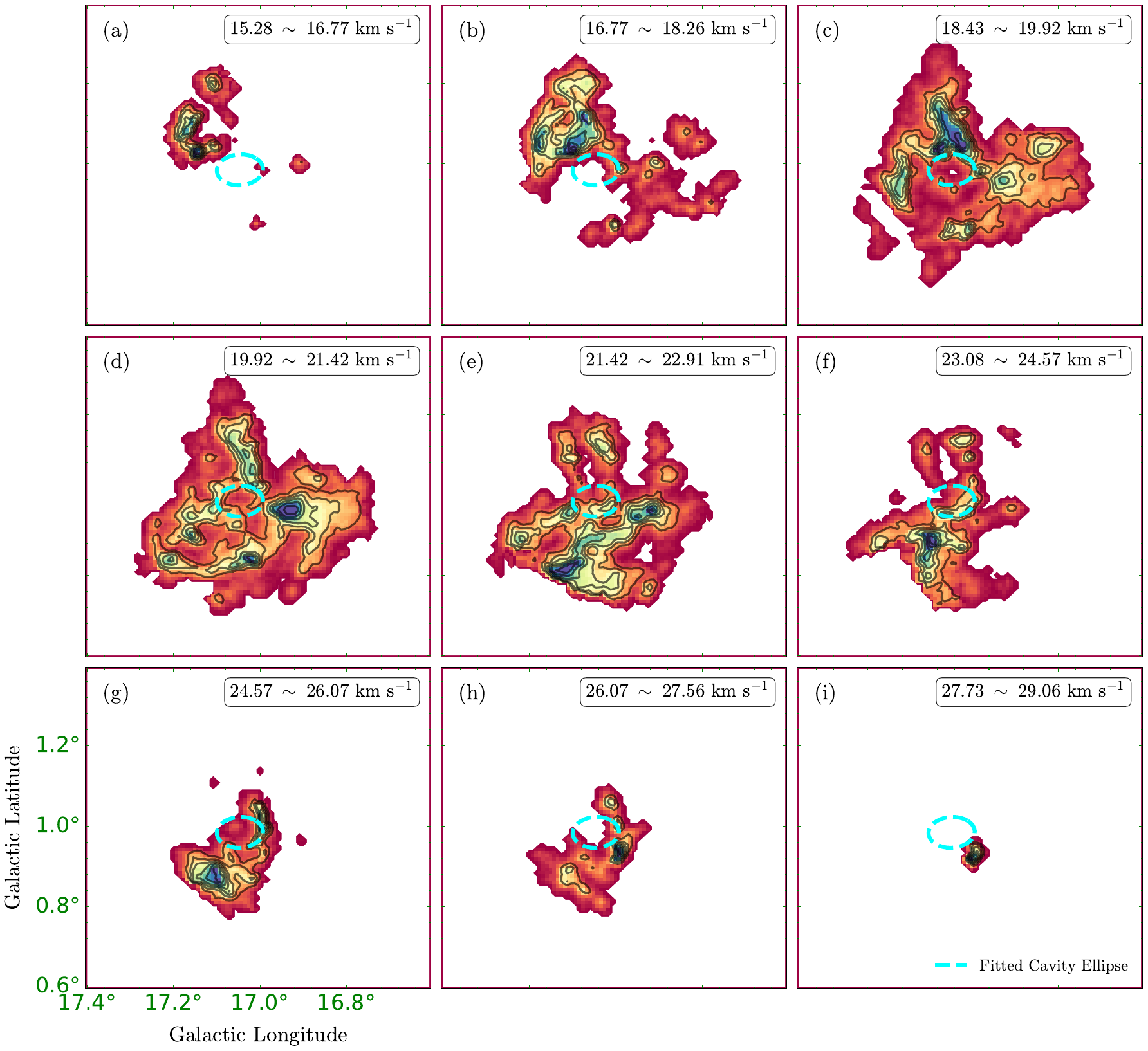}}
\caption{Channel maps of N19--A across consecutive velocity intervals. 
Each panel shows emission integrated over a uniform velocity width, revealing the evolution of gas morphology with velocity. 
The gas exhibits a shell-like structure, with enhanced emission along the shell and a comparatively depleted interior that persists across multiple channels. 
The cyan ellipse marks the centroid-constrained fitted cavity boundary of the bubble.}
\label{Img_Channel_Map_N19A}
\end{figure*}

\subsection{Weight-clump Detection and Associated Molecular Gas}
\label{sec:bubble_identification}

The bubble-weight field $W_{l,b,v}$ defines a continuous measure of the likelihood that each voxel belongs to a bubble cavity. 
The weight typically increases toward the cavity center, reflecting expanding shells with depleted interiors surrounded by diffuse emission. 
This distribution resembles that of molecular clumps, in which dense regions are embedded within extended emission. 
Clump-identification algorithms can therefore be adapted to extract contiguous cavity interiors from the weight cube.

We apply the clump-identification algorithm \texttt{FacetClumps}\footnote{For detailed descriptions of the algorithm and its application in MWISP molecular gas, please refer to \cite{FacetClumps} and \cite{MWISP_Clumps}} to $W_{l,b,v}$, partitioning the weight field into weight-clump candidates. 
Because $W_{l,b,v}$ aggregates evidence across multiple velocity-integration scales, a single cavity may produce several partially overlapping candidates. 
These detections are therefore consolidated through a local merging procedure to recover cavity interiors. 

Merging is restricted to candidates associated with the same weight regions, as delineated by the contours in Figure~\ref{Img_Weighted_Clumps}(b). 
For each weight-clump candidate $C_i$, we define its projected footprint on the $(l,b)$ plane as
\begin{equation}
S_i=\{(l,b)\,|\, (l,b,v)\in C_i\}.
\end{equation}
For a candidate pair $(i,j)$, we estimate the overlap ratio
\begin{equation}
\mathrm{OR}(i,j)=\frac{|S_i \cap S_j|}{\min\left\{ |S_i|, |S_j| \right\}},
\end{equation}
which measures the fractional overlap between the two projected footprints relative to the smaller one.

A candidate $C_i$ is merged into $C_j$ when $\mathrm{OR}(i,j)$ exceeds \texttt{MergeOR}. This criterion suppresses artificial fragmentation while preventing mergers between spatially disjoint structures. The resulting consolidated regions are referred to as weight-clumps, as shown in Figure~\ref{Img_Weighted_Clumps}(c). An example of a cavity (N19--A\footnote{This is an original weight-clump candidate that underwent the merging process but did not merge with any other candidates. Further investigation of N19 will be presented in a future study.}) is shown in Figure~\ref{Img_Weighted_Clump_N19A}, illustrating how the weight-clump effectively captures both the spatial distribution and kinematic range of the cavity.

For each cavity we compute the weighted centroid $(l_{cav},b_{cav},v_{cav})$,
\begin{equation}
(l_{cav},b_{cav},v_{cav})=
\frac{\sum_{(l,b,v)\in\mathcal{B}} (l,b,v)\, W_{l,b,v}}
{\sum_{(l,b,v)\in\mathcal{B}} W_{l,b,v}},
\end{equation}
where $\mathcal{B}$ denotes the voxel set of the corresponding weight-clump. 

To associate cavities with their molecular environment, the three-dimensional cavity mask $\mathcal{B}(l,b,v)$ is cross-matched with the molecular gas clump mask identified by \texttt{FacetClumps}. 
Gas clumps overlapping with $\mathcal{B}$ in PPV space are classified as directly associated (Figure~\ref{Img_Gas_Region_N19A}(a)), whereas those linked through contiguous emission delineate the surrounding envelope (Figure~\ref{Img_Gas_Region_N19A}(b)). 
The systemic velocity of the bubble is defined as the intensity-weighted mean velocity of the associated gas (Figure~\ref{Img_Gas_Region_N19A}(b)).

To describe the projected morphology, a geometric reference for the cavity is obtained by fitting a two-dimensional ellipse to the projected boundary of the weight-clump, with the center fixed at $(l_{cav},b_{cav})$. This fixed-center fitting approach minimizes contour variations due to weight threshold selection. The ellipse is parameterized by semi-major axis $a_{cav_r}$, semi-minor axis $b_{cav_r}$, and position angle $\theta$,
\begin{equation}
\frac{x'^{\,2}}{a_{cav_r}^2}+\frac{y'^{\,2}}{b_{cav_r}^2}=1,
\end{equation}
where $(x',y')$ are coordinates rotated by $\theta$ about $(l_{cav},b_{cav})$. The fitted ellipse (Figure~\ref{Img_Gas_Region_N19A}) defines the characteristic cavity size, elongation, and orientation used in subsequent radial-profile and PV-slice analyses.

Finally, channel maps (Figure~\ref{Img_Channel_Map_N19A}) are inspected to confirm that the cavity persists across the identified velocity interval, supporting the expanding-shell interpretation and helping to distinguish it from neighboring structures in crowded regions.

\begin{figure*}
\centering
\vspace{0cm}
\begin{minipage}[t]{0.4\textwidth}
    \centering
    \centerline{\includegraphics[width=2in]{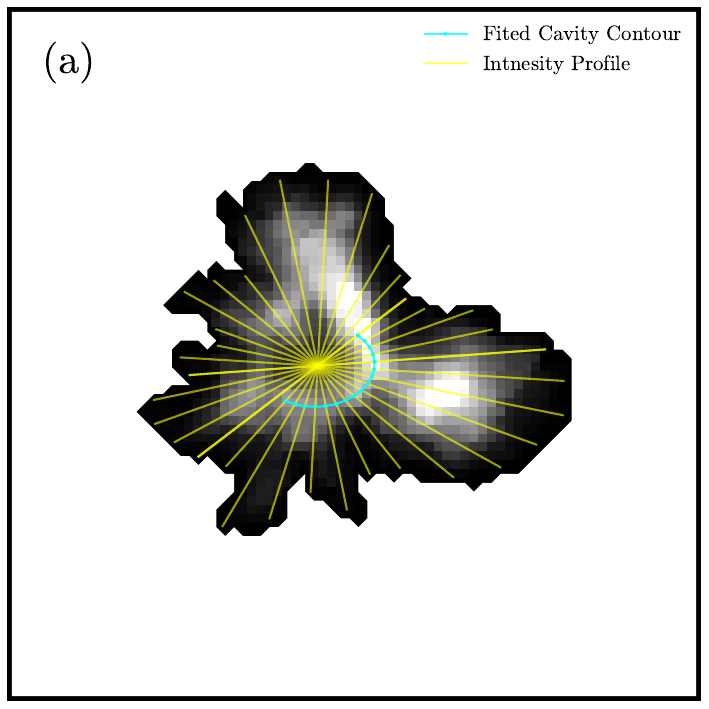}}
    %		\vspace{-4cm}
    %		\centerline{(a)}
\end{minipage}
\begin{minipage}[t]{0.4\textwidth}
    \centering
    \centerline{\includegraphics[width=2.8in]{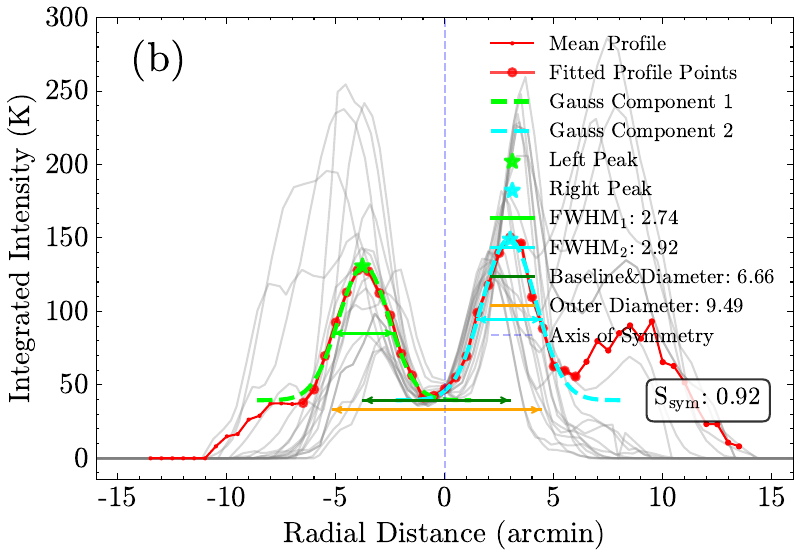}}
    %		\vspace{-4cm}
    %		\centerline{(a)}
\end{minipage}%

\begin{minipage}[t]{0.4\textwidth}
    \centering
    \centerline{\includegraphics[width=2in]{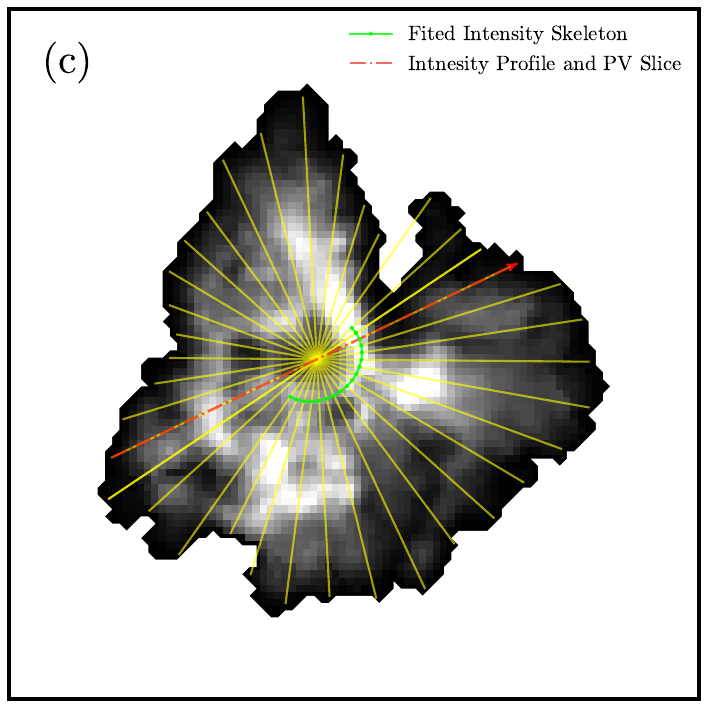}}
    %		\vspace{-4cm}
    %		\centerline{(a)}
\end{minipage}
\begin{minipage}[t]{0.4\textwidth}
    \centering
    \centerline{\includegraphics[width=2.8in]{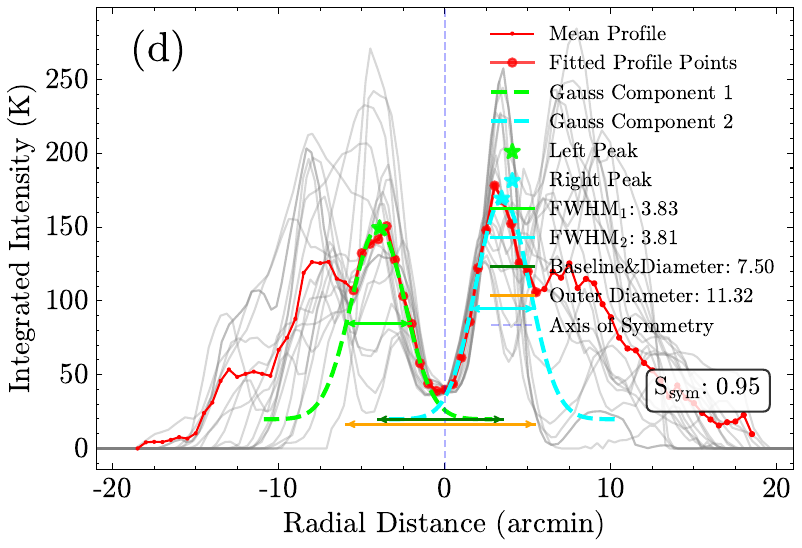}}
    %		\vspace{-4cm}
    %		\centerline{(a)}
\end{minipage}%
\caption{Radial sampling scheme used for the intensity-profile and PV analyses. (a) Radial sampling scheme for the first intensity-profile analysis. Yellow rays represent radial cuts passing through the intensity-weighted centroid of weight-clumps, intersecting reference points along the fitted cavity contour. Intensity profiles are extracted within the gas-clump mask shown in Figure~\ref{Img_Gas_Region_N19A}(a) to reduce contamination. (b) Radial-velocity-integrated intensity profiles and constrained double-Gaussian fitting. Thin gray curves show individual radial profiles from panel (a). The red curve denotes the mean profile, with red circles indicating the sampled points used for fitting. Cyan and light-blue dashed curves represent the two Gaussian components of the best-fit model, and green stars mark the positions of the two shell peaks. Horizontal bars indicate the FWHM (thickness) of each component, and arrows denote the fitted inner (radius and baseline) and outer shell diameters. The boxed values report the overall symmetry score derived from the fitted shell descriptors. 
(c) Radial sampling scheme for the second radial-profile analysis and PV analysis, based on the fitted ellipse from the intensity skeleton as described in Section \ref{sec:unwrap}. Intensity profiles are extracted within the extended gas-clump mask shown in Figure~\ref{Img_Gas_Region_N19A}(b). A representative PV slice is shown in light red, and the arrow represents the direction of the slice.
(d) Radial-velocity-integrated intensity profiles and constrained double-Gaussian fitting from the second analysis. The labeling and fitting procedure are identical to those in panel (b). This second profile refines the shell characteristics, enabling updated estimates of the radius and thickness for the kinematic analysis.
}
\label{Img_Intensity_Profile_N19A}
\end{figure*}

\subsection{Radial Cuts and Constrained Double-Gaussian Fitting}
\label{sec:double_gauss_profiles}

To quantify the shell brightness distribution and derive initial morphological parameters, we construct radial intensity profiles from the velocity-integrated emission. These profiles are averaged and modeled with a constrained double-Gaussian function, enabling measurements of the characteristic shell radius, thickness, and symmetry.

Radial cuts are defined as illustrated in Figure~\ref{Img_Intensity_Profile_N19A}(a). 
Each cut passes through the intensity-weighted centroid $(l_{cav},b_{cav})$ and intersects reference points along the fitted cavity ellipse. 
Sampling begins at the maximum local intensity within a $3\times3$ neighborhood around each ellipse reference point and proceeds with a step size of 1 pixel. 
The total number of cuts is half the ellipse circumference, ensuring approximately uniform azimuthal coverage. 
To suppress contamination from weakly related diffuse emission, the profiles are restricted to the gas-clump mask shown in Figure~\ref{Img_Gas_Region_N19A}(a).

All individual profiles are resampled onto a common radial grid and aligned such that $x=0$ corresponds to the central intensity minimum between the two shell peaks. 
The mean profile $I_{\rm mean}(x)$ is then computed, representing the global radial structure of the shell, while the dispersion among individual profiles reflects azimuthal variations in brightness and thickness.

The mean profile is modeled as the sum of two Gaussian components plus a constant baseline:
\begin{equation}
\begin{aligned}
I(x) = &A_1 \exp\!\left[-\frac{1}{2}\left(\frac{x-\mu_1}{\sigma_1}\right)^2\right]\\
+ &A_2 \exp\!\left[-\frac{1}{2}\left(\frac{x-\mu_2}{\sigma_2}\right)^2\right]
+ C ,
\end{aligned}
\label{eq:double_gauss}
\end{equation}
where $\mu_i$, $\sigma_i$, and $A_i$ denote the position, width, and amplitude of the two shell components, and $C$ represents the local baseline.

The fitting is performed on the portions of the mean profile corresponding to the two dominant peaks above the central minimum, thereby focusing on the shell emission. 
During nonlinear least-squares optimization (Levenberg--Marquardt), the following constraints are imposed: the two peaks lie on opposite sides of the center ($\mu_1<0$, $\mu_2>0$), amplitudes are positive, and the widths are smaller than the peak separation.

From the best-fit parameters, we define the characteristic shell radius and thickness as
\[
R_{1} = \frac{1}{2}\left(|\mu_1| + |\mu_2|\right), \qquad
T_{1} = \frac{1}{2}\left({\rm FWHM}_1 + {\rm FWHM}_2\right),
\]
where ${\rm FWHM}_i = 2.355\,\sigma_i$. 
These quantities provide the initial morphological reference for the azimuth–radius unwrapping and intensity skeleton extraction described in Section~\ref{sec:unwrap}.

To quantify the symmetry of the double-Gaussian profile, we define a dimensionless symmetry score for any shell descriptor $q$ measured on the two sides of the shell:
\begin{equation}
{\rm S_{sym_i}}(q_1,q_2) = 1 - \frac{\left| |q_1| - |q_2| \right|}{|q_1| + |q_2|}.
\label{eq:si_general}
\end{equation}

We evaluate ${\rm S_{sym_i}}$ for the fitted peak amplitude, peak position, and shell width, and adopt their arithmetic mean as the overall symmetry score $S_{\rm sym}$. 
If $S_{\rm sym}$ falls below the threshold parameter \texttt{SymScore}, the shell is considered insufficiently symmetric for reliable profile-based characterization. 
In this case, both the characteristic radius and thickness used in subsequent intensity-based unwrapping are equal to the semi-major axis of the fitted cavity ellipse, $a_{\rm cav}$.

Uncertainties in the derived shell parameters are estimated from the covariance matrix of the double-Gaussian fit (Equation~\ref{eq:double_gauss}) and propagated to $R_1$ and $T_1$ through the fitted peak positions and widths. The uncertainties in the individual descriptors are further propagated to the symmetry score $S_{\rm sym}$.

\begin{figure*}
\centering
\vspace{0cm}
\begin{minipage}[t]{0.4\textwidth}
    \centering
    \centerline{\includegraphics[width=2.5in]{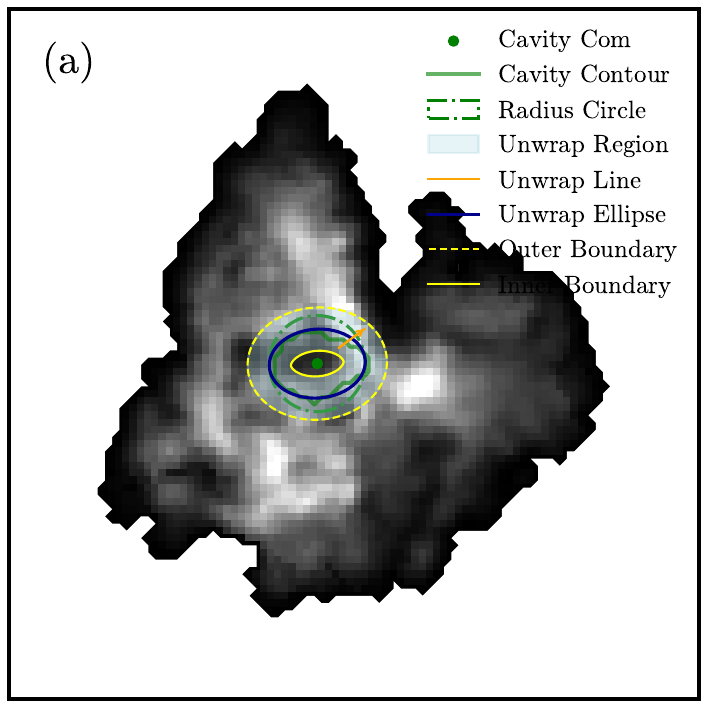}}
    %		\vspace{-4cm}
    %		\centerline{(a)}
\end{minipage}%
\begin{minipage}[t]{0.4\textwidth}
    \centering
    \centerline{\includegraphics[width=2.5in]{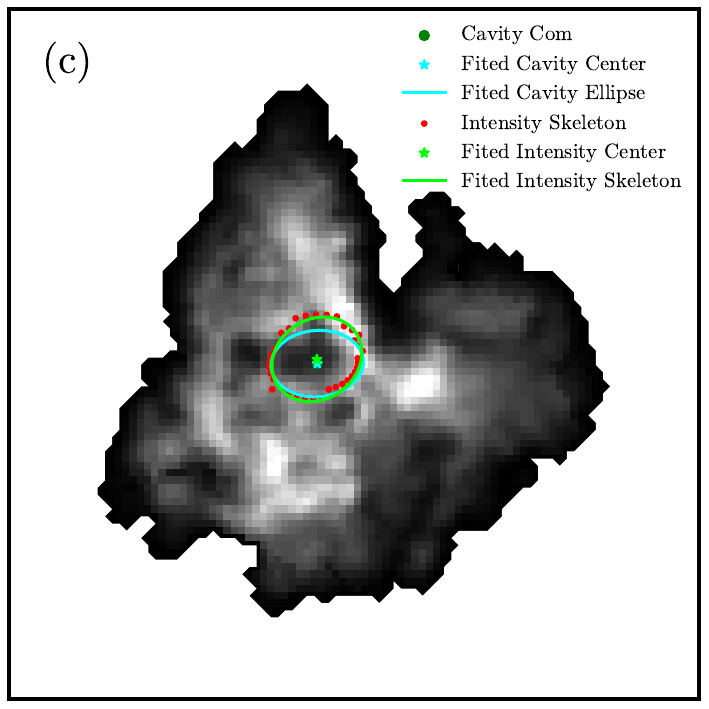}}
    %		\vspace{-4cm}
    %		\centerline{(a)}
\end{minipage}

\begin{minipage}[t]{0.3\textwidth}
    \centering
    \centerline{\includegraphics[width=6in]{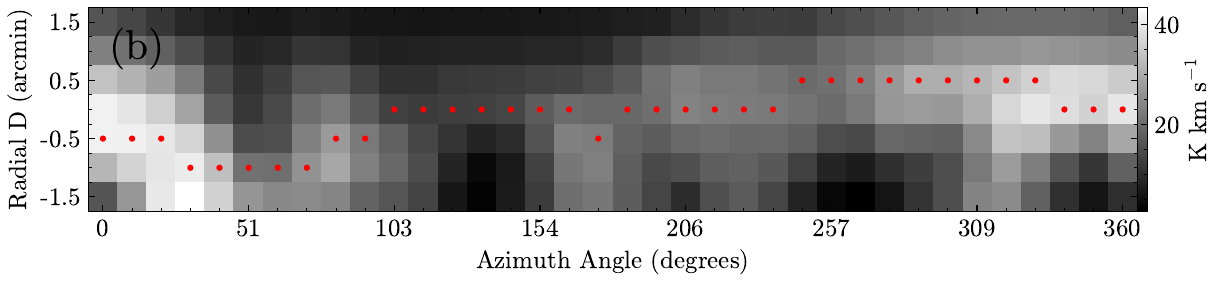}}
    %		\vspace{-4cm}
    %		\centerline{(a)}
\end{minipage}%
\caption{Azimuth-radius unwrapping and intensity-based skeleton extraction for a representative molecular bubble.
(a) Definition of the unwrapping annulus centerd on the intensity-weighted centroid of the weight-clump.
The unwrapping central line is defined by a reference ellipse whose semi-major axis equals the characteristic bubble radius, and whose axis ratio is fixed to that of the fitted ellipse shown in Figure~\ref{Img_Gas_Region_N19A}.
The radial width of the unwrapping annulus is equal to the shell thickness inferred from the radial-profile analysis, as indicated by the shaded region.
The yellow curves mark the inner and outer boundaries of the unwrapping annulus.
(b) Unwrapped velocity-integrated intensity map in azimuth--radius coordinates.
Red points trace the intensity skeleton, which follows the brightest ridge of emission around the shell.
(c) Reprojection of the skeleton into the sky plane.
The red points mark the recovered emission-defined shell intensity skeleton, and the cyan ellipse shows the best-fitting ellipse to the skeleton.}
\label{Img_Intensity_Skt_N19A}
\end{figure*}

\subsection{Azimuth-Radius Unwrapping and Intensity Skeleton Extraction}
\label{sec:unwrap}

To quantify azimuthal variations of molecular shells beyond a parametric ellipse description, each bubble is transformed into an elliptical unwrapped representation, from which the \emph{intensity skeleton} is extracted as an emission-defined ridge tracing the brightest locus of the velocity-integrated CO emission. 
Unlike the cavity ellipse inferred from the weight-clump, which describes the interior depression, the skeleton follows the shell crest and captures azimuthal asymmetry and shell deformation (Figure~\ref{Img_Intensity_Skt_N19A}).

An elliptical unwrapping band is defined around the intensity-weighted centroid $(l_{cav},b_{cav})$ (Figure~\ref{Img_Intensity_Skt_N19A}(a)). 
The reference ellipse has a semi-major axis equal to the characteristic bubble radius, with the axis ratio and position angle inherited from the fitted cavity ellipse:
\[
a_{\rm unwrap} = R_{\rm 1}, \qquad 
b_{\rm unwrap} = R_{\rm 1}\,(b_{cav_r}/a_{cav_r}).
\]
The half-width of the unwrapping band is set by the shell thickness inferred from the radial profile,
\[
s \in [-T_{\rm 1}/2,\,T_{\rm 1}/2],
\]
where $s$ is the signed offset along the ellipse normal.

Parameterizing the ellipse by $t\in[0,2\pi)$ with position angle $\theta$, the reference ellipse is
\begin{equation}
\begin{aligned}
x_{\rm e}(t) &= l_{cav} + a_{\rm unwrap}\cos t \cos\theta - b_{\rm unwrap}\sin t \sin\theta, \\
y_{\rm e}(t) &= b_{cav} + a_{\rm unwrap}\cos t \sin\theta + b_{\rm unwrap}\sin t \cos\theta.
\end{aligned}
\end{equation}

The velocity-integrated intensity map $I(l,b)$ is sampled along the local normal direction to form the unwrapped representation:
\begin{equation}
I_{\rm unwrap}(t,s) = I\!\bigl(x_{\rm e}(t)+s\,\hat n_x(t),\,y_{\rm e}(t)+s\,\hat n_y(t)\bigr),
\end{equation}
where $\hat{\mathbf n}(t)$ is the unit normal of the ellipse. Sampling is performed approximately uniformly along the ellipse perimeter and the normal direction using bilinear interpolation. The starting azimuth is offset to align with the locally brightest point on the ellipse boundary.

Figure~\ref{Img_Intensity_Skt_N19A}(b) shows the resulting unwrapped intensity map. 
Ridge extraction is formulated as a graph-optimisation problem \citep{DPConCFil}, in which non-zero pixels are treated as nodes and neighboring pixels with $1 \le d_{ij} \le \sqrt{2}$ are connected by edges. 
Each edge $(i,j)$ is assigned a cost
\begin{equation}
w_{ij} = \frac{d_{ij}}{\tilde I(\mathbf{x}_i) + \tilde I(\mathbf{x}_j)},
\end{equation}
where $d_{ij}$ is the Euclidean distance between the neighboring pixels, and $\tilde I(\mathbf{x}_i)$ and $\tilde I(\mathbf{x}_j)$ are the integrated intensities at positions $\mathbf{x}_i$ and $\mathbf{x}_j$, respectively. 
Accordingly, brighter and more continuous emission is assigned lower cost. 
A minimum spanning tree (MST) is then constructed to preserve connectivity while minimising the total cost. 

If the unwrapped intensity map contains multiple connected regions, ridge extraction is performed independently within each region that intersects the central radial line ($s=0$). 
Candidate ridge paths are searched in the MST between boundary node pairs located on the two azimuthal edges of the region (corresponding to $t=0$ and $t=2\pi$). 
These node pairs are required to share the same radial index (i.e., the same row in the unwrapped image), which enforces continuity across the azimuthal seam and promotes the formation of a nearly closed skeleton upon reprojection.

For each candidate pair $(s,t)$, the corresponding path $P_{st}$ is evaluated using the cumulative brightness merit
\begin{equation}
\Phi(P_{st}) = \sum_{(i,j)\in P_{st}} \frac{\tilde I(\mathbf{x}_i) + \tilde I(\mathbf{x}_j)}{d_{ij}}.
\end{equation}
The path that maximizes $\Phi$ is selected as the skeleton. 
The resulting path is further refined by removing locally redundant nodes while preserving connectivity, and by bridging short gaps using nearby pixels where necessary. 

The skeleton is finally reprojected into the original sky plane through the inverse normal mapping
\begin{equation}
\mathbf r_{\rm sk}(t)=\mathbf r_{\rm e}(t)+s_{\rm sk}(t)\,\hat{\mathbf n}(t),
\end{equation}
where $s_{\rm sk}(t)$ is the radial offset of the ridge in the unwrapped domain. 

Figure~\ref{Img_Intensity_Skt_N19A}(c) shows the reconstructed skeleton and its best-fitting ellipse. 
This intensity-based unwrapping and skeleton-extraction method couples emission contrast with spatial continuity, enabling a unified characterization of the shell structure, including the intensity centroid $(l_{sk},b_{sk})$, the semi-major ($a_{sk}$) and semi-minor ($b_{sk}$) axes, and the position angle. 
Derived directly from the observed emission ridge, these parameters capture the intrinsic shell morphology and provide a quantitative basis for analyzing molecular-bubble azimuthal symmetry, curvature, and feedback-driven deformation.

\subsection{Refinement of Morphological Parameters}

After the initial shell characterization based on radial intensity profiles and constrained double-Gaussian fitting, we further refine the morphological parameters using the emission-defined intensity skeleton extracted in Section~\ref{sec:unwrap}. 
Because the intensity skeleton traces the brightest ridge of the shell emission, its fitted ellipse defines a more relevant geometric reference for the shell crest than the cavity-based ellipse, which describes the interior depression. 
This skeleton-based ellipse is therefore adopted as the reference geometry for a second radial-profile analysis.

The revised sampling scheme is shown in Figure~\ref{Img_Intensity_Profile_N19A}(c). 
Compared with the first analysis, the second set of radial cuts is defined with respect to the fitted skeleton ellipse, and the intensity profiles are extracted within the more extended gas-clump mask shown in Figure~\ref{Img_Gas_Region_N19A}(b). 
This mask captures shell-associated gas over a broader spatial and velocity range, allowing the full extent of the affected gas to be traced.

Figure~\ref{Img_Intensity_Profile_N19A}(d) presents the refined radial intensity profiles and their double-Gaussian fitting results. 
Compared with the first-pass profiles in Figure~\ref{Img_Intensity_Profile_N19A}(b), the refined profiles exhibit a larger dispersion among individual cuts and a less-prominent double-peaked structure. 
This behavior reflects the inclusion of more extended and structurally complex shell-associated emission. 
If such extended molecular gas were used directly in the first radial-profile analysis, the resulting morphological reference parameters would be more strongly affected by irregular outer emission and would therefore be less robust.

The second double-Gaussian fitting follows the same procedure described in Section~\ref{sec:double_gauss_profiles}. 
However, in this refinement step, both the initial parameter guesses and the fitting bounds are guided by the results of the first-pass fit. 
This strategy stabilizes the optimisation and allows reliable recovery of the shell parameters even when the refined mean profile is more asymmetric or when the two peaks are less clearly separated.

From the refined fit, we derive updated estimates of the shell radius $R_2$, shell thickness $T_2$, and symmetry score $S_{\rm sym_2}$. 
The corresponding outer radius is defined as
\[
R_{\rm out} = R_2 + T_2/2.
\]

These refined morphological parameters are adopted in the kinematic analysis of Section~\ref{sec:pv_method}, where anchoring the PV diagnostics to the emission-defined shell geometry yields more reliable measurements, particularly in crowded or irregular environments.

\begin{figure*}
\centering
\vspace{0cm}
\begin{minipage}[t]{0.32\textwidth}
    \centering
    \centerline{\includegraphics[width=2.2in]{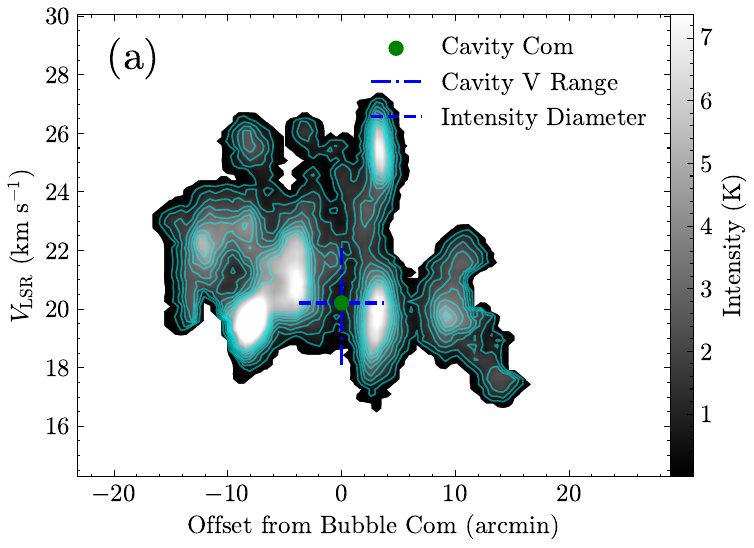}}
    %		\vspace{-4cm}
    %		\centerline{(a)}
\end{minipage}
\begin{minipage}[t]{0.32\textwidth}
    \centering
    \centerline{\includegraphics[width=2.2in]{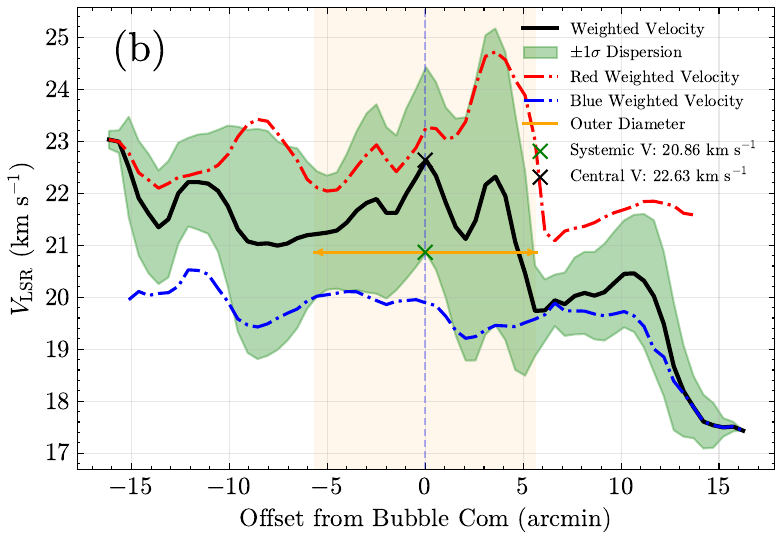}}
    %		\vspace{-4cm}
    %		\centerline{(a)}
\end{minipage}%
\begin{minipage}[t]{0.32\textwidth}
    \centering
    \centerline{\includegraphics[width=2.2in]{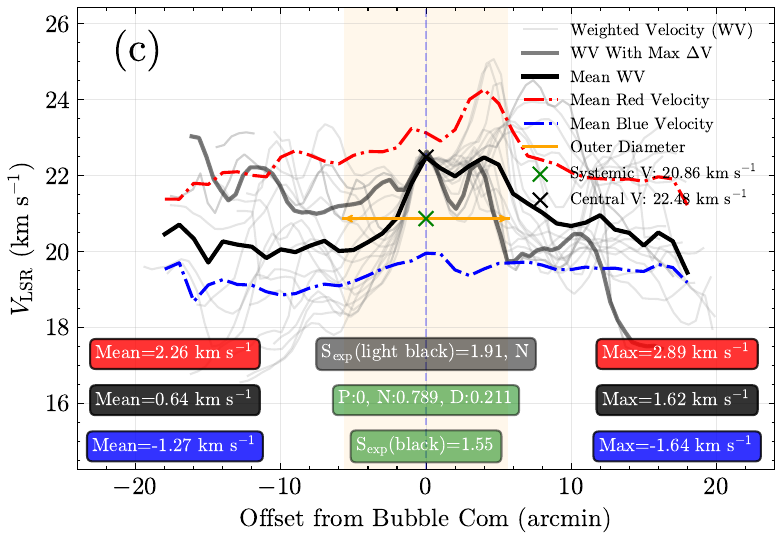}}
    %		\vspace{-4cm}
    %		\centerline{(a)}
\end{minipage}
\caption{PV analysis for N19--A. (a) PV diagram along the radial cut showing the largest deviation of the central velocity from the systemic value; contours trace $^{13}$CO emission, and the green dot marks the bubble centroid. 
(b) Intensity-weighted velocity profile from panel (a); the black curve is the mean velocity, the shaded region is the velocity dispersion, and red/blue dash-dotted curves are the mean redshifted/blueshifted components. Arrows indicate the outer radius adopted for expansion analysis. 
(c) Comparison of intensity-weighted mean-velocity profiles from all cuts (thin gray) and their azimuthal mean (thick black); the light-black curve reproduces panel (b). Colored boxes report mean and maximum velocity offsets for total (black), redshifted (red), and blueshifted (blue) components within the outer radius. The light-black boxed annotation gives $S_{\rm exp}$ and its expansion-velocity sign classification (N). Fractions $P$, $N$, and $D$ quantify azimuthal consistency of velocity signs, while $S_{\rm exp}$ summarizes the overall strength of the expansion-like pattern.
}
\label{Img_PV_Analysis_N19A}
\end{figure*}

\subsection{PV-slice Construction and Expansion Analysis}
\label{sec:pv_method}

To characterize the kinematics of each bubble and assess expansion-like signatures, we extract PV slices along the same radial directions used for the intensity-profile analysis (Figure~\ref{Img_Gas_Region_N19A}(c)). The extraction adopts the extended gas-clump mask in Figure~\ref{Img_Gas_Region_N19A}(b), tracing molecular material associated with the cavity and its immediate surroundings.

Each PV slice is constructed by sampling the CO cube along the radial path and averaging over a finite transverse width. 
The spatial coordinate is expressed as a signed offset $x$, with $x=0$ at the projected cavity centroid, defining a common reference frame for azimuthal comparisons.

From each slice, we derive the intensity-weighted mean velocity $\bar{v}(x)$ and velocity dispersion $\sigma_v(x)$. 
The emission is then decomposed relative to the systemic velocity $v_{\rm sys}$ into redshifted and blueshifted components, yielding $\bar{v}_{\rm red}(x)$ and $\bar{v}_{\rm blue}(x)$.

To obtain a representative shell velocity, all slice profiles are interpolated onto a common offset grid and azimuthally averaged, suppressing slice-to-slice fluctuations while retaining large-scale velocity trends. 
An example PV slice and velocity profile are shown in Figures~\ref{Img_PV_Analysis_N19A}(a) and (b), and the ensemble with azimuthal mean in Figure~\ref{Img_PV_Analysis_N19A}(c).

Expansion-like velocity offsets are quantified within $|x|\le R_{\rm out}$. 
For each slice, we derive the mean velocity on the near ($x\le 0$) and far ($x\ge 0$) sides relative to $v_{\rm sys}$, and summarize azimuthal consistency using three fractions: $P$ (both sides positive), $N$ (both negative), and $D$ (opposite signs). 
These fractions quantify the sign coherence around the cavity in a model-independent manner.

The expansion velocity is defined as
\begin{equation}
v_{\rm exp} = \max_{|x|\le R_{\rm out}}
\left(
|\bar{v}_{\rm red}(x)-v_{\rm sys}|,\;
|\bar{v}_{\rm blue}(x)-v_{\rm sys}|
\right),
\end{equation}
capturing the largest line-of-sight velocity separation while remaining robust to local irregularities and azimuthal variations.

To quantify the statistical significance of an expansion-like pattern relative to the local background, we define a turbulence-normalised expansion coefficient $S_{\rm exp}$ combining two diagnostics. The first term,
\begin{equation}
E_V = \frac{V_{\rm env}-V_c}{\sigma_{\rm env}},
\end{equation}
measures the depth of the central velocity depression relative to the surrounding shell. Here, $V_c$ is the mean velocity at the cavity center, $V_{\rm env}$ is the mean velocity beyond $R_{\rm out}$, and $\sigma_{\rm env}$ is the ambient dispersion, defined as the median of the intensity-weighted $\sigma_v$ along each radial cut in the outer region. 

The second term,
\begin{equation}
E_c = \frac{|c|\,R_{\rm out}^2}{\sigma_{\rm env}},
\end{equation}
quantifies the global curvature of the mean-velocity profile within $|x|\le R_{\rm out}$, where $c$ is the quadratic coefficient from fitting $V(x)=a+bx+cx^2$.  

The combined coefficient is
\begin{equation}
S_{\rm exp} = \sqrt{E_V^2+E_c^2},
\end{equation}
evaluated for the total, redshifted, and blueshifted profiles, with the maximum adopted as the final expansion significance.

Figure~\ref{Img_PV_Analysis_N19A}(c) summarizes these diagnostics, showing the azimuthally averaged velocity structure, $v_{\rm exp}$, the sign-consistency fractions ($P$, $N$, $D$), and $S_{\rm exp}$. 
Together, they enable a directly measurable characterization of bubble kinematics across the full sample.

The uncertainty in $v_{\rm exp}$ is derived from the velocity bin where the maximum offset occurs and the uncertainty in $v_{\rm sys}$. Since $S_{\rm exp}$ is normalised by the ambient velocity dispersion, it is sensitive to local turbulence and should be interpreted as a relative indicator of expansion-like behavior rather than a direct physical measure.

\subsection{Limitations of the BWFields}
\label{sec:limitations}

BWFields has several inherent limitations. First, the present implementation is primarily optimised for enclosed cavities with well-defined interior depressions. Partially open, fragmented, highly deformed, or strongly asymmetric shells may therefore be underweighted, fragmented into multiple candidates, or missed altogether.

Second, the morphological analysis assumes that cavities and shells can be reasonably described by ellipse-based reference geometries and radial profiles. While effective for relatively regular structures, this assumption may oversimplify irregular, broken, or multicomponent shells. In particular, the constrained double-Gaussian model is best suited to cases with two clearly defined radial emission peaks, and may become unreliable when the shell emission is strongly asymmetric, multipeaked, or significantly contaminated by diffuse background emission.

Finally, the kinematic diagnostics quantify expansion-like behavior rather than providing a unique confirmation of physical expansion. Velocity gradients, turbulence, cloud--cloud overlap, shear, or unrelated line-of-sight components can produce similar PV signatures. The derived $S_{\rm exp}$ should therefore be interpreted as a relative indicator of expansion strength, and ideally assessed in conjunction with channel-map inspection or multiwavelength observations.

Despite these limitations, BWFields is applicable to bubbles across a wide range of spatial scales. The identification relies on detecting cavity-like structures in PPV space; as long as a bubble exhibits a sufficiently coherent cavity signature, it can be captured. In addition, the intensity-skeleton extraction and PV-based kinematic analysis are not restricted to perfectly closed shells, and can be applied to partially fragmented or deformed structures\footnote{See the GitHub manual for applications to fragmented shells: \href{https://github.com/JiangYuTS/BWFields/blob/master/Manuals}{https://github.com/JiangYuTS/BWFields/Manuals}}. The performance of the method depends on the characteristics of the input data, including tracer, resolution, and sensitivity, and the algorithm can be further adapted and optimized to accommodate different datasets.

% Both aspects are demonstrated in complex regions such as M16 (Section~\ref{sec:application_g17}). 

\begin{table*}
\centering
\caption{Adopted input parameters of BWFields for the G17-region application.}
\begin{tabular}{lll}
\hline\hline
Parameter & Description & Value \\
\hline
$RMS$ & RMS noise level of the data (K) & 0.22 \\
$Threshold$ & Minimum intensity used to truncate the signals (K)& $5\times RMS$ \\
$SliceDV$ & Maximum velocity-slab width (pixels)& 12 ($\sim2~\mathrm{km\,s^{-1}}$) \\
$BubWeight$ & Minimum bubble-weight for retained weight-clump candidates& 4 \\
$BubSizeLBV$ & Minimum weight-clump size in the spatial direction and velocity channels (pixels, channels)& [16,\,5] \\
$MergeOR$ & Minimum footprint-overlap ratio for merging neighboring candidates& 0.4 \\
$SymScore$ & Minimum symmetry score of the radial intensity profile& 0.8 \\
$ExpSign$ & Threshold separating high and low turbulence-normalised expansion significance& 1 \\
\hline
\end{tabular}
\tablecomments{
Additional optional BWFields parameters and usage details are available in the
\href{https://github.com/JiangYuTS/BWFields/tree/master/Manuals}{manual}.
The \texttt{FacetClumps} parameters are adopted at their default values;
see \citealt{FacetClumps} for details.
}
\label{InputPar}
\end{table*}

\section{Application of BWFields in the G17 Region}
\label{sec:application_g17}

We apply the BWFields to the MWISP $^{13}$CO data cube of the G17 region, which serves as a representative testbed encompassing both relatively isolated shells and structurally complex feedback environments. 
This section summarizes the adopted parameter configuration and presents the spatial and kinematic distribution of the resulting bubble candidates.

\subsection{Adopted BWFields Parameters}
\label{sec:bwfields_parameters}

Table~\ref{InputPar} summarizes the principal parameters adopted for the BWFields analysis of the G17 region. 
These values define a stable working configuration for MWISP $^{13}$CO data, balancing sensitivity to genuine cavity structures against robustness to noise and projection effects. 
The adopted parameter set is further supported by the benchmark cases presented in Appendix~\ref{sec:appendix_bubbles}.

The first group of parameters controls the construction of the bubble-weight field and the segmentation of candidate cavities. 
The noise level (\texttt{RMS}) and intensity threshold (\texttt{Threshold}) define the minimum signal included in the weight-field accumulation. 
The velocity-slab width (\texttt{SliceDV}) sets the maximum integration range used to identify cavity-like depressions across velocity channels. 
The bubble-weight threshold (\texttt{BubWeight}) and minimum weight-clump size (\texttt{BubSizeLBV}) suppress noise-driven detections by requiring candidates to extend over both spatial and velocity dimensions. 
The merging parameter (\texttt{MergeOR}) combines partially overlapping weight-clumps that likely correspond to the same cavity.

The second group of parameters governs the morphological and kinematic filtering of the candidates. 
The symmetry threshold (\texttt{SymScore}) selects shells with approximately symmetric radial intensity profiles, while the turbulence-normalised expansion significance (\texttt{ExpSign}) identifies candidates whose velocity structure is distinguishable from the surrounding turbulent background.

An illustrative catalog of the identified candidates is presented in Appendix~\ref{sec:catalog}, including their morphological and kinematic properties and a confidence metric.

\begin{figure*}
\centering
\vspace{0cm}
\begin{minipage}[t]{0.4\textwidth}
    \centering
    \centerline{\includegraphics[width=6.5in]{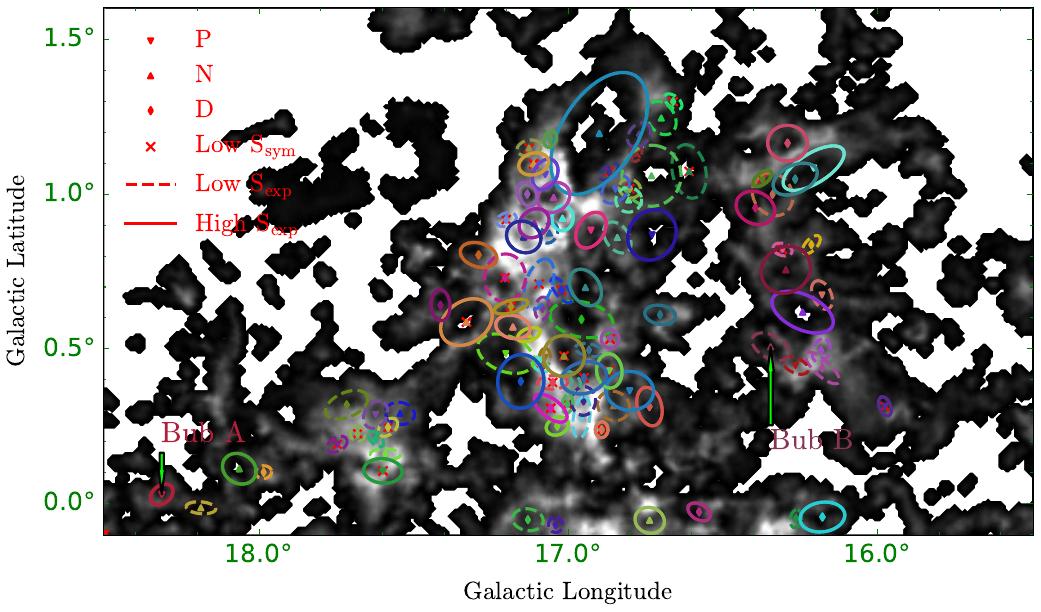}}
\end{minipage}
\caption{Bubble candidates identified by BWFields in the G17 region. A total of 93 candidates are detected. 
The background shows the velocity-integrated molecular emission, while colored contours trace the fitted intensity skeleton of each candidate. 
Central symbols mark the centroids of the fitted ellipses. 
Symbol shapes encode the PV sign-consistency class and whether the symmetry parameter $S_{\rm sym}$ falls below the adopted threshold. 
Different skeleton line styles distinguish candidates with relatively high and low $S_{\rm exp}$ values.
Bub~A and Bub~B denote two relatively independent bubble candidates, whose detailed analyses are presented in Appendix~\ref{sec:example_bubbles} (Figures~\ref{Img_Appendix_M16_New_MKSP} and \ref{Img_Appendix_M16_Without_MKSP}).}
\label{Img_Bubbles_M16}
\end{figure*}

\subsection{Distribution of Bubble Candidates in G17}
\label{sec:all_bubbles}

Figure~\ref{Img_Bubbles_M16} shows the bubble candidates identified in the G17 region. 
The grayscale background represents the velocity-integrated molecular emission, while colored contours trace the fitted intensity skeletons. 
Central symbols mark the ellipse centroids. 
Symbol shapes indicate the symmetry and PV sign-consistency class, and line styles distinguish candidates with relatively high or low turbulence-normalised expansion significance.

The identified candidates span a wide range of morphologies, including compact cavities, elongated shells, and partially fragmented structures. 
This diversity reflects the varied physical conditions in the G17 field, from relatively isolated systems to crowded regions with overlapping shell-like features.

The EoC region illustrates how the PPV-based analysis separates structures that appear similar in projection. 
In velocity-integrated emission, the two cavities appear nearly symmetric (Figure~\ref{Img_Origin_Data}), whereas the BWFields analysis reveals distinct PPV structures. 
The northern component is largely recovered as a single candidate, while the southern component is decomposed into multiple candidates with different geometries.

A similarly complex configuration is found around the infrared bubble N19, where the surrounding molecular gas exhibits multiple cavity-like depressions. 
BWFields identifies several associated candidates, allowing individual shell structures to be distinguished despite strong projection effects. 
These systems may represent either bubble groups composed of multiple interacting cavities, or single bubbles with internal kinematics more complex than typically assumed. 
In this work, the N19 and EoC regions serve as methodological demonstrations; a detailed analysis will be presented in a future study.

Beyond these crowded regions, Figure~\ref{Img_Bubbles_M16} also includes relatively independent candidates. 
Two examples, Bub~A and Bub~B, are marked in the figure and examined in Appendix~\ref{sec:example_bubbles}. 
Both lack associated O-, B-, or A-type stars within their identified boundaries, but they differ in their multiwavelength appearance.

Overall, the G17 application demonstrates that BWFields recovers a heterogeneous population of bubble candidates across a realistic Galactic-plane field, including isolated systems, projected overlaps separable in velocity space, and shells embedded in structured molecular gas. 

\section{Kinematic Interpretation of the BWFields Candidate Population}
\label{sec:kinematic_interpretation}

The bubble candidates identified in the G17 field exhibit a wide range of velocity structures. 
Their kinematic behavior is characterized using the PV sign-consistency classes ($P$, $N$, and $D$) together with the turbulence-normalised expansion significance $S_{\rm exp}$. 
These quantities describe how the shell velocity field is organised relative to the systemic velocity and how prominently it stands out against the local turbulent background.

The $P$ and $N$ classes are typically associated with classical expansion-like signatures in PV space, often appearing as U- or V-shaped velocity structures. 
In these cases, both sides of the shell lie on the same side of the systemic velocity, producing predominantly blueshifted ($P$) or redshifted ($N$) offsets. 
Such patterns are consistent with expanding-shell geometries in which blueshifted and redshifted emission correspond to expansion centers located on the far and near side of the molecular cloud, respectively.

The $D$ class includes shells whose two sides exhibit opposite velocity signs relative to the systemic velocity. 
Although such patterns may arise from shell expansion, they are more sensitive to asymmetric shell geometry, one-sided expansion, or large-scale velocity gradients in the surrounding molecular gas. 
In practice, $D$-type candidates often trace environments where shell dynamics are coupled to the broader kinematic structure of the cloud.

The expansion significance $S_{\rm exp}$ complements the PV classification by quantifying the strength of velocity organisation relative to the ambient turbulent dispersion. 
High-$S_{\rm exp}$ candidates correspond to systems with dynamically organised velocity structures, regardless of whether their PV morphology appears symmetric ($P/N$) or asymmetric ($D$). 
Conversely, low-$S_{\rm exp}$ values may reflect weak shells, strong environmental confusion, or the influence of large-scale velocity gradients.

Because PV morphology alone does not uniquely constrain the origin of a cavity, interpreting individual candidates requires additional environmental context, including infrared or radio counterparts, associations with H\,{\sc ii} regions or OB stars, large-scale velocity gradients, and potential line-of-sight superposition of unrelated structures.

The benchmark bubbles presented in Appendix~\ref{sec:appendix_bubbles} illustrate these considerations. 
For example, N4 ($P$) and N75 ($N$) exhibit well-defined shell morphology together with independent radio counterparts, supporting a feedback-driven origin, whereas N74 ($D$) shows velocity patterns more strongly influenced by asymmetric or gradient-like kinematics. 
Taken together, these examples indicate that the $P/N/D$ classes and $S_{\rm exp}$ act as complementary descriptors of shell kinematics rather than direct indicators of origin. Robust physical interpretation generally requires complementary multiwavelength information.

\section{Conclusions}
\label{sec:conclusions}
We present the BWFields framework, a PPV-based method for identifying and characterizing molecular-bubble candidates directly from spectral-line data cubes. 
The method constructs a three-dimensional bubble-weight field that encodes topological evidence for cavity interiors and segments connected structures as weight-clumps. 
By associating cavities with surrounding molecular gas and integrating radial intensity profiles, emission-defined intensity skeletons, and azimuthally sampled PV diagnostics, BWFields enables a unified and physically interpretable description of shell morphology and kinematics. 

Applied to MWISP $^{13}$CO observations of the G17 region, BWFields recovers a diverse population of bubble candidates, including isolated cavities, projected overlaps that can be disentangled in velocity space, and shell-like structures embedded in complex molecular environments. 
These results demonstrate that jointly analysing morphology and kinematics in PPV space enables systematic identification and characterization of bubble candidates in crowded Galactic-plane regions.

BWFields establishes a scalable approach for molecular-bubble candidates in large spectral-line surveys. 
The present work focuses on the methodology and its application to a representative region, while a full Galactic-plane analysis based on MWISP data \citep{CO_Survey_MWISP_3} and the MWISP clump catalog \citep{MWISP_Clumps}, together with the resulting bubble-candidate catalog, will be presented in a companion paper.

\section*{Acknowledgements}
We are grateful to the anonymous referee for the invaluable insights and comments, which enabled us to refine and enhance this work. 
This work is supported by the National Key R\&D Program of China (grant No. 2023YFA1608000) and the National Natural Science Foundation of China (grant No. U2031202). X.C. and K.W. acknowledge support from the Tianshan Talent Training Program (2024TSYCTD0013), and the Tianchi Talent Program of Xinjiang Uygur Autonomous Region. K.W. additionally acknowledges funding from the National SKA Program of China (Grant No. 2025SKA0140100), and the National Natural Science Foundation of China (No. 12573025). This research makes use of the data from the Milky Way Imaging Scroll Painting (MWISP) project, which is a multiline survey in $^{12}$CO/$^{13}$CO/C$^{18}$O along the northern Galactic plane with the PMO 13.7m telescope. We are grateful to all the members of the MWISP working group, particularly the staff members at PMO 13.7m telescope, for their long-term support. MWISP was sponsored by the National Key R\&D Program of China with grants 2023YFA1608000 and 2017YFA0402701, and the CAS Key Research Program of Frontier Sciences with grant QYZDJSSW-SLH047. 

$Software$: NumPy \citep{Numpy}, Matplotlib \citep{Matplotlib}, Networkx \citep{Networkx}, Astropy \citep{Astropy_1,Astropy_2,Astropy_3}, Scikit-learn \citep{scikit-learn}, Scikit-image \citep{scikit-image}, SciPy \citep{SciPy}, FacetClumps \citep{FacetClumps_Sfot}, DPConCFil \citep{DPConCFil_Sfot}, BWFields \citep{BWFieldsZenodo}. 

\bibliography{MWISP_Bubbles.bib}

@INPROCEEDINGS{Review_2,
	author = {{Pineda}, J.~E. and {Arzoumanian}, D. and {Andre}, P. and {Friesen}, R.~K. and {Zavagno}, A. and {Clarke}, S.~D. and {Inoue}, T. and {Chen}, C. and {Lee}, Y. and {Soler}, J.~D. and {Kuffmeier}, M.},
	title = "{From Bubbles and Filaments to Cores and Disks: Gas Gathering and Growth of Structure Leading to the Formation of Stellar Systems}",
	booktitle = {Protostars and Planets VII},
	year = 2023,
	editor = {{Inutsuka}, S. and {Aikawa}, Y. and {Muto}, T. and {Tomida}, K. and {Tamura}, M.},
	series = {Astronomical Society of the Pacific Conference Series},
	volume = {534},
	month = jul,
	pages = {233},
	doi = {10.48550/arXiv.2205.03935},
	archivePrefix = {arXiv},
	eprint = {2205.03935},
	primaryClass = {astro-ph.GA},
	adsurl = {https://ui.adsabs.harvard.edu/abs/2023ASPC..534..233P}
}

@ARTICLE{Review_5,
       author = {{McKee}, Christopher F. and {Ostriker}, Eve C.},
        title = "{Theory of Star Formation}",
      journal = {\araa},
         year = 2007,
        month = sep,
       volume = {45},
       number = {1},
        pages = {565-687},
          doi = {10.1146/annurev.astro.45.051806.110602},
archivePrefix = {arXiv},
       eprint = {0707.3514},
 primaryClass = {astro-ph},
       adsurl = {https://ui.adsabs.harvard.edu/abs/2007ARA&A..45..565M}
}

@ARTICLE{Review_6,
       author = {{Krumholz}, Mark R.},
        title = "{The big problems in star formation: The star formation rate, stellar clustering, and the initial mass function}",
      journal = {\physrep},
         year = 2014,
        month = jun,
       volume = {539},
        pages = {49-134},
          doi = {10.1016/j.physrep.2014.02.001},
archivePrefix = {arXiv},
       eprint = {1402.0867},
 primaryClass = {astro-ph.GA},
       adsurl = {https://ui.adsabs.harvard.edu/abs/2014PhR...539...49K}
}

@ARTICLE{Review_13,
       author = {{Heyer}, Mark and {Dame}, T.~M.},
        title = "{Molecular Clouds in the Milky Way}",
      journal = {\araa},
         year = 2015,
        month = aug,
       volume = {53},
        pages = {583-629},
          doi = {10.1146/annurev-astro-082214-122324},
       adsurl = {https://ui.adsabs.harvard.edu/abs/2015ARA&A..53..583H}
}

@ARTICLE{Deharveng2010,
       author = {{Deharveng}, L. and {Schuller}, F. and {Anderson}, L.~D. and {Zavagno}, A. and {Wyrowski}, F. and {Menten}, K.~M. and {Bronfman}, L. and {Testi}, L. and {Walmsley}, C.~M. and {Wienen}, M.},
        title = "{A gallery of bubbles. The nature of the bubbles observed by Spitzer and what ATLASGAL tells us about the surrounding neutral material}",
      journal = {\aap},
         year = 2010,
        month = nov,
       volume = {523},
          eid = {A6},
        pages = {A6},
          doi = {10.1051/0004-6361/201014422},
archivePrefix = {arXiv},
       eprint = {1008.0926},
 primaryClass = {astro-ph.GA},
       adsurl = {https://ui.adsabs.harvard.edu/abs/2010A&A...523A...6D}
}

@ARTICLE{Beuther2016,
       author = {{Beuther}, H. and {Bihr}, S. and {Rugel}, M. and {Johnston}, K. and {Wang}, Y. and {Walter}, F. and {Brunthaler}, A. and {Walsh}, A.~J. and {Ott}, J. and {Stil}, J. and {Henning}, Th. and {Schierhuber}, T. and {Kainulainen}, J. and {Heyer}, M. and {Goldsmith}, P.~F. and {Anderson}, L.~D. and {Longmore}, S.~N. and {Klessen}, R.~S. and {Glover}, S.~C.~O. and {Urquhart}, J.~S. and {Plume}, R. and {Ragan}, S.~E. and {Schneider}, N. and {McClure-Griffiths}, N.~M. and {Menten}, K.~M. and {Smith}, R. and {Roy}, N. and {Shanahan}, R. and {Nguyen-Luong}, Q. and {Bigiel}, F.},
        title = "{The HI/OH/Recombination line survey of the inner Milky Way (THOR). Survey overview and data release 1}",
      journal = {\aap},
         year = 2016,
        month = oct,
       volume = {595},
          eid = {A32},
        pages = {A32},
          doi = {10.1051/0004-6361/201629143},
archivePrefix = {arXiv},
       eprint = {1609.03329},
 primaryClass = {astro-ph.GA},
       adsurl = {https://ui.adsabs.harvard.edu/abs/2016A&A...595A..32B}
}

@ARTICLE{Arce2011,
       author = {{Arce}, H{\'e}ctor G. and {Borkin}, Michelle A. and {Goodman}, Alyssa A. and {Pineda}, Jaime E. and {Beaumont}, Christopher N.},
        title = "{A Bubbling Nearby Molecular Cloud: COMPLETE Shells in Perseus}",
      journal = {\apj},
         year = 2011,
        month = dec,
       volume = {742},
       number = {2},
          eid = {105},
        pages = {105},
          doi = {10.1088/0004-637X/742/2/105},
archivePrefix = {arXiv},
       eprint = {1109.3368},
 primaryClass = {astro-ph.SR},
       adsurl = {https://ui.adsabs.harvard.edu/abs/2011ApJ...742..105A}
}

@article{Li2015,
doi = {10.1088/0067-0049/219/2/20},
url = {https://doi.org/10.1088/0067-0049/219/2/20},
year = {2015},
month = {aug},
publisher = {The American Astronomical Society},
volume = {219},
number = {2},
pages = {20},
author = {Li, Huixian and Li, Di and Qian, Lei and Xu, Duo and Goldsmith, Paul F. and Noriega-Crespo, Alberto and Wu, Yuefang and Song, Yuzhe and Nan, Rendong},
title = {OUTFLOWS AND BUBBLES IN TAURUS: STAR-FORMATION FEEDBACK SUFFICIENT TO MAINTAIN TURBULENCE},
journal = {\apjs}
}

@ARTICLE{Liu2024,
       author = {{Liu}, Dejian and {Xu}, Ye and {Li}, YingJie and {Lin}, Zehao and {Hao}, Chaojie and {Yang}, WenJin and {Li}, Jingjing and {Liu}, Xinrong and {Dong}, Yiwei and {Bian}, Shuaibo and {Kong}, Deyun},
        title = "{Molecular Bubble and Outflow in S Mon Revealed by Multiband Data Sets}",
      journal = {\apj},
         year = 2024,
        month = mar,
       volume = {964},
       number = {1},
          eid = {93},
        pages = {93},
          doi = {10.3847/1538-4357/ad24e0},
archivePrefix = {arXiv},
       eprint = {2401.17525},
 primaryClass = {astro-ph.GA},
       adsurl = {https://ui.adsabs.harvard.edu/abs/2024ApJ...964...93L}
}

@ARTICLE{Beaumont2014,
       author = {{Beaumont}, Christopher N. and {Goodman}, Alyssa A. and {Kendrew}, Sarah and {Williams}, Jonathan P. and {Simpson}, Robert},
        title = "{The Milky Way Project: Leveraging Citizen Science and Machine Learning to Detect Interstellar Bubbles}",
      journal = {\apjs},
         year = 2014,
        month = sep,
       volume = {214},
       number = {1},
          eid = {3},
        pages = {3},
          doi = {10.1088/0067-0049/214/1/3},
archivePrefix = {arXiv},
       eprint = {1406.2692},
 primaryClass = {astro-ph.GA},
       adsurl = {https://ui.adsabs.harvard.edu/abs/2014ApJS..214....3B}
}

@ARTICLE{Xu2017,
       author = {{Xu}, Duo and {Offner}, Stella S.~R.},
        title = "{Assessing the Performance of a Machine Learning Algorithm in Identifying Bubbles in Dust Emission}",
      journal = {\apj},
         year = 2017,
        month = dec,
       volume = {851},
       number = {2},
          eid = {149},
        pages = {149},
          doi = {10.3847/1538-4357/aa9a42},
archivePrefix = {arXiv},
       eprint = {1711.03480},
 primaryClass = {astro-ph.GA},
       adsurl = {https://ui.adsabs.harvard.edu/abs/2017ApJ...851..149X}
}

@ARTICLE{Xu2020,
       author = {{Xu}, Duo and {Offner}, Stella S.~R. and {Gutermuth}, Robert and {Oort}, Colin Van},
        title = "{Application of Convolutional Neural Networks to Identify Stellar Feedback Bubbles in CO Emission}",
      journal = {\apj},
         year = 2020,
        month = feb,
       volume = {890},
       number = {1},
          eid = {64},
        pages = {64},
          doi = {10.3847/1538-4357/ab6607},
archivePrefix = {arXiv},
       eprint = {2001.04506},
 primaryClass = {astro-ph.GA},
       adsurl = {https://ui.adsabs.harvard.edu/abs/2020ApJ...890...64X}
}

@ARTICLE{Infrared_Bubble_Recognition,
       author = {{Nishimoto}, Shimpei and {Onishi}, Toshikazu and {Nishimura}, Atsushi and {Fujita}, Shinji and {Kawanishi}, Yasutomo and {Nakatani}, Shuyo and {Tokuda}, Kazuki and {Shimajiri}, Yoshito and {Kaneko}, Hiroyuki and {Miyamoto}, Yusuke and {Inoue}, Tsuyoshi and {Ito}, Atsushi M.},
        title = "{Infrared bubble recognition in the Milky Way and beyond using deep learning}",
      journal = {\pasj},
         year = 2025,
        month = apr,
       volume = {77},
       number = {2},
        pages = {403-424},
          doi = {10.1093/pasj/psaf008},
archivePrefix = {arXiv},
       eprint = {2504.03367},
 primaryClass = {astro-ph.GA},
       adsurl = {https://ui.adsabs.harvard.edu/abs/2025PASJ...77..403N}
}

@ARTICLE{Anderson2014,
       author = {{Anderson}, L.~D. and {Bania}, T.~M. and {Balser}, Dana S. and {Cunningham}, V. and {Wenger}, T.~V. and {Johnstone}, B.~M. and {Armentrout}, W.~P.},
        title = "{The WISE Catalog of Galactic H II Regions}",
      journal = {\apjs},
         year = 2014,
        month = may,
       volume = {212},
       number = {1},
          eid = {1},
        pages = {1},
          doi = {10.1088/0067-0049/212/1/1},
archivePrefix = {arXiv},
       eprint = {1312.6202},
 primaryClass = {astro-ph.GA},
       adsurl = {https://ui.adsabs.harvard.edu/abs/2014ApJS..212....1A}
}

@article{CO_Survey_MWISP_1,
	author = {{Su}, Yang and {Yang}, Ji and {Zhang}, Shaobo and {Gong}, Yan and {Wang}, Hongchi and {Zhou}, Xin and {Wang}, Min and {Chen}, Zhiwei and {Sun}, Yan and {Chen}, Xuepeng and {Xu}, Ye and {Jiang}, Zhibo},
	title = "{The Milky Way Imaging Scroll Painting (MWISP): Project Details and Initial Results from the Galactic Longitudes of 25.{\textdegree}8-49.{\textdegree}7}",
	journal = {\apjs},
	year = 2019,
	month = jan,
	volume = {240},
	number = {1},
	eid = {9},
	pages = {9},
	doi = {10.3847/1538-4365/aaf1c8},
	archivePrefix = {arXiv},
	eprint = {1901.00285},
	primaryClass = {astro-ph.GA},
	adsurl = {https://ui.adsabs.harvard.edu/abs/2019ApJS..240....9S}
}

@ARTICLE{CO_Survey_MWISP_2,
       author = {{Shan}, Wenlei and {Yang}, Ji and {Shi}, Shengcai and {Yao}, Qijun and {Zuo}, Yingxi and {Lin}, Zhenhui and {Chen}, Shanhuai and {Zhang}, Xuguo and {Duan}, Wenying and {Cao}, Aiqing and {Li}, Sheng and {Li}, Zhenqiang and {Liu}, Jie and {Zhong}, Jiaqiang},
        title = "{Development of Superconducting Spectroscopic Array Receiver: A Multibeam 2SB SIS Receiver for Millimeter-Wave Radio Astronomy}",
      journal = {IEEE Transactions on Terahertz Science and Technology},
         year = 2012,
        month = nov,
       volume = {2},
       number = {6},
        pages = {593-604},
          doi = {10.1109/TTHZ.2012.2213818},
       adsurl = {https://ui.adsabs.harvard.edu/abs/2012ITTST...2..593S}
}

@ARTICLE{CO_Survey_MWISP_3,
       author = {{Yang}, Ji and {Yan}, Qing-Zeng and {Su}, Yang and {Zhang}, Shaobo and {Zhou}, Xin and {Sun}, Yan and {Ao}, Yiping and {Chen}, Xuepeng and {Chen}, Zhiwei and {Du}, Fujun and {Fang}, Min and {Gong}, Yan and {Jiang}, Zhibo and {Jin}, Shengyu and {Ju}, Binggang and {Li}, Chong and {Li}, Yingjie and {Liu}, Yi and {Lu}, Dengrong and {Luo}, Chunsheng and {Ma}, Yuehui and {Mao}, Ruiqing and {Sun}, Jixian and {Wang}, Chen and {Wang}, Hongchi and {Wang}, Min and {Wang (Qinghai)}, Min and {Wang}, Xindong and {Xu}, Wenting and {Xu}, Ye and {Yan}, Kun and {Yan}, Ping and {Yuan}, Lixia and {Zhang}, Miaomiao and {Zhang}, Yongxing},
        title = "{The Milky Way Imaging Scroll Painting Survey: Data Release 1}",
      journal = {\apjs},
         year = 2026,
        month = feb,
       volume = {282},
       number = {2},
          eid = {65},
        pages = {65},
          doi = {10.3847/1538-4365/ae29e7},
archivePrefix = {arXiv},
       eprint = {2512.08260},
 primaryClass = {astro-ph.GA},
       adsurl = {https://ui.adsabs.harvard.edu/abs/2026ApJS..282...65Y}
}

@ARTICLE{MWISP_Clumps,
       author = {{Jiang}, Yu and {Yan}, Qing-Zeng and {Yang}, Ji and {Zheng}, Sheng and {Chen}, Xuepeng and {Su}, Yang and {Jiang}, Zhibo and {Chen}, Zhiwei and {Zhou}, Xin and {Huang}, Yao and {Luo}, Xiaoyu and {Feng}, Haoran and {Liu}, De-Jian},
        title = "{Investigations of MWISP Clumps: $^{13}$CO Clump Source Catalog and Physical Properties}",
      journal = {\apjs},
         year = 2025,
        month = oct,
       volume = {280},
       number = {2},
          eid = {75},
        pages = {75},
          doi = {10.3847/1538-4365/ae00b9},
archivePrefix = {arXiv},
       eprint = {2509.01955},
 primaryClass = {astro-ph.GA},
       adsurl = {https://ui.adsabs.harvard.edu/abs/2025ApJS..280...75J}
}

@ARTICLE{NGC628_1,
       author = {{Barnes}, Ashley. T. and {Watkins}, Elizabeth J. and {Meidt}, Sharon E. and {Kreckel}, Kathryn and {Sormani}, Mattia C. and {Tre{\ss}}, Robin G. and {Glover}, Simon C.~O. and {Bigiel}, Frank and {Chandar}, Rupali and {Emsellem}, Eric and {Lee}, Janice C. and {Leroy}, Adam K. and {Sandstrom}, Karin M. and {Schinnerer}, Eva and {Rosolowsky}, Erik and {Belfiore}, Francesco and {Blanc}, Guillermo A. and {Boquien}, M{\'e}d{\'e}ric and {Brok}, Jakob den and {Cao}, Yixian and {Chevance}, M{\'e}lanie and {Dale}, Daniel A. and {Egorov}, Oleg V. and {Eibensteiner}, Cosima and {Grasha}, Kathryn and {Groves}, Brent and {Hassani}, Hamid and {Henshaw}, Jonathan D. and {Jeffreson}, Sarah and {Jim{\'e}nez-Donaire}, Mar{\'\i}a J. and {Keller}, Benjamin W. and {Klessen}, Ralf S. and {Koch}, Eric W. and {Kruijssen}, J.~M. Diederik and {Larson}, Kirsten L. and {Li}, Jing and {Liu}, Daizhong and {Lopez}, Laura A. and {Murphy}, Eric J. and {Neumann}, Lukas and {Pety}, J{\'e}r{\^o}me and {Pinna}, Francesca and {Querejeta}, Miguel and {Renaud}, Florent and {Saito}, Toshiki and {Sarbadhicary}, Sumit K. and {Sardone}, Amy and {Smith}, Rowan J. and {Stuber}, Sophia K. and {Sun}, Jiayi and {Thilker}, David A. and {Usero}, Antonio and {Whitmore}, Bradley C. and {Williams}, Thomas G.},
        title = "{PHANGS-JWST First Results: Multiwavelength View of Feedback-driven Bubbles (the Phantom Voids) across NGC 628}",
      journal = {\apjl},
         year = 2023,
        month = feb,
       volume = {944},
       number = {2},
          eid = {L22},
        pages = {L22},
          doi = {10.3847/2041-8213/aca7b9},
archivePrefix = {arXiv},
       eprint = {2212.00812},
 primaryClass = {astro-ph.GA},
       adsurl = {https://ui.adsabs.harvard.edu/abs/2023ApJ...944L..22B}
}

@ARTICLE{NGC628_2,
       author = {{Zhou}, J.~W. and {Han}, A.~A.},
        title = "{Automated void identification by Blendmask: from hierarchical molecular gas to hierarchical voids in NGC 628}",
      journal = {arXiv e-prints},
         year = 2026,
        month = may,
          eid = {arXiv:2605.30533},
        pages = {arXiv:2605.30533},
          doi = {10.48550/arXiv.2605.30533},
archivePrefix = {arXiv},
       eprint = {2605.30533},
 primaryClass = {astro-ph.GA},
       adsurl = {https://ui.adsabs.harvard.edu/abs/2026arXiv260530533Z}
}

@ARTICLE{FacetClumps,
	author = {{Jiang}, Yu and {Chen}, Zhiwei and {Zheng}, Sheng and {Jiang}, Zhibo and {Huang}, Yao and {Zeng}, Shuguang and {Zeng}, Xiangyun and {Luo}, Xiaoyu},
	title = "{FacetClumps: A Facet-based Molecular Clump Detection Algorithm}",
	journal = {\apjs},
	year = 2023,
	month = aug,
	volume = {267},
	number = {2},
	eid = {32},
	pages = {32},
	doi = {10.3847/1538-4365/acda89},
	archivePrefix = {arXiv},
	eprint = {2305.18709},
	primaryClass = {astro-ph.IM},
	adsurl = {https://ui.adsabs.harvard.edu/abs/2023ApJS..267...32J}
}

@software{FacetClumps_Sfot,
       author = {{Jiang}, Yu},
        title = "{FacetClumps: A Facet-based Molecular Clump Detection Algorithm}",
         year = 2023,
        month = may,
          eid = {10.5281/zenodo.7991006},
          doi = {10.5281/zenodo.7991006},
      version = {v0.0.4},
    publisher = {Zenodo},
       adsurl = {https://ui.adsabs.harvard.edu/abs/2023zndo...7991006J}
}

@ARTICLE{DPConCFil,
       author = {{Jiang}, Yu and {Chen}, Xuepeng and {Zheng}, Sheng and {Jiang}, Zhibo and {Chen}, Zhiwei and {Huang}, Yao and {Su}, Yang and {Sun}, Li and {Feng}, Jian-Cheng and {Feng}, Haoran and {Yang}, Ji},
        title = "{Investigations of MWISP Filaments. I. Filament Identification and Analysis Algorithms, and Source Catalog}",
      journal = {\apjs},
         year = 2025,
        month = jan,
       volume = {276},
       number = {1},
          eid = {27},
        pages = {27},
          doi = {10.3847/1538-4365/ad91a8},
archivePrefix = {arXiv},
       eprint = {2412.01238},
 primaryClass = {astro-ph.GA},
       adsurl = {https://ui.adsabs.harvard.edu/abs/2025ApJS..276...27J}
}

@software{DPConCFil_Sfot,
       author = {{Jiang}, Yu},
        title = "{DPConCFil: A Collection of Filament Identification and Analysis Algorithms}",
         year = 2024,
        month = sep,
          eid = {10.5281/zenodo.13853675},
          doi = {10.5281/zenodo.13853675},
    publisher = {Zenodo},
       adsurl = {https://ui.adsabs.harvard.edu/abs/2024zndo..13853675J}
}

@ARTICLE{Churchwell_1,
       author = {{Churchwell}, E. and {Povich}, M.~S. and {Allen}, D. and {Taylor}, M.~G. and {Meade}, M.~R. and {Babler}, B.~L. and {Indebetouw}, R. and {Watson}, C. and {Whitney}, B.~A. and {Wolfire}, M.~G. and {Bania}, T.~M. and {Benjamin}, R.~A. and {Clemens}, D.~P. and {Cohen}, M. and {Cyganowski}, C.~J. and {Jackson}, J.~M. and {Kobulnicky}, H.~A. and {Mathis}, J.~S. and {Mercer}, E.~P. and {Stolovy}, S.~R. and {Uzpen}, B. and {Watson}, D.~F. and {Wolff}, M.~J.},
        title = "{The Bubbling Galactic Disk}",
      journal = {\apj},
         year = 2006,
        month = oct,
       volume = {649},
       number = {2},
        pages = {759-778},
          doi = {10.1086/507015},
       adsurl = {https://ui.adsabs.harvard.edu/abs/2006ApJ...649..759C}
}

@ARTICLE{Churchwell_2,
       author = {{Churchwell}, E. and {Watson}, D.~F. and {Povich}, M.~S. and {Taylor}, M.~G. and {Babler}, B.~L. and {Meade}, M.~R. and {Benjamin}, R.~A. and {Indebetouw}, R. and {Whitney}, B.~A.},
        title = "{The Bubbling Galactic Disk. II. The Inner 20{\textdegree}}",
      journal = {\apj},
         year = 2007,
        month = nov,
       volume = {670},
       number = {1},
        pages = {428-441},
          doi = {10.1086/521646},
       adsurl = {https://ui.adsabs.harvard.edu/abs/2007ApJ...670..428C}
}

@ARTICLE{Bubble_Catalog_1,
       author = {{Simpson}, R.~J. and {Povich}, M.~S. and {Kendrew}, S. and {Lintott}, C.~J. and {Bressert}, E. and {Arvidsson}, K. and {Cyganowski}, C. and {Maddison}, S. and {Schawinski}, K. and {Sherman}, R. and {Smith}, A.~M. and {Wolf-Chase}, G.},
        title = "{The Milky Way Project First Data Release: a bubblier Galactic disc}",
      journal = {\mnras},
         year = 2012,
        month = aug,
       volume = {424},
       number = {4},
        pages = {2442-2460},
          doi = {10.1111/j.1365-2966.2012.20770.x},
archivePrefix = {arXiv},
       eprint = {1201.6357},
 primaryClass = {astro-ph.GA},
       adsurl = {https://ui.adsabs.harvard.edu/abs/2012MNRAS.424.2442S}
}

@ARTICLE{Bubble_Catalog_2,
       author = {{Kohno}, Mikito and {Sofue}, Yoshiaki and {Fukui}, Yasuo and {Tachihara}, Kengo},
        title = "{A catalog of molecular clouds possibly associated with Galactic infrared bubbles. I. The southern Galactic plane}",
      journal = {\pasj},
         year = 2025,
        month = oct,
       volume = {77},
       number = {5},
        pages = {1036-1049},
          doi = {10.1093/pasj/psaf081},
archivePrefix = {arXiv},
       eprint = {2507.01222},
 primaryClass = {astro-ph.GA},
       adsurl = {https://ui.adsabs.harvard.edu/abs/2025PASJ...77.1036K}
}

@ARTICLE{MeerKAT_0,
       author = {{Goedhart}, S. and {Cotton}, W.~D. and {Camilo}, F. and {Thompson}, M.~A. and {Umana}, G. and {Bietenholz}, M. and {Woudt}, P.~A. and {Anderson}, L.~D. and {Bordiu}, C. and {Buckley}, D.~A.~H. and {Buemi}, C.~S. and {Bufano}, F. and {Cavallaro}, F. and {Chen}, H. and {Chibueze}, J.~O. and {Egbo}, D. and {Frank}, B.~S. and {Hoare}, M.~G. and {Ingallinera}, A. and {Irabor}, T. and {Kraan-Korteweg}, R.~C. and {Kurapati}, S. and {Leto}, P. and {Loru}, S. and {Mutale}, M. and {Obonyo}, W.~O. and {Plavin}, A. and {Rajohnson}, S.~H.~A. and {Rigby}, A. and {Riggi}, S. and {Seidu}, M. and {Serra}, P. and {Smart}, B.~M. and {Stappers}, B.~W. and {Steyn}, N. and {Surnis}, M. and {Trigilio}, C. and {Williams}, G.~M. and {Abbott}, T.~D. and {Adam}, R.~M. and {Asad}, K.~M.~B. and {Baloyi}, T. and {Bauermeister}, E.~F. and {Bennet}, T.~G.~H. and {Bester}, H. and {Botha}, A.~G. and {Brederode}, L.~R.~S. and {Buchner}, S. and {Burger}, J.~P. and {Cheetham}, T. and {Cloete}, K. and {de Villiers}, M.~S. and {de Villiers}, D.~I.~L. and {du Toit}, L.~J. and {Esterhuyse}, S.~W.~P. and {Fanaroff}, B.~L. and {Fourie}, D.~J. and {Gamatham}, R.~R.~G. and {Gatsi}, T.~G. and {Geyer}, M. and {Gouws}, M. and {Gumede}, S.~C. and {Heywood}, I. and {Hokwana}, A. and {Hoosen}, S.~W. and {Horn}, D.~M. and {Horrell}, L.~M.~G. and {Hugo}, B.~V. and {Isaacson}, A.~I. and {J{\'o}zsa}, G.~I.~G. and {Jonas}, J.~L. and {Jordaan}, J.~D.~B.~L. and {Joubert}, A.~F. and {Julie}, R.~P.~M. and {Kapp}, F.~B. and {Kriek}, N. and {Kriel}, H. and {Krishnan}, V.~K. and {Kusel}, T.~W. and {Legodi}, L.~S. and {Lehmensiek}, R. and {Lord}, R.~T. and {Macfarlane}, P.~S. and {Magnus}, L.~G. and {Magozore}, C. and {Main}, J.~P.~L. and {Malan}, J.~A. and {Manley}, J.~R. and {Marais}, S.~J. and {Maree}, M.~D.~J. and {Martens}, A. and {Maruping}, P. and {McAlpine}, K. and {Merry}, B.~C. and {Mgodeli}, M. and {Millenaar}, R.~P. and {Mokone}, O.~J. and {Monama}, T.~E. and {New}, W.~S. and {Ngcebetsha}, B. and {Ngoasheng}, K.~J. and {Nicolson}, G.~D. and {Ockards}, M.~T. and {Oozeer}, N. and {Passmoor}, S.~S. and {Patel}, A.~A. and {Peens-Hough}, A. and {Perkins}, S.~J. and {Ramaila}, A.~J.~T. and {Ratcliffe}, S.~M. and {Renil}, R. and {Richter}, L.~L. and {Salie}, S. and {Sambu}, N. and {Schollar}, C.~T.~G. and {Schwardt}, L.~C. and {Schwartz}, R.~L. and {Serylak}, M. and {Siebrits}, R. and {Sirothia}, S.~K. and {Slabber}, M.~J. and {Smirnov}, O.~M. and {Tiplady}, A.~J. and {van Balla}, T.~J. and {van der Byl}, A. and {Van Tonder}, V. and {Venter}, A.~J. and {Venter}, M. and {Welz}, M.~G. and {Williams}, L.~P.},
        title = "{The SARAO MeerKAT 1.3 GHz Galactic Plane Survey}",
      journal = {\mnras},
         year = 2024,
        month = jun,
       volume = {531},
       number = {1},
        pages = {649-681},
          doi = {10.1093/mnras/stae1166},
archivePrefix = {arXiv},
       eprint = {2312.07275},
 primaryClass = {astro-ph.GA},
       adsurl = {https://ui.adsabs.harvard.edu/abs/2024MNRAS.531..649G}
}

@ARTICLE{MeerKAT_Associated_With_Clouds,
       author = {{Langa}, Moses O. and {Thompson}, Mark A. and {Rigby}, Andrew J. and {Williams}, Gwenllian M. and {Mutale}, Mubela and {Baki}, Paul O. and {Chibueze}, James O. and {Obonyo}, Willice O.},
        title = "{H II regions and supernova remnants associated with molecular clouds: a pilot study with the SARAO MeerKAT Galactic Plane Survey}",
      journal = {\mnras},
         year = 2026,
        month = jan,
       volume = {545},
       number = {2},
          eid = {staf2037},
        pages = {staf2037},
          doi = {10.1093/mnras/staf2037},
archivePrefix = {arXiv},
       eprint = {2511.20542},
 primaryClass = {astro-ph.GA},
       adsurl = {https://ui.adsabs.harvard.edu/abs/2026MNRAS.545f2037L}
}

@ARTICLE{GLIMPSE_0,
       author = {{Benjamin}, Robert A. and {Churchwell}, E. and {Babler}, Brian L. and {Bania}, T.~M. and {Clemens}, Dan P. and {Cohen}, Martin and {Dickey}, John M. and {Indebetouw}, R{\'e}my and {Jackson}, James M. and {Kobulnicky}, Henry A. and {Lazarian}, Alex and {Marston}, A.~P. and {Mathis}, John S. and {Meade}, Marilyn R. and {Seager}, Sara and {Stolovy}, S.~R. and {Watson}, C. and {Whitney}, Barbara A. and {Wolff}, Michael J. and {Wolfire}, Mark G.},
        title = "{GLIMPSE. I. An SIRTF Legacy Project to Map the Inner Galaxy}",
      journal = {\pasp},
         year = 2003,
        month = aug,
       volume = {115},
       number = {810},
        pages = {953-964},
          doi = {10.1086/376696},
archivePrefix = {arXiv},
       eprint = {astro-ph/0306274},
 primaryClass = {astro-ph},
       adsurl = {https://ui.adsabs.harvard.edu/abs/2003PASP..115..953B}
}

@ARTICLE{Spitzer_0,
       author = {{Fazio}, G.~G. and {Hora}, J.~L. and {Allen}, L.~E. and {Ashby}, M.~L.~N. and {Barmby}, P. and {Deutsch}, L.~K. and {Huang}, J.-S. and {Kleiner}, S. and {Marengo}, M. and {Megeath}, S.~T. and {Melnick}, G.~J. and {Pahre}, M.~A. and {Patten}, B.~M. and {Polizotti}, J. and {Smith}, H.~A. and {Taylor}, R.~S. and {Wang}, Z. and {Willner}, S.~P. and {Hoffmann}, W.~F. and {Pipher}, J.~L. and {Forrest}, W.~J. and {McMurty}, C.~W. and {McCreight}, C.~R. and {McKelvey}, M.~E. and {McMurray}, R.~E. and {Koch}, D.~G. and {Moseley}, S.~H. and {Arendt}, R.~G. and {Mentzell}, J.~E. and {Marx}, C.~T. and {Losch}, P. and {Mayman}, P. and {Eichhorn}, W. and {Krebs}, D. and {Jhabvala}, M. and {Gezari}, D.~Y. and {Fixsen}, D.~J. and {Flores}, J. and {Shakoorzadeh}, K. and {Jungo}, R. and {Hakun}, C. and {Workman}, L. and {Karpati}, G. and {Kichak}, R. and {Whitley}, R. and {Mann}, S. and {Tollestrup}, E.~V. and {Eisenhardt}, P. and {Stern}, D. and {Gorjian}, V. and {Bhattacharya}, B. and {Carey}, S. and {Nelson}, B.~O. and {Glaccum}, W.~J. and {Lacy}, M. and {Lowrance}, P.~J. and {Laine}, S. and {Reach}, W.~T. and {Stauffer}, J.~A. and {Surace}, J.~A. and {Wilson}, G. and {Wright}, E.~L. and {Hoffman}, A. and {Domingo}, G. and {Cohen}, M.},
        title = "{The Infrared Array Camera (IRAC) for the Spitzer Space Telescope}",
      journal = {\apjs},
         year = 2004,
        month = sep,
       volume = {154},
       number = {1},
        pages = {10-17},
          doi = {10.1086/422843},
archivePrefix = {arXiv},
       eprint = {astro-ph/0405616},
 primaryClass = {astro-ph},
       adsurl = {https://ui.adsabs.harvard.edu/abs/2004ApJS..154...10F}
}

@ARTICLE{MIPSGAL_0,
       author = {{Rieke}, G.~H. and {Young}, E.~T. and {Engelbracht}, C.~W. and {Kelly}, D.~M. and {Low}, F.~J. and {Haller}, E.~E. and {Beeman}, J.~W. and {Gordon}, K.~D. and {Stansberry}, J.~A. and {Misselt}, K.~A. and {Cadien}, J. and {Morrison}, J.~E. and {Rivlis}, G. and {Latter}, W.~B. and {Noriega-Crespo}, A. and {Padgett}, D.~L. and {Stapelfeldt}, K.~R. and {Hines}, D.~C. and {Egami}, E. and {Muzerolle}, J. and {Alonso-Herrero}, A. and {Blaylock}, M. and {Dole}, H. and {Hinz}, J.~L. and {Le Floc'h}, E. and {Papovich}, C. and {P{\'e}rez-Gonz{\'a}lez}, P.~G. and {Smith}, P.~S. and {Su}, K.~Y.~L. and {Bennett}, L. and {Frayer}, D.~T. and {Henderson}, D. and {Lu}, N. and {Masci}, F. and {Pesenson}, M. and {Rebull}, L. and {Rho}, J. and {Keene}, J. and {Stolovy}, S. and {Wachter}, S. and {Wheaton}, W. and {Werner}, M.~W. and {Richards}, P.~L.},
        title = "{The Multiband Imaging Photometer for Spitzer (MIPS)}",
      journal = {\apjs},
         year = 2004,
        month = sep,
       volume = {154},
       number = {1},
        pages = {25-29},
          doi = {10.1086/422717},
       adsurl = {https://ui.adsabs.harvard.edu/abs/2004ApJS..154...25R}
}

@ARTICLE{MIPSGAL_1,
       author = {{Carey}, S.~J. and {Noriega-Crespo}, A. and {Mizuno}, D.~R. and {Shenoy}, S. and {Paladini}, R. and {Kraemer}, K.~E. and {Price}, S.~D. and {Flagey}, N. and {Ryan}, E. and {Ingalls}, J.~G. and {Kuchar}, T.~A. and {Pinheiro Gon{\c{c}}alves}, Daniela and {Indebetouw}, R. and {Billot}, N. and {Marleau}, F.~R. and {Padgett}, D.~L. and {Rebull}, L.~M. and {Bressert}, E. and {Ali}, Babar and {Molinari}, S. and {Martin}, P.~G. and {Berriman}, G.~B. and {Boulanger}, F. and {Latter}, W.~B. and {Miville-Deschenes}, M.~A. and {Shipman}, R. and {Testi}, L.},
        title = "{MIPSGAL: A Survey of the Inner Galactic Plane at 24 and 70 {\ensuremath{\mu}}m}",
      journal = {\pasp},
         year = 2009,
        month = jan,
       volume = {121},
       number = {875},
        pages = {76},
          doi = {10.1086/596581},
       adsurl = {https://ui.adsabs.harvard.edu/abs/2009PASP..121...76C}
}

@INPROCEEDINGS{MIPSGAL_2,
       author = {{Becker}, R.~H. and {White}, R.~L. and {Helfand}, D.~J.},
        title = "{MAGPIS: The Multi-Array Galactic Plane Imaging Survey}",
    booktitle = {The Transient Milky Way: A Perspective for MIRAX},
         year = 2006,
       editor = {{D'Amico}, Flavio and {Braga}, Jo{\~a}o and {Rothschild}, Richard E.},
       series = {American Institute of Physics Conference Series},
       volume = {840},
        month = jun,
    publisher = {AIP},
        pages = {102-104},
          doi = {10.1063/1.2216612},
archivePrefix = {arXiv},
       eprint = {astro-ph/0510468},
 primaryClass = {astro-ph},
       adsurl = {https://ui.adsabs.harvard.edu/abs/2006AIPC..840..102B}
}

@dataset{MIPSGAL_3,
       author = {{Gutermuth}, Robert A. and {Heyer}, Mark},
        title = "{MIPSGAL 24 micron Archive}",
 howpublished = {NASA IPAC DataSet, IRSA259},
         year = 2020,
        month = jan,
          doi = {10.26131/IRSA259},
       adsurl = {https://ui.adsabs.harvard.edu/abs/2020ipac.data.I259G}
}

@ARTICLE{SOFIA_0,
       author = {{Schneider}, N. and {Simon}, R. and {Guevara}, C. and {Buchbender}, C. and {Higgins}, R.~D. and {Okada}, Y. and {Stutzki}, J. and {G{\"u}sten}, R. and {Anderson}, L.~D. and {Bally}, J. and {Beuther}, H. and {Bonne}, L. and {Bontemps}, S. and {Chambers}, E. and {Csengeri}, T. and {Graf}, U.~U. and {Gusdorf}, A. and {Jacobs}, K. and {Justen}, M. and {Kabanovic}, S. and {Karim}, R. and {Luisi}, M. and {Menten}, K. and {Mertens}, M. and {Mookerjea}, B. and {Ossenkopf-Okada}, V. and {Pabst}, C. and {Pound}, M.~W. and {Richter}, H. and {Reyes}, N. and {Ricken}, O. and {R{\"o}llig}, M. and {Russeil}, D. and {S{\'a}nchez-Monge}, {\'A}. and {Sandell}, G. and {Tiwari}, M. and {Wiesemeyer}, H. and {Wolfire}, M. and {Wyrowski}, F. and {Zavagno}, A. and {Tielens}, A.~G.~G.~M.},
        title = "{FEEDBACK: a SOFIA Legacy Program to Study Stellar Feedback in Regions of Massive Star Formation}",
      journal = {\pasp},
         year = 2020,
        month = oct,
       volume = {132},
       number = {1016},
          eid = {104301},
        pages = {104301},
          doi = {10.1088/1538-3873/aba840},
archivePrefix = {arXiv},
       eprint = {2009.08730},
 primaryClass = {astro-ph.GA},
       adsurl = {https://ui.adsabs.harvard.edu/abs/2020PASP..132j4301S}
}

@ARTICLE{SOFIA_Orion_1,
       author = {{Pabst}, C.~H.~M. and {Hacar}, A. and {Goicoechea}, J.~R. and {Teyssier}, D. and {Bern{\'e}}, O. and {Wolfire}, M.~G. and {Higgins}, R.~D. and {Chambers}, E.~T. and {Kabanovic}, S. and {G{\"u}sten}, R. and {Stutzki}, J. and {Kramer}, C. and {Tielens}, A.~G.~G.~M.},
        title = "{[C II] 158 {\ensuremath{\mu}}m line emission from Orion A I. A template for extragalactic studies?}",
      journal = {\aap},
         year = 2021,
        month = jul,
       volume = {651},
          eid = {A111},
        pages = {A111},
          doi = {10.1051/0004-6361/202140804},
archivePrefix = {arXiv},
       eprint = {2105.03735},
 primaryClass = {astro-ph.GA},
       adsurl = {https://ui.adsabs.harvard.edu/abs/2021A&A...651A.111P}
}

@ARTICLE{SOFIA_Orion_2,
       author = {{Pabst}, C.~H.~M. and {Goicoechea}, J.~R. and {Hacar}, A. and {Teyssier}, D. and {Bern{\'e}}, O. and {Wolfire}, M.~G. and {Higgins}, R.~D. and {Chambers}, E.~T. and {Kabanovic}, S. and {G{\"u}sten}, R. and {Stutzki}, J. and {Kramer}, C. and {Tielens}, A.~G.~G.~M.},
        title = "{[C II] 158 {\ensuremath{\mu}}m line emission from Orion A. II. Photodissociation region physics}",
      journal = {\aap},
         year = 2022,
        month = feb,
       volume = {658},
          eid = {A98},
        pages = {A98},
          doi = {10.1051/0004-6361/202140805},
archivePrefix = {arXiv},
       eprint = {2111.12363},
 primaryClass = {astro-ph.GA},
       adsurl = {https://ui.adsabs.harvard.edu/abs/2022A&A...658A..98P}
}

@ARTICLE{SOFIA_Orion_3,
       author = {{Kavak}, {\"U}. and {Goicoechea}, J.~R. and {Pabst}, C.~H.~M. and {Bally}, J. and {van der Tak}, F.~F.~S. and {Tielens}, A.~G.~G.~M.},
        title = "{Breaking Orion's Veil with fossil outflows}",
      journal = {\aap},
         year = 2022,
        month = apr,
       volume = {660},
          eid = {A109},
        pages = {A109},
          doi = {10.1051/0004-6361/202141367},
archivePrefix = {arXiv},
       eprint = {2202.04711},
 primaryClass = {astro-ph.GA},
       adsurl = {https://ui.adsabs.harvard.edu/abs/2022A&A...660A.109K}
}

@ARTICLE{SOFIA_Orion_4,
       author = {{Kavak}, {\"U}. and {Bally}, J. and {Goicoechea}, J.~R. and {Pabst}, C.~H.~M. and {van der Tak}, F.~F.~S. and {Tielens}, A.~G.~G.~M.},
        title = "{Dents in the Veil: protostellar feedback in Orion}",
      journal = {\aap},
         year = 2022,
        month = jul,
       volume = {663},
          eid = {A117},
        pages = {A117},
          doi = {10.1051/0004-6361/202243332},
archivePrefix = {arXiv},
       eprint = {2203.12025},
 primaryClass = {astro-ph.SR},
       adsurl = {https://ui.adsabs.harvard.edu/abs/2022A&A...663A.117K}
}

@ARTICLE{SOFIA_Cygnus_1,
       author = {{Dannhauer}, Simon M. and {Vider}, Sebastian and {Schneider}, Nicola and {Simon}, Robert and {Comeron}, Fernando and {Keilmann}, Eduard and {Walch}, Stefanie and {Bonne}, Lars and {Kabanovic}, Slawa and {Ossenkopf-Okada}, Volker and {Seifried}, Daniel and {Csengeri}, Timea and {Djupvik}, Amanda and {Gong}, Yan and {Brunthaler}, Andreas and {Rugel}, Michael and {Riechers}, Dominik A. and {Bontemps}, Sylvain and {Honingh}, Netty and {Graf}, Urs U. and {Tielens}, Alexander G.~G.~M.},
        title = "{The Diamond Ring in Cygnus X: Advanced stage of an expanding bubble of ionised carbon}",
      journal = {\aap},
         year = 2025,
        month = nov,
       volume = {703},
          eid = {A197},
        pages = {A197},
          doi = {10.1051/0004-6361/202556159},
archivePrefix = {arXiv},
       eprint = {2509.22427},
 primaryClass = {astro-ph.GA},
       adsurl = {https://ui.adsabs.harvard.edu/abs/2025A&A...703A.197D}
}

@ARTICLE{PHANGS_Bub_1,
       author = {{Watkins}, E.~J. and {Kreckel}, K. and {Groves}, B. and {Glover}, S.~C.~O. and {Whitmore}, B.~C. and {Leroy}, A.~K. and {Schinnerer}, E. and {Meidt}, S.~E. and {Egorov}, O.~V. and {Barnes}, A.~T. and {Lee}, J.~C. and {Bigiel}, F. and {Boquien}, M. and {Chandar}, R. and {Chevance}, M. and {Dale}, D.~A. and {Grasha}, K. and {Klessen}, R.~S. and {Kruijssen}, J.~M.~D. and {Larson}, K.~L. and {Li}, J. and {M{\'e}ndez-Delgado}, J.~E. and {Pessa}, I. and {Saito}, T. and {Sanchez-Blazquez}, P. and {Sarbadhicary}, S.~K. and {Scheuermann}, F. and {Thilker}, D.~A. and {Williams}, T.~G.},
        title = "{Quantifying the energetics of molecular superbubbles in PHANGS galaxies}",
      journal = {\aap},
         year = 2023,
        month = aug,
       volume = {676},
          eid = {A67},
        pages = {A67},
          doi = {10.1051/0004-6361/202346075},
archivePrefix = {arXiv},
       eprint = {2302.03699},
 primaryClass = {astro-ph.GA},
       adsurl = {https://ui.adsabs.harvard.edu/abs/2023A&A...676A..67W}
}

@ARTICLE{ALMA_Bub_1,
       author = {{Lin}, Shuting and {Feng}, Siyi and {Xu}, Fengwei and {Wang}, Ke and {Sanhueza}, Patricio and {Wang}, Junzhi and {Zhang}, Zhi-Yu and {Zhang}, Yichen and {Morii}, Kaho and {Liu}, Hauyu Baobab and {Liu}, Sheng-Yuan and {Wang}, Lile and {Sabatini}, Giovanni and {Li}, Hui and {Baan}, Willem and {Zhu}, Zhi-Kai and {Li}, Shanghuo},
        title = "{A Dense Molecular Ringlike Structure in Gaseous CO Depletion Region G34.74{\ensuremath{-}}0.12}",
      journal = {\apjl},
         year = 2025,
        month = oct,
       volume = {992},
       number = {1},
          eid = {L15},
        pages = {L15},
          doi = {10.3847/2041-8213/ae0579},
archivePrefix = {arXiv},
       eprint = {2509.11475},
 primaryClass = {astro-ph.GA},
       adsurl = {https://ui.adsabs.harvard.edu/abs/2025ApJ...992L..15L}
}

@ARTICLE{PoC,
       author = {{Hester}, J.~J. and {Scowen}, P.~A. and {Sankrit}, R. and {Lauer}, T.~R. and {Ajhar}, E.~A. and {Baum}, W.~A. and {Code}, A. and {Currie}, D.~G. and {Danielson}, G.~E. and {Ewald}, S.~P. and {Faber}, S.~M. and {Grillmair}, C.~J. and {Groth}, E.~J. and {Holtzman}, J.~A. and {Hunter}, D.~A. and {Kristian}, J. and {Light}, R.~M. and {Lynds}, C.~R. and {Monet}, D.~G. and {O'Neil}, Jr., E.~J. and {Shaya}, E.~J. and {Seidelmann}, P.~K. and {Westphal}, J.~A.},
        title = "{Hubble Space Telescope WFPC2 Imaging of M16: Photoevaporation and Emerging Young Stellar Objects}",
      journal = {\aj},
         year = 1996,
        month = jun,
       volume = {111},
        pages = {2349},
          doi = {10.1086/117968},
       adsurl = {https://ui.adsabs.harvard.edu/abs/1996AJ....111.2349H}
}

@ARTICLE{M16_1,
       author = {{Hill}, T. and {Motte}, F. and {Didelon}, P. and {White}, G.~J. and {Marston}, A.~P. and {Nguy{\^e}n Luong}, Q. and {Bontemps}, S. and {Andr{\'e}}, Ph. and {Schneider}, N. and {Hennemann}, M. and {Sauvage}, M. and {Di Francesco}, J. and {Minier}, V. and {Anderson}, L.~D. and {Bernard}, J.~P. and {Elia}, D. and {Griffin}, M.~J. and {Li}, J.~Z. and {Peretto}, N. and {Pezzuto}, S. and {Polychroni}, D. and {Roussel}, H. and {Rygl}, K.~L.~J. and {Schisano}, E. and {Sousbie}, T. and {Testi}, L. and {Thompson}, D. Ward and {Zavagno}, A.},
        title = "{The M 16 molecular complex under the influence of NGC 6611. Herschel's perspective of the heating effect on the Eagle Nebula}",
      journal = {\aap},
         year = 2012,
        month = jun,
       volume = {542},
          eid = {A114},
        pages = {A114},
          doi = {10.1051/0004-6361/201219009},
archivePrefix = {arXiv},
       eprint = {1204.6317},
 primaryClass = {astro-ph.SR},
       adsurl = {https://ui.adsabs.harvard.edu/abs/2012A&A...542A.114H}
}

@ARTICLE{M16_2,
       author = {{Zhan}, Xiao-Liang and {Jiang}, Zhi-Bo and {Chen}, Zhi-Wei and {Zhang}, Miao-Miao and {Song}, Chao},
        title = "{Structures of GMC W 37}",
      journal = {Research in Astronomy and Astrophysics},
         year = 2016,
        month = apr,
       volume = {16},
       number = {4},
          eid = {56},
        pages = {56},
          doi = {10.1088/1674-4527/16/4/056},
archivePrefix = {arXiv},
       eprint = {1511.01243},
 primaryClass = {astro-ph.GA},
       adsurl = {https://ui.adsabs.harvard.edu/abs/2016RAA....16...56Z}
}

@ARTICLE{M16_3,
       author = {{Nishimura}, Atsushi and {Costes}, Jean and {Inaba}, Tetsuta and {Tachihara}, Kengo and {Hattori}, Yusuke and {Kohno}, Mikito and {Ohama}, Akio and {Torii}, Kazufumi and {Sano}, Hidetoshi and {Yamamoto}, Hiroaki and {Hasegawa}, Yutaka and {Kimura}, Kimihiro and {Ogawa}, Hideo and {Fukui}, Yasuo},
        title = "{A new view of the giant molecular cloud M16 (Eagle Nebula) in 12CO J=1-0 and 2-1 transitions with NANTEN2}",
      journal = {arXiv e-prints},
         year = 2017,
        month = jun,
          eid = {arXiv:1706.06002},
        pages = {arXiv:1706.06002},
          doi = {10.48550/arXiv.1706.06002},
archivePrefix = {arXiv},
       eprint = {1706.06002},
 primaryClass = {astro-ph.GA},
       adsurl = {https://ui.adsabs.harvard.edu/abs/2017arXiv170606002N}
}

@ARTICLE{M16_4,
       author = {{Xu}, Jin-Long and {Zavagno}, Annie and {Yu}, Naiping and {Liu}, Xiao-Lan and {Xu}, Ye and {Yuan}, Jinghua and {Zhang}, Chuan-Peng and {Zhang}, Si-Ju and {Zhang}, Guo-Yin and {Ning}, Chang-Chun and {Ju}, Bing-Gang},
        title = "{The effects of ionization feedback on star formation: a case study of the M 16 H II region}",
      journal = {\aap},
         year = 2019,
        month = jul,
       volume = {627},
          eid = {A27},
        pages = {A27},
          doi = {10.1051/0004-6361/201935024},
archivePrefix = {arXiv},
       eprint = {1905.08030},
 primaryClass = {astro-ph.SR},
       adsurl = {https://ui.adsabs.harvard.edu/abs/2019A&A...627A..27X}
}

@ARTICLE{M16_SOFIA_1,
       author = {{Karim}, Ramsey L. and {Pound}, Marc W. and {Tielens}, Alexander G.~G.~M. and {Tiwari}, Maitraiyee and {Bonne}, Lars and {Wolfire}, Mark G. and {Schneider}, Nicola and {Kavak}, {\"U}mit and {Mundy}, Lee G. and {Simon}, Robert and {G{\"u}sten}, Rolf and {Stutzki}, J{\"u}rgen and {Wyrowski}, Friedrich and {Honingh}, Netty},
        title = "{SOFIA FEEDBACK Survey: The Pillars of Creation in [C II] and Molecular Lines}",
      journal = {\aj},
         year = 2023,
        month = dec,
       volume = {166},
       number = {6},
          eid = {240},
        pages = {240},
          doi = {10.3847/1538-3881/acff6c},
archivePrefix = {arXiv},
       eprint = {2309.14637},
 primaryClass = {astro-ph.GA},
       adsurl = {https://ui.adsabs.harvard.edu/abs/2023AJ....166..240K}
}

@ARTICLE{M16_SOFIA_2,
       author = {{Mutale}, M. and {Thompson}, M.~A. and {Williams}, G.~M. and {Rigby}, A.~J. and {Hoare}, M.~G. and {Urquhart}, J.~S. and {Bietenholz}, M.~F. and {Bordiu}, C. and {Camilo}, F. and {Cotton}, W.~D. and {Goedhart}, S. and {Obonyo}, W.~O. and {Riggi}, S. and {Yang}, A.~Y.},
        title = "{The SARAO MeerKAT Galactic Plane Survey compact source catalogue}",
      journal = {\mnras},
         year = 2025,
        month = oct,
          doi = {10.1093/mnras/staf1849},
archivePrefix = {arXiv},
       eprint = {2510.23707},
 primaryClass = {astro-ph.GA},
       adsurl = {https://ui.adsabs.harvard.edu/abs/2025MNRAS.tmp.1749M}
}

@ARTICLE{M16_SOFIA_3,
       author = {{Karim}, Ramsey L. and {Pound}, Marc W. and {Tielens}, Alexander G.~G.~M. and {Kaastra}, Jelle S. and {Townsley}, Leisa K. and {Broos}, Patrick S. and {Tiwari}, Maitraiyee and {Bonne}, Lars and {Kavak}, {\"U}mit and {Wolfire}, Mark G. and {Schneider}, Nicola and {Simon}, Robert and {G{\"u}sten}, Rolf and {Stutzki}, J{\"u}rgen and {Mertens}, Marc and {Ricken}, Oliver and {Wyrowski}, Friedrich and {Mundy}, Lee G.},
        title = "{SOFIA FEEDBACK Survey: The Eagle Nebula in [C II] and Molecular Lines}",
      journal = {\apj},
         year = 2025,
        month = dec,
       volume = {995},
       number = {2},
          eid = {196},
        pages = {196},
          doi = {10.3847/1538-4357/ae17cc},
archivePrefix = {arXiv},
       eprint = {2511.03978},
 primaryClass = {astro-ph.GA},
       adsurl = {https://ui.adsabs.harvard.edu/abs/2025ApJ...995..196K}
}

@ARTICLE{N74_1,
       author = {{Beaumont}, Christopher N. and {Williams}, Jonathan P.},
        title = "{Molecular Rings Around Interstellar Bubbles and the Thickness of Star-Forming Clouds}",
      journal = {\apj},
         year = 2010,
        month = feb,
       volume = {709},
       number = {2},
        pages = {791-800},
          doi = {10.1088/0004-637X/709/2/791},
archivePrefix = {arXiv},
       eprint = {0912.1852},
 primaryClass = {astro-ph.GA},
       adsurl = {https://ui.adsabs.harvard.edu/abs/2010ApJ...709..791B}
}

@ARTICLE{N74_2,
       author = {{Sherman}, Reid A.},
        title = "{Investigation of Molecular Cloud Structure around Infrared Bubbles: CARMA Observations of N14, N22, and N74}",
      journal = {\apj},
         year = 2012,
        month = nov,
       volume = {760},
       number = {1},
          eid = {58},
        pages = {58},
          doi = {10.1088/0004-637X/760/1/58},
archivePrefix = {arXiv},
       eprint = {1210.3614},
 primaryClass = {astro-ph.GA},
       adsurl = {https://ui.adsabs.harvard.edu/abs/2012ApJ...760...58S}
}

@ARTICLE{N74_3,
       author = {{Alexander}, Michael J. and {Kobulnicky}, Henry A. and {Kerton}, Charles R. and {Arvidsson}, Kim},
        title = "{The Interstellar Bubbles of G38.9-0.4 and the Impact of Stellar Feedback on Star Formation}",
      journal = {\apj},
         year = 2013,
        month = jun,
       volume = {770},
       number = {1},
          eid = {1},
        pages = {1},
          doi = {10.1088/0004-637X/770/1/1},
archivePrefix = {arXiv},
       eprint = {1304.7251},
 primaryClass = {astro-ph.GA},
       adsurl = {https://ui.adsabs.harvard.edu/abs/2013ApJ...770....1A}
}

@ARTICLE{N74_4,
       author = {{Yan}, Qing-zeng and {Xu}, Ye and {Zhang}, Bo and {Lu}, Deng-rong and {Chen}, Xi and {Tang}, Zheng-hong},
        title = "{Molecular Lines of 13 Galactic Infrared Bubble Regions}",
      journal = {\aj},
         year = 2016,
        month = nov,
       volume = {152},
       number = {5},
          eid = {117},
        pages = {117},
          doi = {10.3847/0004-6256/152/5/117},
archivePrefix = {arXiv},
       eprint = {1609.03051},
 primaryClass = {astro-ph.GA},
       adsurl = {https://ui.adsabs.harvard.edu/abs/2016AJ....152..117Y}
}

@misc{BWFieldsZenodo,
  author       = {Jiang, Yu},
  title        = {BWFields: Identification and Analysis of Enclosed
                   Molecular Bubbles by Weight Fields
                  },
  month        = jun,
  year         = 2026,
  publisher    = {Zenodo},
  version      = {BWFields V0.0.2},
  doi          = {10.5281/zenodo.20621543},
  url          = {https://doi.org/10.5281/zenodo.20621543},
}

@ARTICLE{Numpy,
	author = {{Oliphant}, Travis E.},
	title = "{Python for Scientific Computing}",
	journal = {Computing in Science and Engineering},
	year = 2007,
	month = jan,
	volume = {9},
	number = {3},
	pages = {10-20},
	doi = {10.1109/MCSE.2007.58},
	adsurl = {https://ui.adsabs.harvard.edu/abs/2007CSE.....9c..10O}
}

@ARTICLE{Astropy_1,
       author = {{Astropy Collaboration} and {Robitaille}, Thomas P. and {Tollerud}, Erik J. and {Greenfield}, Perry and {Droettboom}, Michael and {Bray}, Erik and {Aldcroft}, Tom and {Davis}, Matt and {Ginsburg}, Adam and {Price-Whelan}, Adrian M. and {Kerzendorf}, Wolfgang E. and {Conley}, Alexander and {Crighton}, Neil and {Barbary}, Kyle and {Muna}, Demitri and {Ferguson}, Henry and {Grollier}, Fr{\'e}d{\'e}ric and {Parikh}, Madhura M. and {Nair}, Prasanth H. and {Unther}, Hans M. and {Deil}, Christoph and {Woillez}, Julien and {Conseil}, Simon and {Kramer}, Roban and {Turner}, James E.~H. and {Singer}, Leo and {Fox}, Ryan and {Weaver}, Benjamin A. and {Zabalza}, Victor and {Edwards}, Zachary I. and {Azalee Bostroem}, K. and {Burke}, D.~J. and {Casey}, Andrew R. and {Crawford}, Steven M. and {Dencheva}, Nadia and {Ely}, Justin and {Jenness}, Tim and {Labrie}, Kathleen and {Lim}, Pey Lian and {Pierfederici}, Francesco and {Pontzen}, Andrew and {Ptak}, Andy and {Refsdal}, Brian and {Servillat}, Mathieu and {Streicher}, Ole},
        title = "{Astropy: A community Python package for astronomy}",
      journal = {\aap},
         year = 2013,
        month = oct,
       volume = {558},
          eid = {A33},
        pages = {A33},
          doi = {10.1051/0004-6361/201322068},
archivePrefix = {arXiv},
       eprint = {1307.6212},
 primaryClass = {astro-ph.IM},
       adsurl = {https://ui.adsabs.harvard.edu/abs/2013A&A...558A..33A}
}

@ARTICLE{Astropy_2,
	author = {{Astropy Collaboration} and {Price-Whelan}, A.~M. and {Sip{\H{o}}cz}, B.~M. and {G{\"u}nther}, H.~M. and {Lim}, P.~L. and {Crawford}, S.~M. and {Conseil}, S. and {Shupe}, D.~L. and {Craig}, M.~W. and {Dencheva}, N. and {Ginsburg}, A. and {VanderPlas}, J.~T. and {Bradley}, L.~D. and {P{\'e}rez-Su{\'a}rez}, D. and {de Val-Borro}, M. and {Aldcroft}, T.~L. and {Cruz}, K.~L. and {Robitaille}, T.~P. and {Tollerud}, E.~J. and {Ardelean}, C. and {Babej}, T. and {Bach}, Y.~P. and {Bachetti}, M. and {Bakanov}, A.~V. and {Bamford}, S.~P. and {Barentsen}, G. and {Barmby}, P. and {Baumbach}, A. and {Berry}, K.~L. and {Biscani}, F. and {Boquien}, M. and {Bostroem}, K.~A. and {Bouma}, L.~G. and {Brammer}, G.~B. and {Bray}, E.~M. and {Breytenbach}, H. and {Buddelmeijer}, H. and {Burke}, D.~J. and {Calderone}, G. and {Cano Rodr{\'\i}guez}, J.~L. and {Cara}, M. and {Cardoso}, J.~V.~M. and {Cheedella}, S. and {Copin}, Y. and {Corrales}, L. and {Crichton}, D. and {D'Avella}, D. and {Deil}, C. and {Depagne}, {\'E}. and {Dietrich}, J.~P. and {Donath}, A. and {Droettboom}, M. and {Earl}, N. and {Erben}, T. and {Fabbro}, S. and {Ferreira}, L.~A. and {Finethy}, T. and {Fox}, R.~T. and {Garrison}, L.~H. and {Gibbons}, S.~L.~J. and {Goldstein}, D.~A. and {Gommers}, R. and {Greco}, J.~P. and {Greenfield}, P. and {Groener}, A.~M. and {Grollier}, F. and {Hagen}, A. and {Hirst}, P. and {Homeier}, D. and {Horton}, A.~J. and {Hosseinzadeh}, G. and {Hu}, L. and {Hunkeler}, J.~S. and {Ivezi{\'c}}, {\v{Z}}. and {Jain}, A. and {Jenness}, T. and {Kanarek}, G. and {Kendrew}, S. and {Kern}, N.~S. and {Kerzendorf}, W.~E. and {Khvalko}, A. and {King}, J. and {Kirkby}, D. and {Kulkarni}, A.~M. and {Kumar}, A. and {Lee}, A. and {Lenz}, D. and {Littlefair}, S.~P. and {Ma}, Z. and {Macleod}, D.~M. and {Mastropietro}, M. and {McCully}, C. and {Montagnac}, S. and {Morris}, B.~M. and {Mueller}, M. and {Mumford}, S.~J. and {Muna}, D. and {Murphy}, N.~A. and {Nelson}, S. and {Nguyen}, G.~H. and {Ninan}, J.~P. and {N{\"o}the}, M. and {Ogaz}, S. and {Oh}, S. and {Parejko}, J.~K. and {Parley}, N. and {Pascual}, S. and {Patil}, R. and {Patil}, A.~A. and {Plunkett}, A.~L. and {Prochaska}, J.~X. and {Rastogi}, T. and {Reddy Janga}, V. and {Sabater}, J. and {Sakurikar}, P. and {Seifert}, M. and {Sherbert}, L.~E. and {Sherwood-Taylor}, H. and {Shih}, A.~Y. and {Sick}, J. and {Silbiger}, M.~T. and {Singanamalla}, S. and {Singer}, L.~P. and {Sladen}, P.~H. and {Sooley}, K.~A. and {Sornarajah}, S. and {Streicher}, O. and {Teuben}, P. and {Thomas}, S.~W. and {Tremblay}, G.~R. and {Turner}, J.~E.~H. and {Terr{\'o}n}, V. and {van Kerkwijk}, M.~H. and {de la Vega}, A. and {Watkins}, L.~L. and {Weaver}, B.~A. and {Whitmore}, J.~B. and {Woillez}, J. and {Zabalza}, V. and {Astropy Contributors}},
	title = "{The Astropy Project: Building an Open-science Project and Status of the v2.0 Core Package}",
	journal = {\aj},
	year = 2018,
	month = sep,
	volume = {156},
	number = {3},
	eid = {123},
	pages = {123},
	doi = {10.3847/1538-3881/aabc4f},
	archivePrefix = {arXiv},
	eprint = {1801.02634},
	primaryClass = {astro-ph.IM},
	adsurl = {https://ui.adsabs.harvard.edu/abs/2018AJ....156..123A}
}

@ARTICLE{Astropy_3,
       author = {{Astropy Collaboration} and {Price-Whelan}, Adrian M. and {Lim}, Pey Lian and {Earl}, Nicholas and {Starkman}, Nathaniel and {Bradley}, Larry and {Shupe}, David L. and {Patil}, Aarya A. and {Corrales}, Lia and {Brasseur}, C.~E. and {N{\"o}the}, Maximilian and {Donath}, Axel and {Tollerud}, Erik and {Morris}, Brett M. and {Ginsburg}, Adam and {Vaher}, Eero and {Weaver}, Benjamin A. and {Tocknell}, James and {Jamieson}, William and {van Kerkwijk}, Marten H. and {Robitaille}, Thomas P. and {Merry}, Bruce and {Bachetti}, Matteo and {G{\"u}nther}, H. Moritz and {Aldcroft}, Thomas L. and {Alvarado-Montes}, Jaime A. and {Archibald}, Anne M. and {B{\'o}di}, Attila and {Bapat}, Shreyas and {Barentsen}, Geert and {Baz{\'a}n}, Juanjo and {Biswas}, Manish and {Boquien}, M{\'e}d{\'e}ric and {Burke}, D.~J. and {Cara}, Daria and {Cara}, Mihai and {Conroy}, Kyle E. and {Conseil}, Simon and {Craig}, Matthew W. and {Cross}, Robert M. and {Cruz}, Kelle L. and {D'Eugenio}, Francesco and {Dencheva}, Nadia and {Devillepoix}, Hadrien A.~R. and {Dietrich}, J{\"o}rg P. and {Eigenbrot}, Arthur Davis and {Erben}, Thomas and {Ferreira}, Leonardo and {Foreman-Mackey}, Daniel and {Fox}, Ryan and {Freij}, Nabil and {Garg}, Suyog and {Geda}, Robel and {Glattly}, Lauren and {Gondhalekar}, Yash and {Gordon}, Karl D. and {Grant}, David and {Greenfield}, Perry and {Groener}, Austen M. and {Guest}, Steve and {Gurovich}, Sebastian and {Handberg}, Rasmus and {Hart}, Akeem and {Hatfield-Dodds}, Zac and {Homeier}, Derek and {Hosseinzadeh}, Griffin and {Jenness}, Tim and {Jones}, Craig K. and {Joseph}, Prajwel and {Kalmbach}, J. Bryce and {Karamehmetoglu}, Emir and {Ka{\l}uszy{\'n}ski}, Miko{\l}aj and {Kelley}, Michael S.~P. and {Kern}, Nicholas and {Kerzendorf}, Wolfgang E. and {Koch}, Eric W. and {Kulumani}, Shankar and {Lee}, Antony and {Ly}, Chun and {Ma}, Zhiyuan and {MacBride}, Conor and {Maljaars}, Jakob M. and {Muna}, Demitri and {Murphy}, N.~A. and {Norman}, Henrik and {O'Steen}, Richard and {Oman}, Kyle A. and {Pacifici}, Camilla and {Pascual}, Sergio and {Pascual-Granado}, J. and {Patil}, Rohit R. and {Perren}, Gabriel I. and {Pickering}, Timothy E. and {Rastogi}, Tanuj and {Roulston}, Benjamin R. and {Ryan}, Daniel F. and {Rykoff}, Eli S. and {Sabater}, Jose and {Sakurikar}, Parikshit and {Salgado}, Jes{\'u}s and {Sanghi}, Aniket and {Saunders}, Nicholas and {Savchenko}, Volodymyr and {Schwardt}, Ludwig and {Seifert-Eckert}, Michael and {Shih}, Albert Y. and {Jain}, Anany Shrey and {Shukla}, Gyanendra and {Sick}, Jonathan and {Simpson}, Chris and {Singanamalla}, Sudheesh and {Singer}, Leo P. and {Singhal}, Jaladh and {Sinha}, Manodeep and {Sip{\H{o}}cz}, Brigitta M. and {Spitler}, Lee R. and {Stansby}, David and {Streicher}, Ole and {{\v{S}}umak}, Jani and {Swinbank}, John D. and {Taranu}, Dan S. and {Tewary}, Nikita and {Tremblay}, Grant R. and {de Val-Borro}, Miguel and {Van Kooten}, Samuel J. and {Vasovi{\'c}}, Zlatan and {Verma}, Shresth and {de Miranda Cardoso}, Jos{\'e} Vin{\'\i}cius and {Williams}, Peter K.~G. and {Wilson}, Tom J. and {Winkel}, Benjamin and {Wood-Vasey}, W.~M. and {Xue}, Rui and {Yoachim}, Peter and {Zhang}, Chen and {Zonca}, Andrea and {Astropy Project Contributors}},
        title = "{The Astropy Project: Sustaining and Growing a Community-oriented Open-source Project and the Latest Major Release (v5.0) of the Core Package}",
      journal = {\apj},
         year = 2022,
        month = aug,
       volume = {935},
       number = {2},
          eid = {167},
        pages = {167},
          doi = {10.3847/1538-4357/ac7c74},
archivePrefix = {arXiv},
       eprint = {2206.14220},
 primaryClass = {astro-ph.IM},
       adsurl = {https://ui.adsabs.harvard.edu/abs/2022ApJ...935..167A}
}

@ARTICLE{Matplotlib,
	author = {{Hunter}, John D.},
	title = "{Matplotlib: A 2D Graphics Environment}",
	journal = {Computing in Science and Engineering},
	year = 2007,
	month = may,
	volume = {9},
	number = {3},
	pages = {90-95},
	doi = {10.1109/MCSE.2007.55},
	adsurl = {https://ui.adsabs.harvard.edu/abs/2007CSE.....9...90H}
}

@Article{scikit-image,
	title		= {scikit-image: image processing in {P}ython},
	author	= {van der Walt, {S}t\'efan and {S}ch\"onberger, {J}ohannes
	{L}. and {Nunez-Iglesias}, {J}uan and {B}oulogne,
	{F}ran\c{c}ois and {W}arner, {J}oshua {D}. and {Y}ager,
	{N}eil and {G}ouillart, {E}mmanuelle and {Y}u, {T}ony and
	the scikit-image contributors},
	year		= {2014},
	month		= {6},
	volume	= {2},
	pages		= {e453},
	journal	= {PeerJ},
	issn		= {2167-8359},
	url		= {https://doi.org/10.7717/peerj.453},
	doi		= {10.7717/peerj.453}
}

@Article{scikit-learn,
	title		= {Scikit-learn: Machine Learning in {P}ython},
	author	= {Pedregosa, F. and Varoquaux, G. and Gramfort, A. and
	Michel, V. and Thirion, B. and Grisel, O. and Blondel, M.
	and Prettenhofer, P. and Weiss, R. and Dubourg, V. and
	Vanderplas, J. and Passos, A. and Cournapeau, D. and
	Brucher, M. and Perrot, M. and Duchesnay, E.},
	journal	= {Journal of Machine Learning Research},
	volume	= {12},
	pages		= {2825--2830},
	year		= {2011}
}

@ARTICLE{SciPy,
       author = {{Virtanen}, Pauli and {Gommers}, Ralf and {Oliphant}, Travis E. and {Haberland}, Matt and {Reddy}, Tyler and {Cournapeau}, David and {Burovski}, Evgeni and {Peterson}, Pearu and {Weckesser}, Warren and {Bright}, Jonathan and {van der Walt}, St{\'e}fan J. and {Brett}, Matthew and {Wilson}, Joshua and {Millman}, K. Jarrod and {Mayorov}, Nikolay and {Nelson}, Andrew R.~J. and {Jones}, Eric and {Kern}, Robert and {Larson}, Eric and {Carey}, C.~J. and {Polat}, {\.I}lhan and {Feng}, Yu and {Moore}, Eric W. and {VanderPlas}, Jake and {Laxalde}, Denis and {Perktold}, Josef and {Cimrman}, Robert and {Henriksen}, Ian and {Quintero}, E.~A. and {Harris}, Charles R. and {Archibald}, Anne M. and {Ribeiro}, Ant{\^o}nio H. and {Pedregosa}, Fabian and {van Mulbregt}, Paul and {SciPy 1. 0 Contributors}},
        title = "{SciPy 1.0: fundamental algorithms for scientific computing in Python}",
      journal = {Nature Methods},
         year = 2020,
        month = feb,
       volume = {17},
        pages = {261-272},
          doi = {10.1038/s41592-019-0686-2},
archivePrefix = {arXiv},
       eprint = {1907.10121},
 primaryClass = {cs.MS},
       adsurl = {https://ui.adsabs.harvard.edu/abs/2020NatMe..17..261V}
}

@article{Networkx,
	title = {Exploring network structure, dynamics, and function using NetworkX},
	author = {{Hagberg}, Aric and {Swart}, Pieter J. and {Schult}, Daniel A.},
	journal = {Proceedings of the 7th Python in Science Conference (SciPy2008)},
	abstractNote = {NetworkX is a Python language package for exploration and analysis of networks and network algorithms. The core package provides data structures for representing many types of networks, or graphs, including simple graphs, directed graphs, and graphs with parallel edges and self loops. The nodes in NetworkX graphs can be any (hashable) Python object and edges can contain arbitrary data; this flexibility mades NetworkX ideal for representing networks found in many different scientific fields. In addition to the basic data structures many graph algorithms are implemented for calculating network properties and structure measures: shortest paths, betweenness centrality, clustering, and degree distribution and many more. NetworkX can read and write various graph formats for eash exchange with existing data, and provides generators for many classic graphs and popular graph models, such as the Erdoes-Renyi, Small World, and Barabasi-Albert models, are included. The ease-of-use and flexibility of the Python programming language together with connection to the SciPy tools make NetworkX a powerful tool for scientific computations. We discuss some of our recent work studying synchronization of coupled oscillators to demonstrate how NetworkX enables research in the field of computational networks.},
	doi = {},
	url = {https://www.osti.gov/biblio/960616},
	place = {United States},
	year = {2008},
	month = {1}
}
\bibliographystyle{aasjournal}

\appendix
\begin{appendices}
\renewcommand{\thetable}{\thesection.\arabic{table}}
\renewcommand{\thefigure}{\thesection.\arabic{figure}}

\section{Verified Bubble Cases}
\label{sec:appendix_bubbles}
\setcounter{figure}{0}

The verified bubbles presented here serve as methodological benchmarks for assessing the reliability of the BWFields. 
These examples show that the combined use of the bubble-weight field, emission-defined intensity skeletons, and PV-based diagnostics consistently recovers well-defined shell structures and enables direct measurement of their geometric and kinematic properties from CO spectral-line data, while also supporting the symmetry and kinematic-significance thresholds adopted in the main analysis.

\begin{figure*}
\centering
\vspace{0cm}
\begin{minipage}[t]{0.32\textwidth}
    \centering
    \centerline{\includegraphics[width=2.25in]{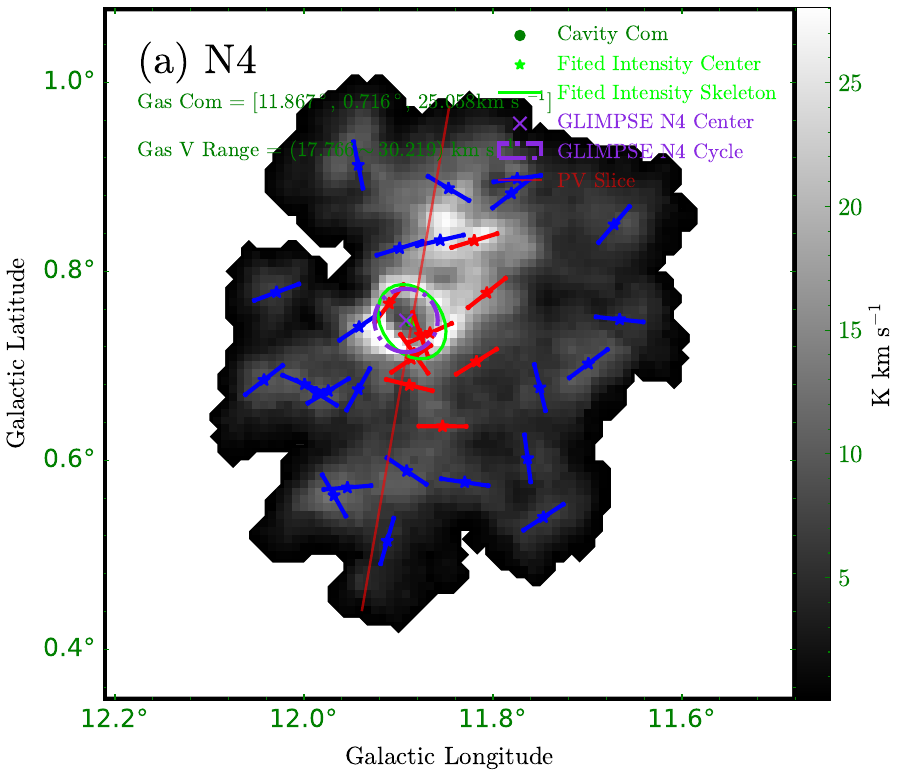}}
\end{minipage}%
\begin{minipage}[t]{0.32\textwidth}
    \centering
    \centerline{\includegraphics[width=2.05in]{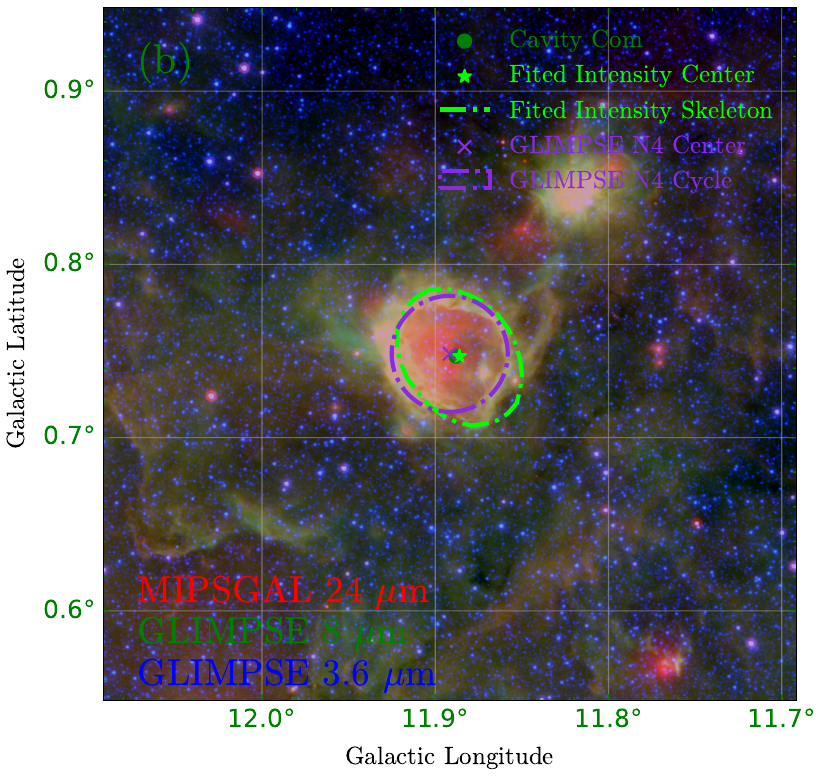}}
    %		\vspace{-4cm}
    %		\centerline{(a)}
\end{minipage}
\begin{minipage}[t]{0.32\textwidth}
    \centering
    \centerline{\includegraphics[width=2.25in]{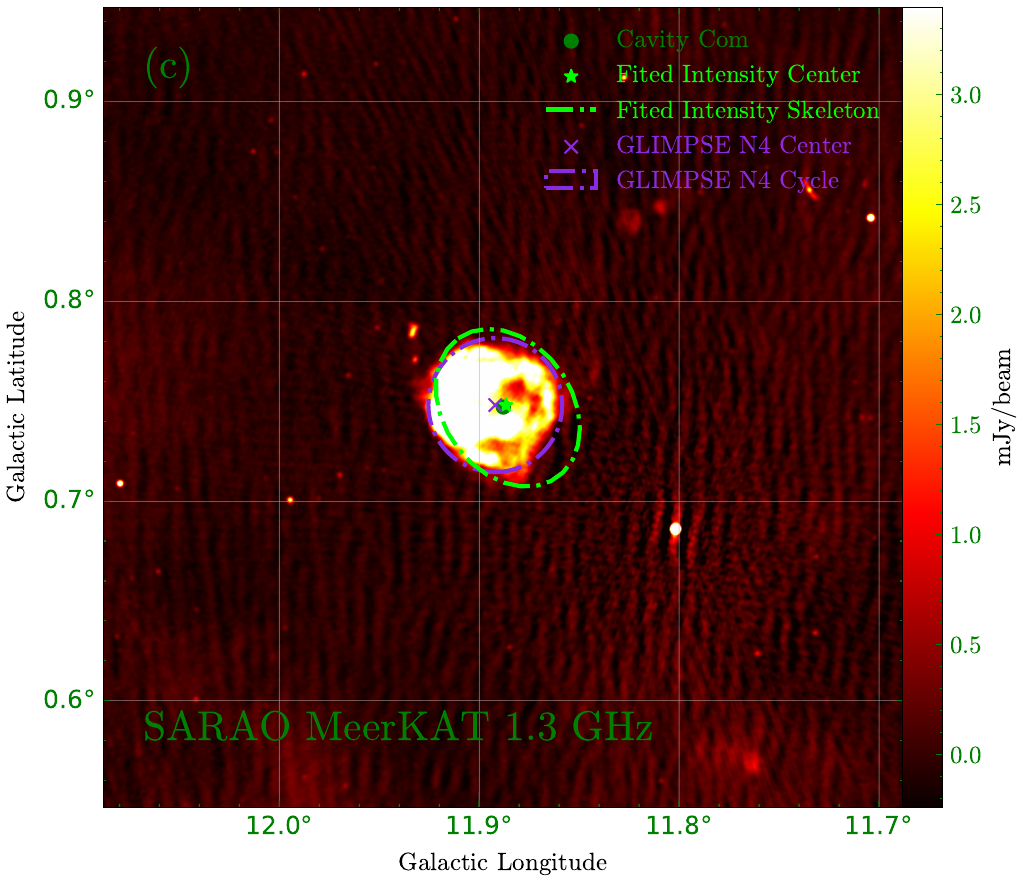}}
\end{minipage}

\begin{minipage}[t]{0.32\textwidth}
    \centering
    \centerline{\includegraphics[width=2.2in]{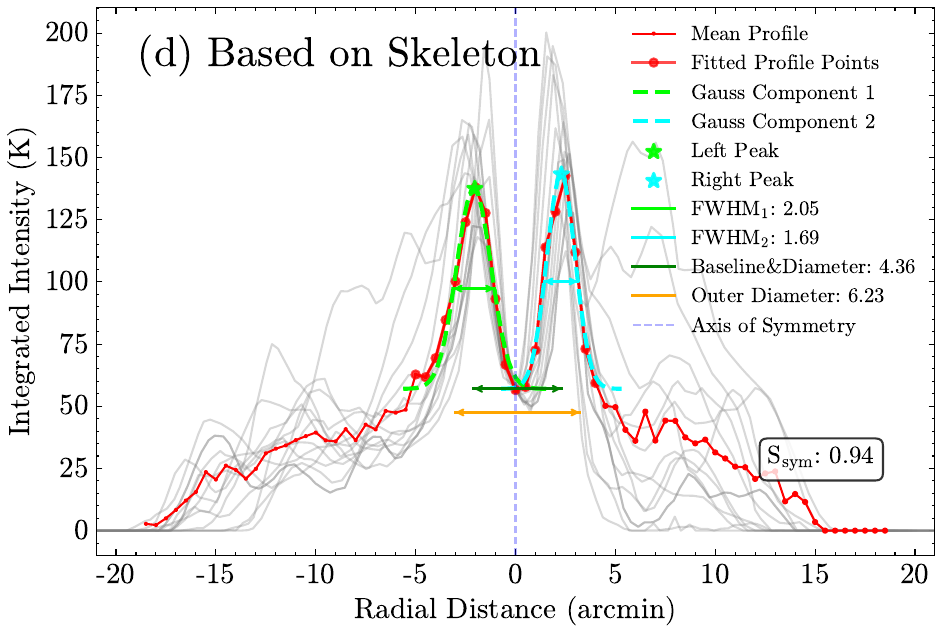}}
\end{minipage}%
\begin{minipage}[t]{0.32\textwidth} 
    \centering
    \centerline{\includegraphics[width=2.1in]{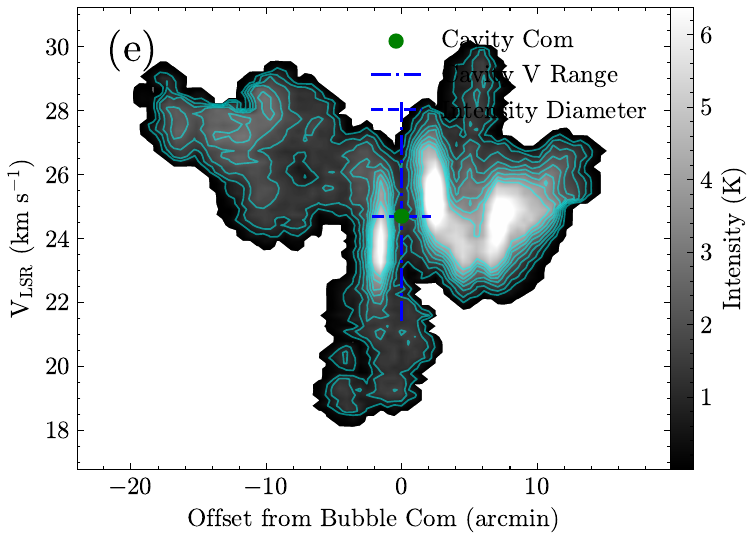}}
    %		\vspace{-4cm}
    %		\centerline{(a)}
\end{minipage}%
\begin{minipage}[t]{0.32\textwidth}
    \centering
    \centerline{\includegraphics[width=2.2in]{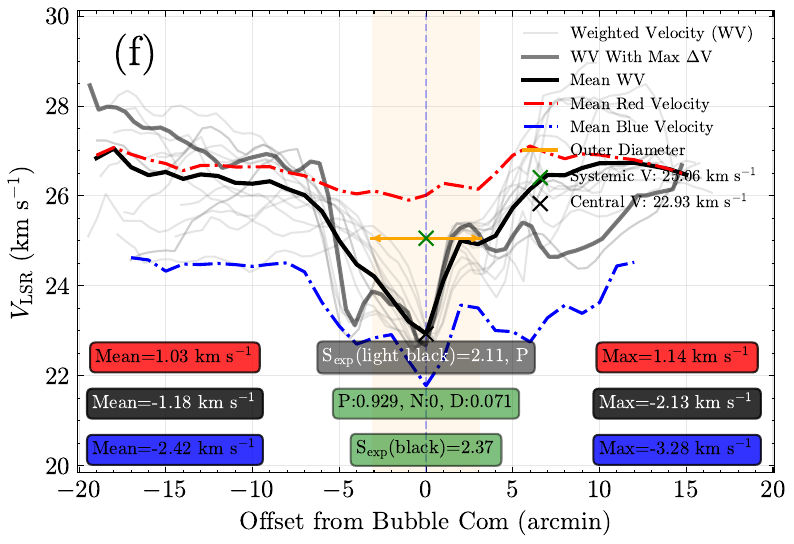}}
    %		\vspace{-4cm}
    %		\centerline{(a)}
\end{minipage}%
\caption{Example of morphological, kinematic, and multiwavelength analysis for the bubble N4  \citep{Churchwell_1}. 
(a) Velocity-integrated $^{13}$CO intensity map. The green circle marks the cavity centroid, and the cyan curve traces the emission-defined intensity skeleton with its fitted center (cyan star). Red markers denote clumps directly associated with the inner weight-clump, while blue markers indicate connected clumps in the surrounding envelope. The blue-violet cross and circle mark the GLIMPSE N4 center and radius. Annotated values indicate the centroid and velocity range of the associated molecular gas. 
(b) Mid-infrared three-color image of N4, composed of MIPSGAL 24\,$\mu$m (red), GLIMPSE 8.0\,$\mu$m (green), and GLIMPSE 3.6\,$\mu$m (blue), with the fitted intensity skeleton overlaid. The spatial correspondence between the molecular shell and the infrared bubble is evident. 
(c) MeerKAT 1.3\,GHz continuum image of the same region, with the fitted intensity skeleton overlaid, showing the spatial correspondence between the radio emission and the molecular shell. 
(d) Ensemble of radial intensity profiles from the second analysis, sampled along the ellipse fitted to the intensity skeleton, together with the mean profile and its constrained double-Gaussian fit. The fitted shell thickness, diameters, and symmetry score are indicated. 
(e) Representative PV diagram extracted along a selected slice. The $^{13}$CO emission is shown in contours, with the cavity centroid, velocity range, and intensity diameter marked. 
(f) Ensemble of intensity-weighted mean-velocity profiles from all radial cuts and their azimuthal mean, together with the redshifted and blueshifted components. Colored boxes summarize the velocity offsets within the outer radius, the PV classification, and the turbulence-normalised expansion significance.
}
\label{Img_Appendix_N4}
\end{figure*}

N4 (Figure~\ref{Img_Appendix_N4}) is an isolated, morphologically well-defined molecular bubble associated with MeerKAT 1.3\,GHz emission \citep{MeerKAT_0}, which traces ionized gas within H\,{\sc ii} regions. 
The mid-infrared three-color image (Figure~\ref{Img_Appendix_N4}(b)) reveals a clear ring structure: the 8.0\,$\mu$m emission (dominated by PAH features) delineates the photodissociation region (PDR), while the 24\,$\mu$m emission traces warm dust within the bubble interior \citep{GLIMPSE_0,MIPSGAL_0,MIPSGAL_1,MIPSGAL_2,MIPSGAL_3}. 
The 3.6\,$\mu$m emission is dominated by stellar continuum and primarily traces the stellar environment \citep{Spitzer_0}. 
This infrared morphology closely coincides with the molecular shell traced by the $^{13}$CO emission and the fitted intensity skeleton, demonstrating the physical correspondence between the molecular cavity, the PDR shell, and the ionized interior. 

The well-defined shell, high radial symmetry, and clear red--blue velocity separation provide strong evidence for an expanding structure. 
While broadly consistent with the expanding bubble model \citep{Arce2011}, the PV structure does not form a smooth elliptical or arc-like pattern, instead exhibiting clear intensity gaps. 
This source serves as a reference case for illustrating the BWFields workflow, including skeleton extraction, radial-profile fitting, and the derivation of expansion diagnostics, and represents a high-confidence bubble within the candidate population.

\begin{figure*}
\centering
\vspace{0cm}
\begin{minipage}[t]{0.32\textwidth}
    \centering
    \centerline{\includegraphics[width=2.25in]{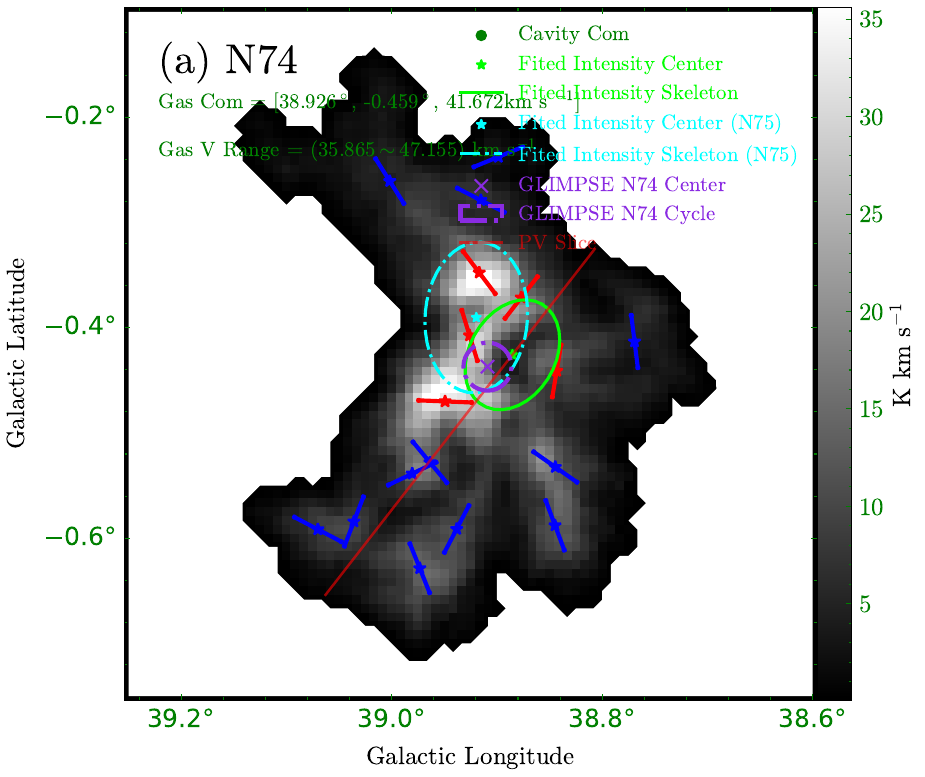}}
    %		\vspace{-4cm}
    %		\centerline{(a)}
\end{minipage}%
\begin{minipage}[t]{0.32\textwidth}
    \centering
    \centerline{\includegraphics[width=2in]{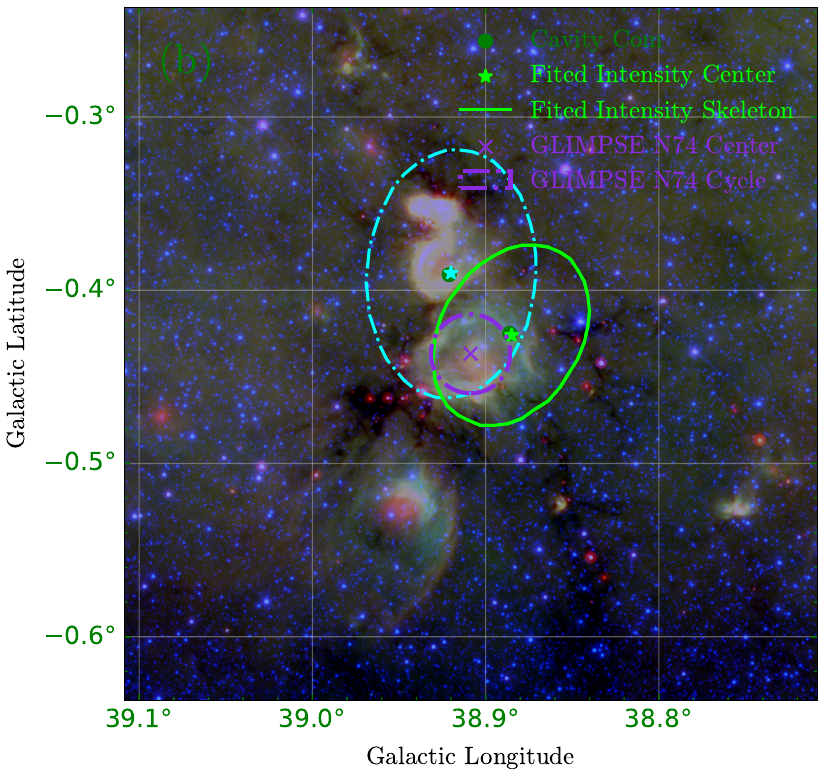}}
    %		\vspace{-4cm}
    %		\centerline{(a)}
\end{minipage}
\begin{minipage}[t]{0.32\textwidth}
    \centering
    \centerline{\includegraphics[width=2.25in]{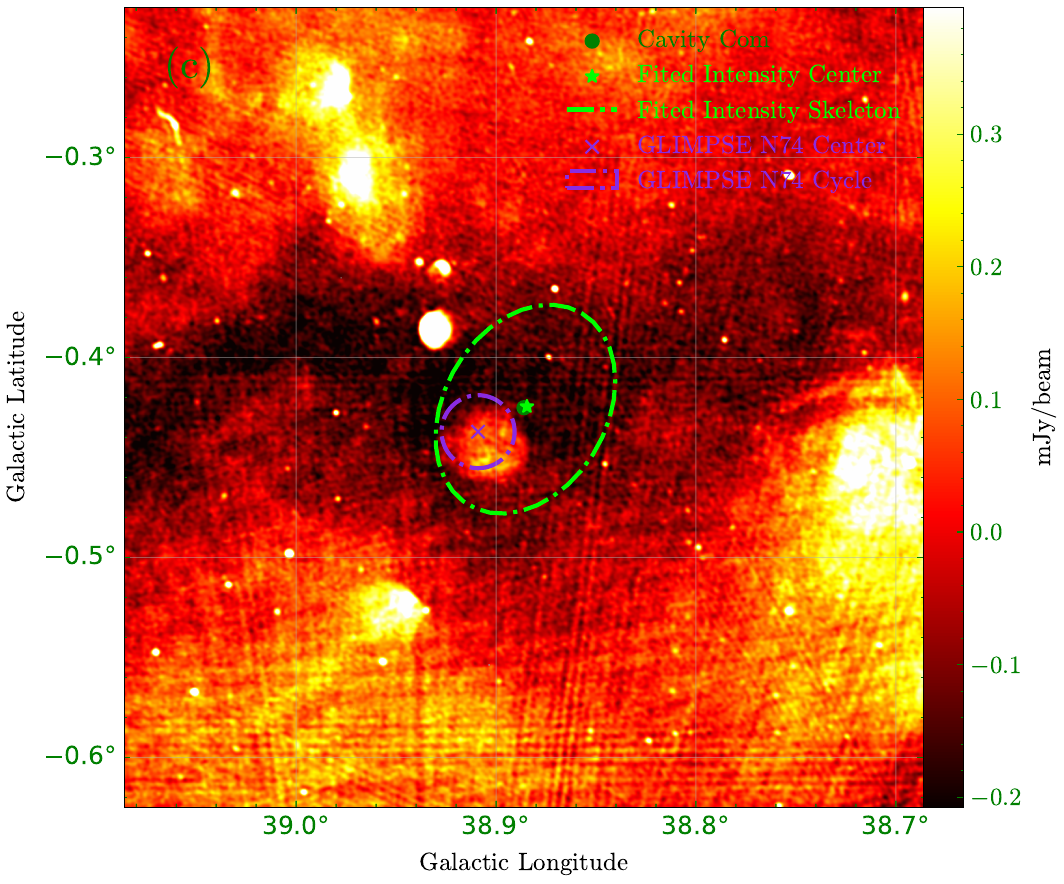}}
    %		\vspace{-4cm}
    %		\centerline{(a)}
\end{minipage}

\begin{minipage}[t]{0.32\textwidth}
    \centering
    \centerline{\includegraphics[width=2.2in]{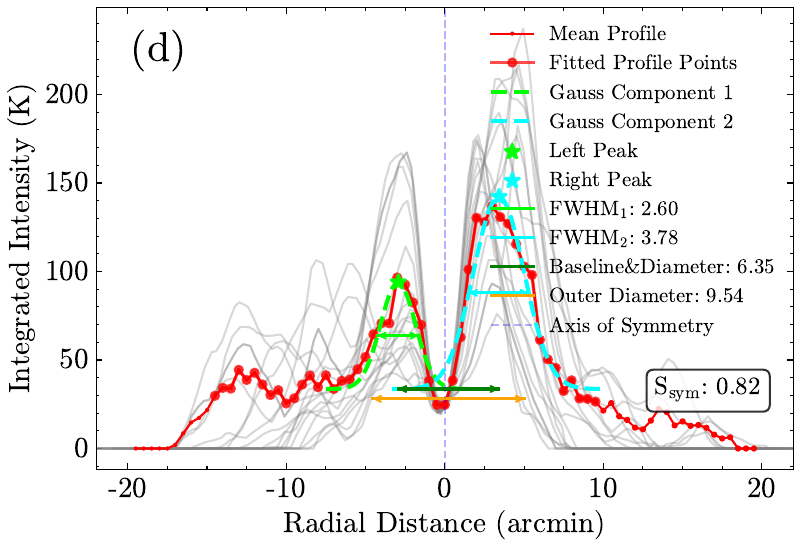}}
    %		\vspace{-4cm}
    %		\centerline{(a)}
\end{minipage}%
\begin{minipage}[t]{0.32\textwidth} 
    \centering
    \centerline{\includegraphics[width=2.15in]{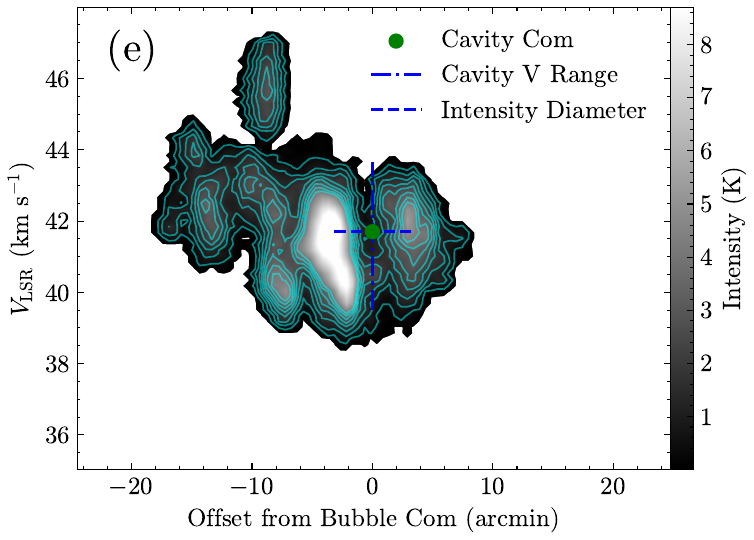}}
    %		\vspace{-4cm}
    %		\centerline{(a)}
\end{minipage}%
\begin{minipage}[t]{0.32\textwidth}
    \centering
    \centerline{\includegraphics[width=2.2in]{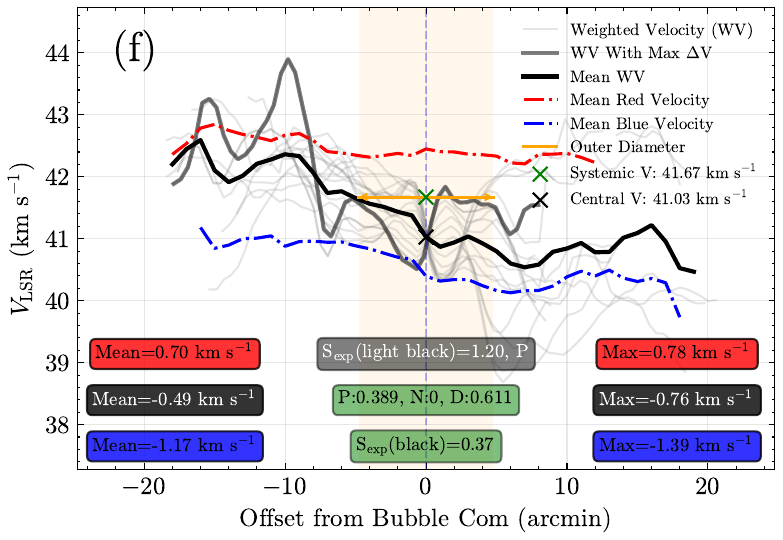}}
    %		\vspace{-4cm}
    %		\centerline{(a)}
\end{minipage}%
\caption{Morphological, kinematic, and multiwavelength views of the molecular bubble N74 (panels as in Figure~\ref{Img_Appendix_N4}). In panels~(a) and (b), the fitted intensity skeletons of both N74 and N75 are overplotted to highlight their projected overlap. N74 has a systemic velocity $V_{\rm sys}=41.69~{\rm km~s^{-1}}$, substantially lower than that of N75 (Figure~\ref{Img_Appendix_N75}), indicating that the two bubbles correspond to distinct molecular layers along the line of sight.
}
\label{Img_Appendix_N74}
\end{figure*}

\begin{figure*}
\centering
\vspace{0cm}
\begin{minipage}[t]{0.32\textwidth}
    \centering
    \centerline{\includegraphics[width=2.25in]{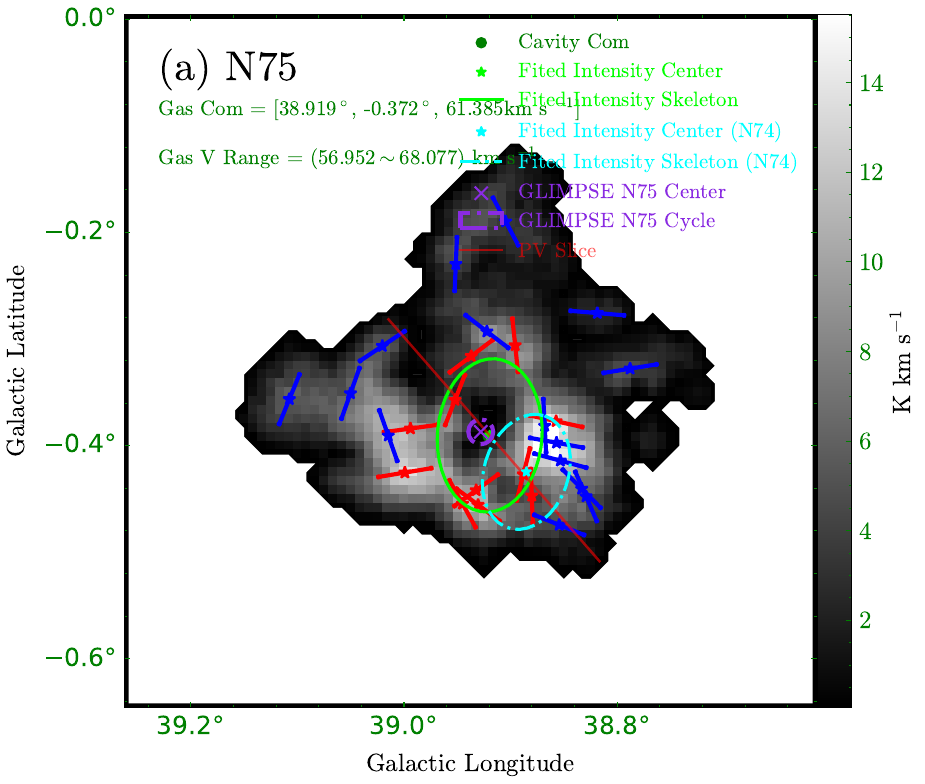}}
    %		\vspace{-4cm}
    %		\centerline{(a)}
\end{minipage}%
\begin{minipage}[t]{0.32\textwidth}
    \centering
    \centerline{\includegraphics[width=2in]{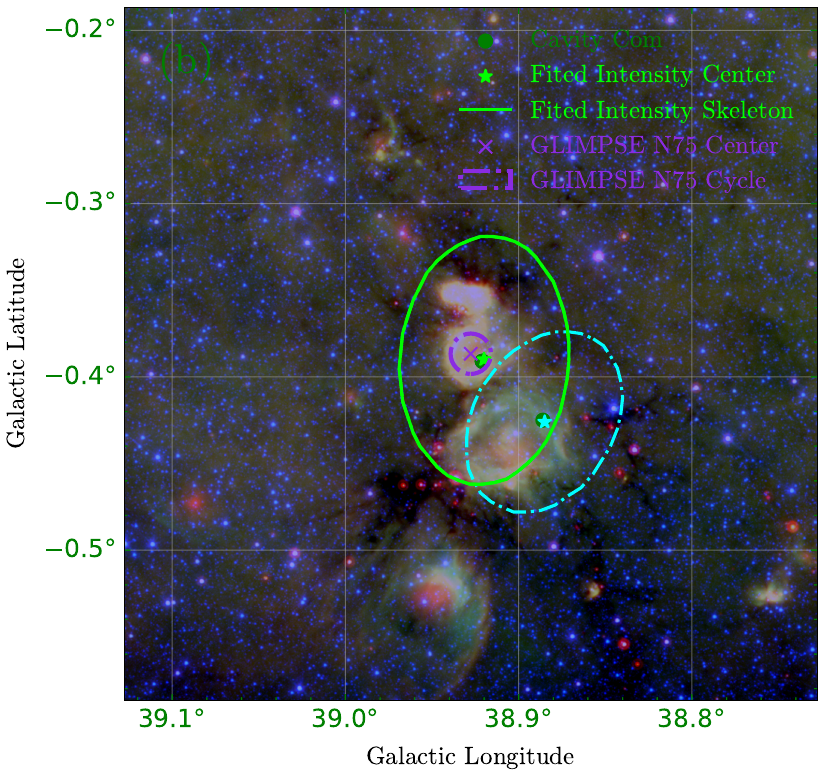}}
    %		\vspace{-4cm}
    %		\centerline{(a)}
\end{minipage}
\begin{minipage}[t]{0.32\textwidth}
    \centering
    \centerline{\includegraphics[width=2.25in]{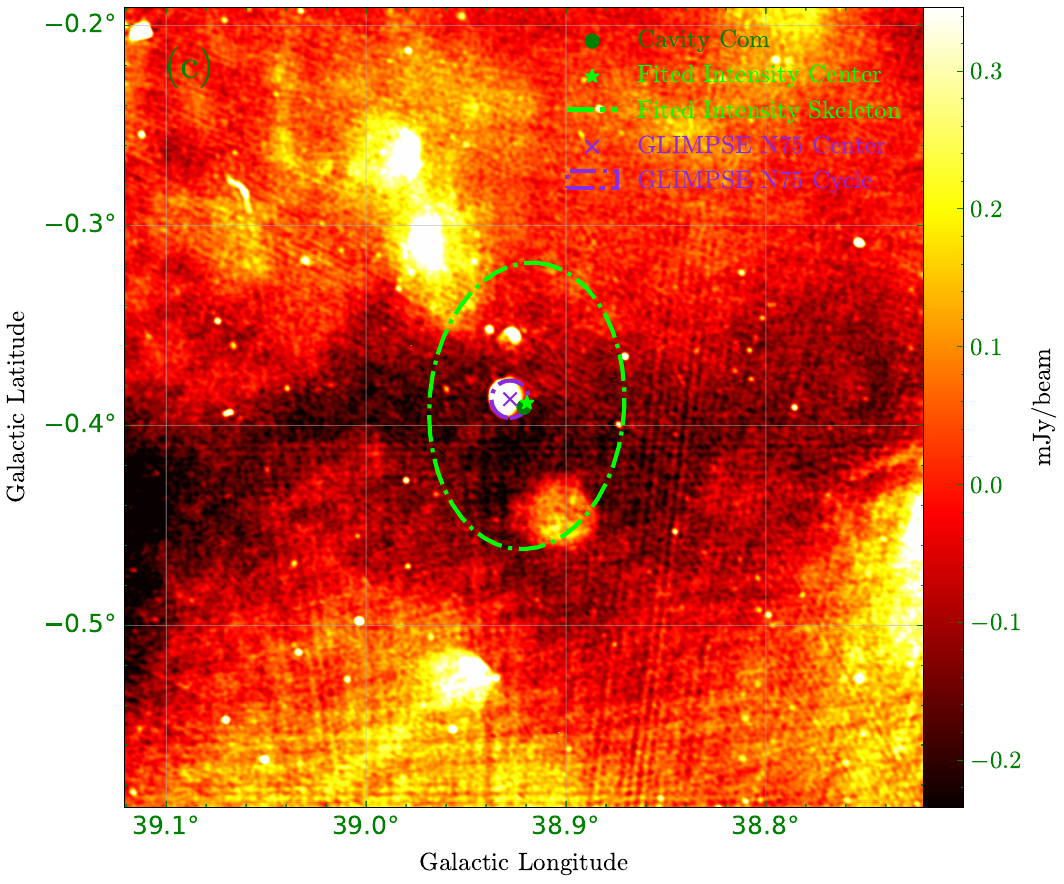}}
    %		\vspace{-4cm}
    %		\centerline{(a)}
\end{minipage}

\begin{minipage}[t]{0.32\textwidth}
    \centering
    \centerline{\includegraphics[width=2.2in]{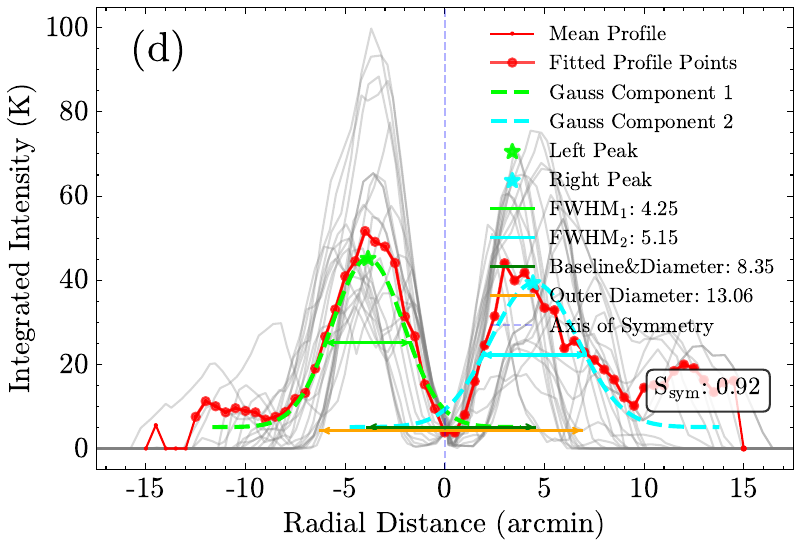}}
    %		\vspace{-4cm}
    %		\centerline{(a)}
\end{minipage}%
\begin{minipage}[t]{0.32\textwidth} 
    \centering
    \centerline{\includegraphics[width=2.15in]{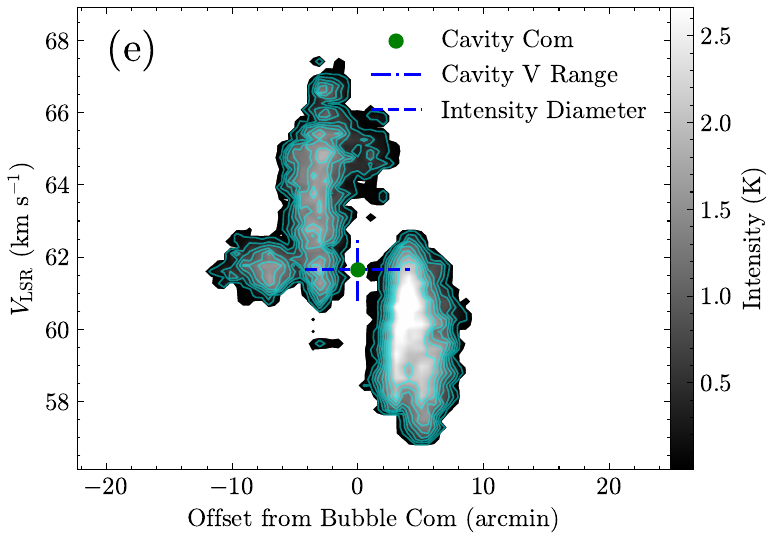}}
    %		\vspace{-4cm}
    %		\centerline{(a)}
\end{minipage}%
\begin{minipage}[t]{0.32\textwidth}
    \centering
    \centerline{\includegraphics[width=2.2in]{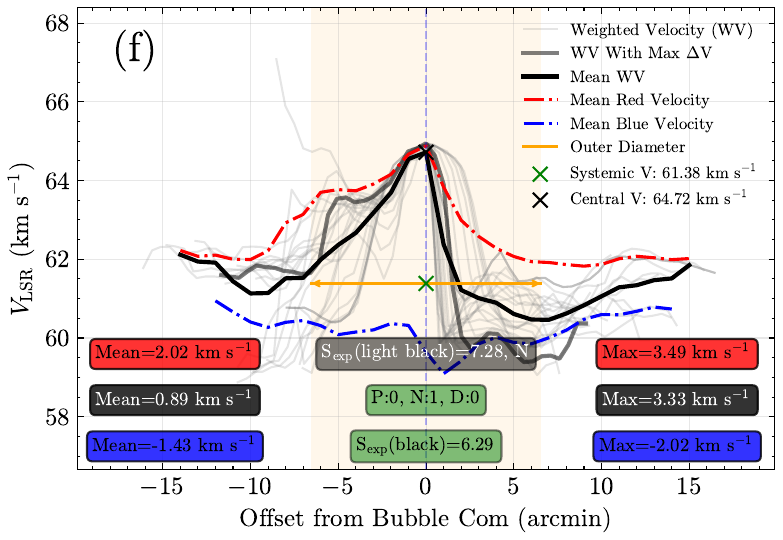}}
    %		\vspace{-4cm}
    %		\centerline{(a)}
\end{minipage}%
\caption{Morphological, kinematic, and multiwavelength views of the molecular bubble N75 (panels as in Figure~\ref{Img_Appendix_N4}). 
Panel~(a) and (b) also show the fitted intensity skeleton of N74 for comparison. 
The systemic velocity of N75 ($V_{\rm sys}=61.39~{\rm km~s^{-1}}$) differs from that of N74 (Figure~\ref{Img_Appendix_N74}) by $\sim 20~{\rm km~s^{-1}}$, indicating that the two spatially coincident bubbles arise from distinct molecular layers.}
\label{Img_Appendix_N75}
\end{figure*}

N74 and N75 (Figures~\ref{Img_Appendix_N74} and \ref{Img_Appendix_N75}) provide a more demanding test of the method. 
The mid-infrared images (Figures~\ref{Img_Appendix_N74}(b) and \ref{Img_Appendix_N75}(b)) show ring-like emission structures.
These infrared shells broadly follow the molecular structures identified in $^{13}$CO, although their apparent centers and morphologies partially overlap in projection. 
Although the two bubbles overlap in projection and exhibit partially superposed intensity skeletons, they are clearly separated in velocity space, with systemic velocities of $V_{\rm sys}=41.7$ and $61.4~{\rm km~s^{-1}}$, respectively. 
The PPV-based analysis, therefore, identifies them as two distinct velocity-separated bubbles, each characterized by symmetric radial intensity profiles and organized expansion-like velocity patterns. 
While previous studies have noted their apparent association within the same projected molecular complex (Figure~\ref{Img_Appendix_N74}(a); \citealt{N74_1,N74_2,N74_3,N74_4}), the present results demonstrate that velocity-resolved diagnostics are essential for disentangling such line-of-sight superpositions. 

Both bubbles also exhibit clear counterparts in the MeerKAT 1.3\,GHz continuum images (Figures~\ref{Img_Appendix_N74}(c) and \ref{Img_Appendix_N75}(c)), providing independent evidence that the identified molecular cavities correspond to ionized interiors. 
Notably, the CO emission associated with N74 is offset from the central position of the dust and ionizing source, whereas the latter shows a closer morphological correspondence with a faint cavity located at the lower-right edge of N75. 
As this cavity is not sufficiently prominent to be identified as an independent bubble, BWFields recognizes it as part of N75. 
This configuration suggests that both N74 and N75 are more likely linked to the same molecular cloud associated with N75 (Figure~\ref{Img_Appendix_N75}(a)), rather than to a distinct molecular structure traditionally attributed to N74 (Figure~\ref{Img_Appendix_N74}(a)). 
More generally, these results highlight that establishing consistent associations between molecular gas, PDR structures, and ionized emission across different wavelengths remains a nontrivial challenge \citep{MeerKAT_Associated_With_Clouds}. 

In addition, N74 exhibits an extended arm-like substructure in its spatial morphology (Figure~\ref{Img_Appendix_N74}(a)), indicating a significant nonaxisymmetric component. The PV diagram reveals a central structure reminiscent of a Keplerian disk (Figure~\ref{Img_Appendix_N74}(e)). Furthermore, subsets of the velocity profiles show an S-shaped velocity pattern within the outer radial extent of the shell (Figure~\ref{Img_Appendix_N74}(f)). These kinematic features indicate clear deviations from simple spherical expansion and are consistent with a scenario in which N74 may represent a rotating or dynamically sheared bubble, warranting further investigation.

Taken together, these benchmark cases demonstrate that BWFields robustly recovers well-defined shell structures across a range of environments, including isolated bubbles and overlapping systems along the line of sight. 
The alignment between shell morphology, emission-defined geometry, and organised velocity structure supports the robustness of the identification procedure and motivates its application to the full bubble sample analysed in the main text, while also indicating that BWFields can be further refined to accommodate more complex environments and kinematic configurations.

\section{Example Bubbles in the G17 Region}
\label{sec:example_bubbles}
\setcounter{figure}{0}

Bub~A and Bub~B are two relatively isolated bubble candidates in the G17 region (Figure~\ref{Img_Bubbles_M16}), selected to illustrate the diversity of structures recovered by BWFields.  
Neither candidate contains known O-, B-, or A-type stars within its identified boundary. 
Bub~A exhibits a small but distinct bubble-like feature in the mid-infrared three-color image, together with a corresponding central ionizing source, although it is not included in existing infrared bubble catalogs \citep{Churchwell_1}. 
Its molecular shell is clearly recovered in $^{13}$CO, and the associated radial profiles and PV diagnostics indicate an expanding cavity--shell structure. 

In contrast, Bub~B shows no obvious counterpart in infrared or ionized-gas emission and lacks a detectable GLIMPSE 8.0\,$\mu$m emission. 
Nevertheless, BWFields identifies a molecular cavity with measurable shell geometry and velocity structure in the CO data. 
These two examples demonstrate that the G17 bubble candidates are heterogeneous: some have clear multiwavelength counterparts, whereas others are primarily detected through their molecular morphology and kinematics. 
A detailed multiwavelength interpretation of these systems is beyond the scope of this work.

\begin{figure*}
\centering
\vspace{0cm}
\begin{minipage}[t]{0.32\textwidth}
    \centering
    \centerline{\includegraphics[width=2.25in]{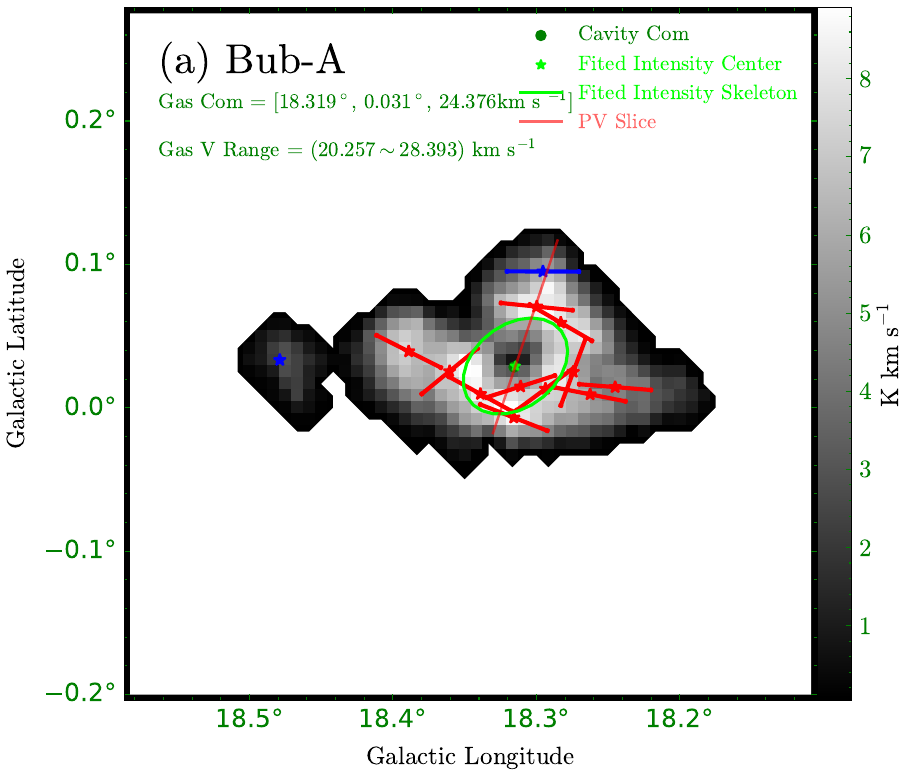}}
    %		\vspace{-4cm}
    %		\centerline{(a)}
\end{minipage}%
\begin{minipage}[t]{0.32\textwidth}
    \centering
    \centerline{\includegraphics[width=2in]{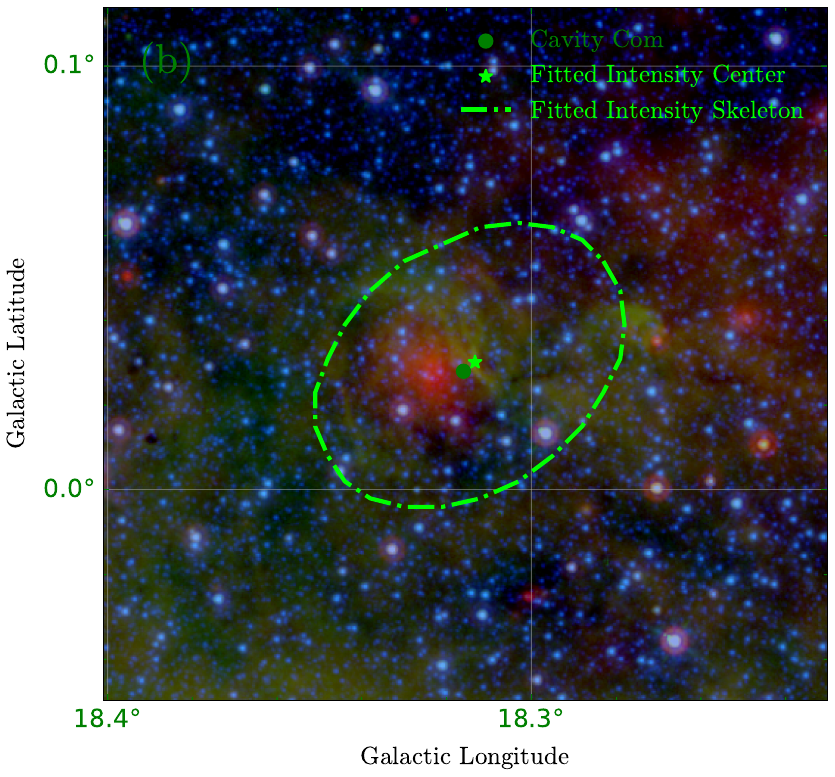}}
    %		\vspace{-4cm}
    %		\centerline{(a)}
\end{minipage}
\begin{minipage}[t]{0.32\textwidth}
    \centering
    \centerline{\includegraphics[width=2.25in]{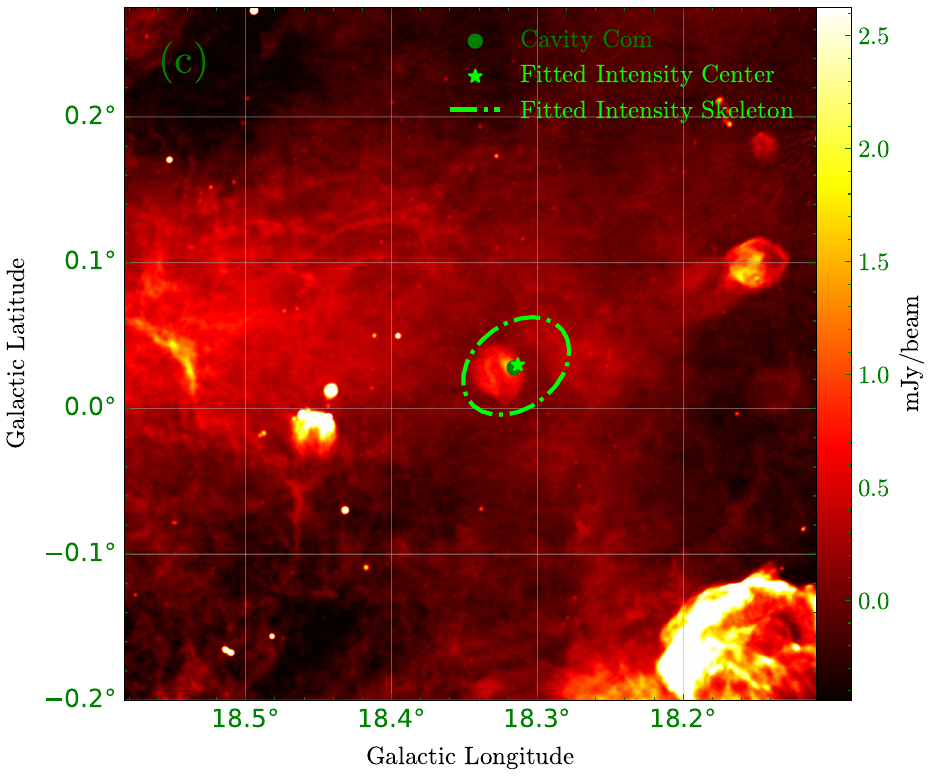}}
    %		\vspace{-4cm}
    %		\centerline{(a)}
\end{minipage}

\begin{minipage}[t]{0.32\textwidth}
    \centering
    \centerline{\includegraphics[width=2.2in]{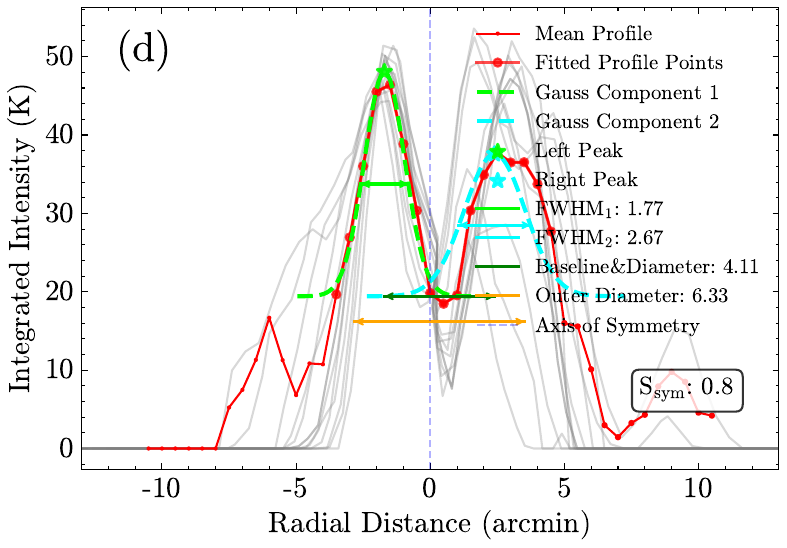}}
    %		\vspace{-4cm}
    %		\centerline{(a)}
\end{minipage}%
\begin{minipage}[t]{0.32\textwidth} 
    \centering
    \centerline{\includegraphics[width=2.15in]{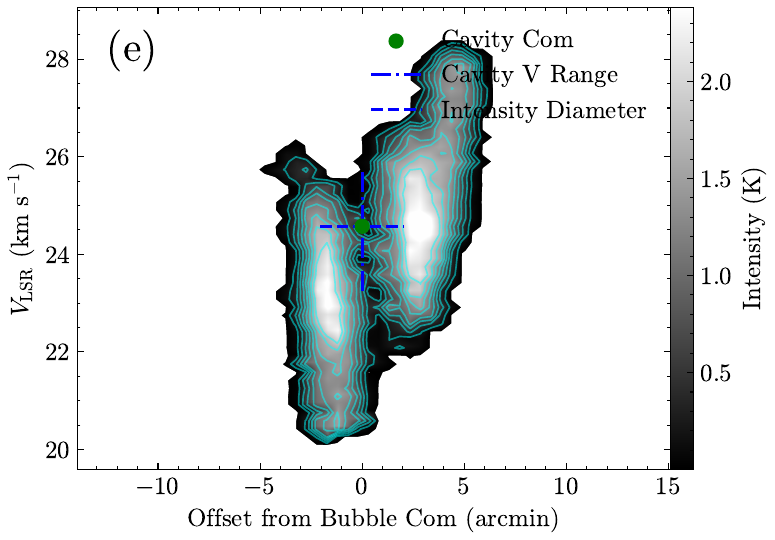}}
    %		\vspace{-4cm}
    %		\centerline{(a)}
\end{minipage}%
\begin{minipage}[t]{0.32\textwidth}
    \centering
    \centerline{\includegraphics[width=2.2in]{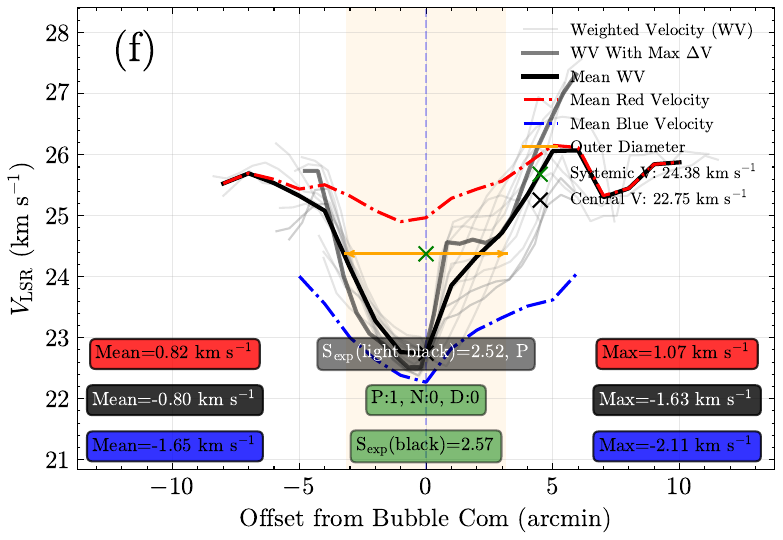}}
    %		\vspace{-4cm}
    %		\centerline{(a)}
\end{minipage}%
\caption{Morphological, kinematic, and multiwavelength views of Bub~A (panels as in Figure~\ref{Img_Appendix_N4}). 
Bub~A shows a compact molecular cavity with a corresponding small but distinct mid-infrared bubble-like feature and a central ionising source; although it is not listed in existing infrared bubble catalogs.
}
\label{Img_Appendix_M16_New_MKSP}
\end{figure*}

\begin{figure*}
\centering
\vspace{0cm}
\begin{minipage}[t]{0.32\textwidth}
    \centering
    \centerline{\includegraphics[width=2.25in]{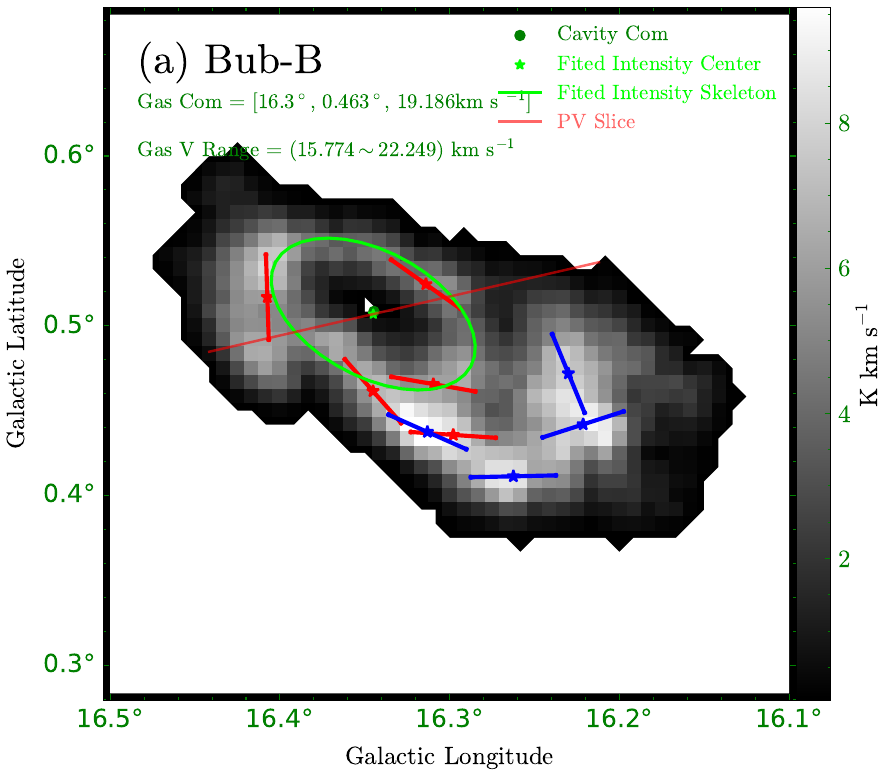}}
    %		\vspace{-4cm}
    %		\centerline{(a)}
\end{minipage}%
\begin{minipage}[t]{0.32\textwidth}
    \centering
    \centerline{\includegraphics[width=2in]{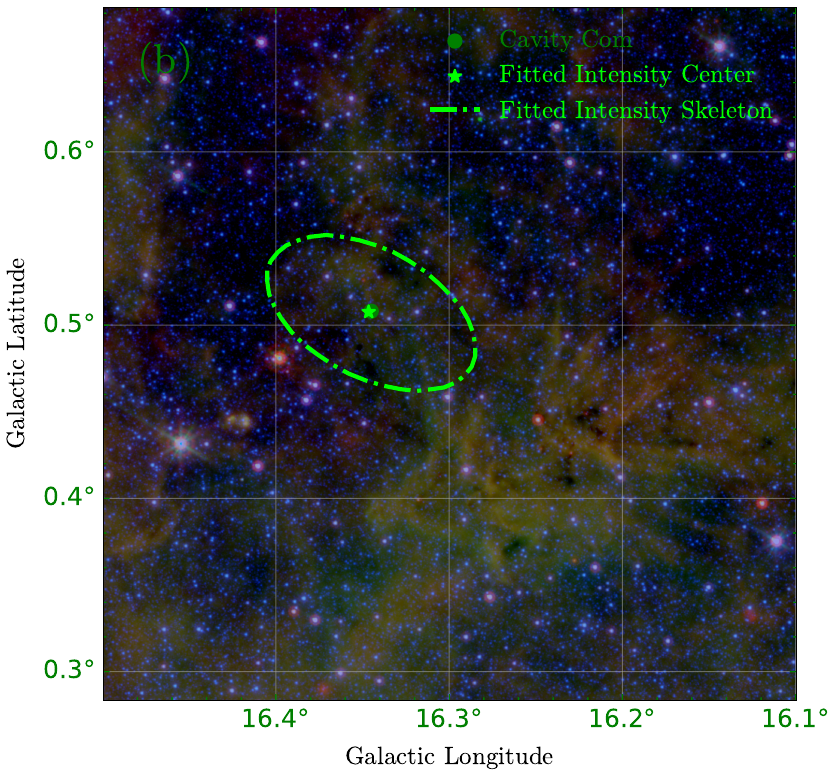}}
    %		\vspace{-4cm}
    %		\centerline{(a)}
\end{minipage}
\begin{minipage}[t]{0.32\textwidth}
    \centering
    \centerline{\includegraphics[width=2.25in]{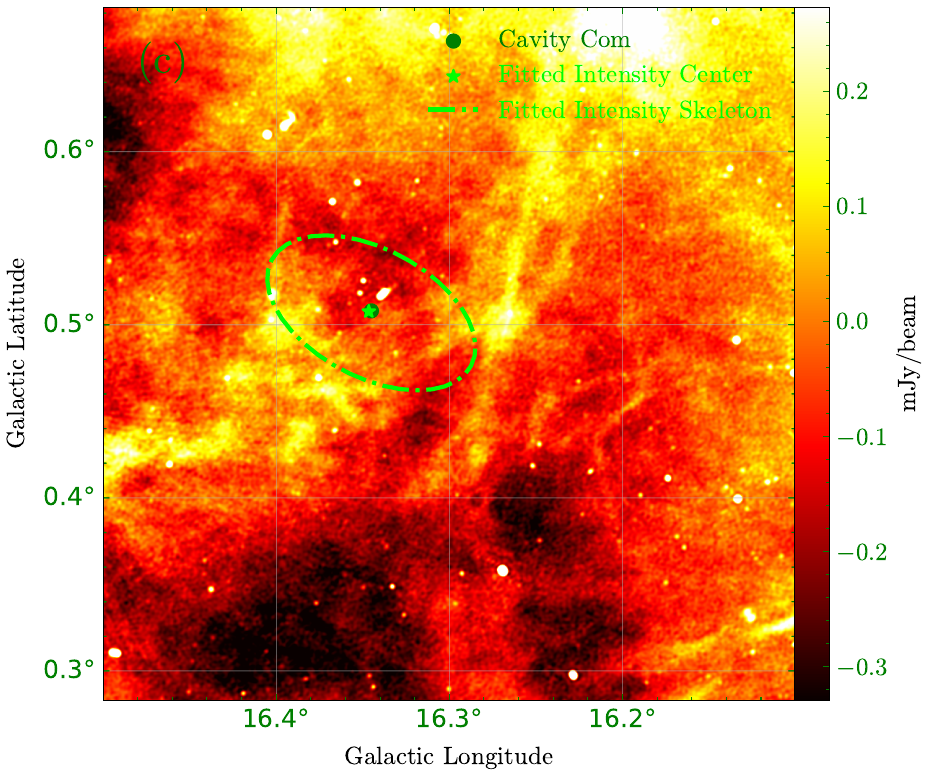}}
    %		\vspace{-4cm}
    %		\centerline{(a)}
\end{minipage}

\begin{minipage}[t]{0.32\textwidth}
    \centering
    \centerline{\includegraphics[width=2.2in]{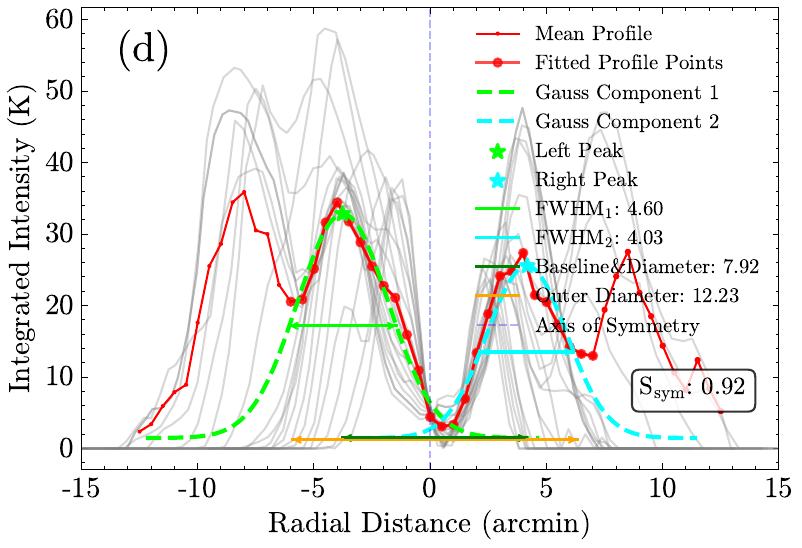}}
    %		\vspace{-4cm}
    %		\centerline{(a)}
\end{minipage}%
\begin{minipage}[t]{0.32\textwidth} 
    \centering
    \centerline{\includegraphics[width=2.15in]{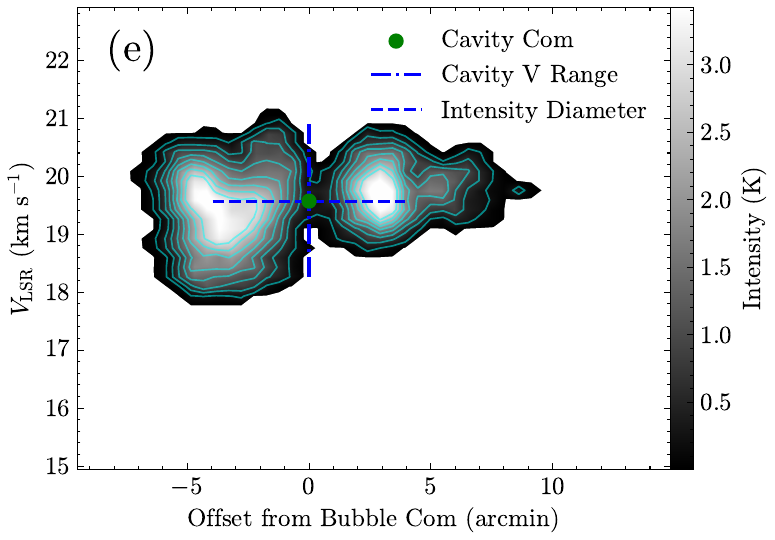}}
    %		\vspace{-4cm}
    %		\centerline{(a)}
\end{minipage}%
\begin{minipage}[t]{0.32\textwidth}
    \centering
    \centerline{\includegraphics[width=2.2in]{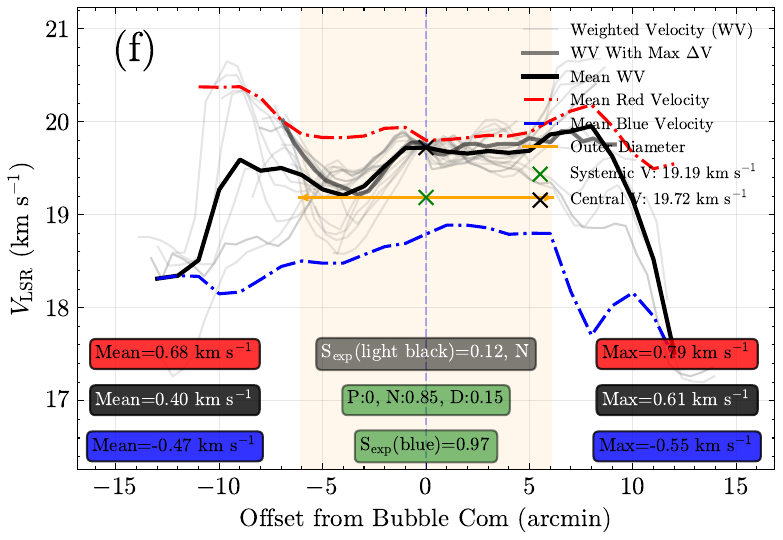}}
    %		\vspace{-4cm}
    %		\centerline{(a)}
\end{minipage}%
\caption{Morphological, kinematic, and multiwavelength views of Bub~B (panels as in Figure~\ref{Img_Appendix_N4}). 
Unlike Bub~A, Bub~B has no obvious infrared or ionized-gas counterpart and lacks a detectable GLIMPSE 8.0\,$\mu$m emission, but it is recovered as a bubble in the $^{13}$CO data.
}
\label{Img_Appendix_M16_Without_MKSP}
\end{figure*}

\section{Source catalog} \label{sec:catalog}
\setcounter{table}{0} 

Table~\ref{Catalogue_Table_Detection} presents an illustrative catalog of BWFields candidates. Each entry corresponds to a cavity candidate traced by a weight-clump in the bubble-weight field and subsequently characterized using the morphological and kinematic diagnostics described in the main text. For each candidate, we list the position of the cavity interior, the geometric parameters of the emission-defined shell skeleton, and the velocity quantities derived from the PV analysis. 

In addition, we include a confidence metric that quantifies detection robustness. Defined as the mean bubble-weight value within the corresponding weight-clump, it reflects the degree to which a candidate is consistently supported across spatial and velocity scales, and serves as a relative indicator for ranking and reliability assessment within the catalog.

\begin{table*}[htbp]
\centering
\rotatebox{90}{
\begin{minipage}{\textheight}
\centering
\caption{Parameters of the bubble candidates.}
\begin{tabular}{lcccccccccccccccc}
\toprule
M BID & $l$ & $b$ & $V_{\rm LSR}$ & $l_{\rm sk}$ & $b_{\rm sk}$ & $V_{\rm sys}$ & $V_{\rm low}$ & $V_{\rm up}$ & Angle & $a_{\rm sk}$ & $b_{\rm sk}$ & $T$ & $V_{\rm exp}$ & $S_{\rm sym}$ & $S_{\rm exp}$ & Conf \\
\midrule
 & (deg) & (deg) & (km\,s$^{-1}$) & (deg) & (deg) & (km\,s$^{-1}$) & (km\,s$^{-1}$) & (km\,s$^{-1}$) & (deg) & (arcmin) & (arcmin) & (arcmin) & (km\,s$^{-1}$) &  &  &  \\
 (1)&(2)&(3)&(4)&(5)&(6)&(7)&(8)&(9)&(10)&(11)&(12)&(13)&(14)&(15)&(16)&(17)\\
\midrule
MB001 & 17.047 & 0.985 & 20.209 & 17.047 & 0.990 & 20.864 & 18.098 & 22.083 & 56.16 & 3.27 & 2.77 & 3.82 & 4.527 & 0.95 & 1.55 & 79.88 \\
MB002 & 17.131 & 1.004 & 19.421 & 17.130 & 1.000 & 20.104 & 18.098 & 21.419 & -12.41 & 2.41 & 1.71 & 3.03 & 3.291 & 0.92 & 1.24 & 66.94 \\
MB003 & 17.011 & 0.928 & 20.586 & 17.017 & 0.922 & 22.022 & 18.098 & 25.092 & 5.73 & 2.45 & 2.07 & 3.54 & 5.500 & 0.87 & 2.55 & 80.06 \\
MB004 & 17.111 & 0.903 & 20.504 & 17.109 & 0.909 & 21.315 & 18.098 & 22.249 & 82.19 & 2.78 & 2.39 & 5.60 & 4.546 & 0.92 & 3.07 & 112.75 \\
MB005 & 17.062 & 0.878 & 20.729 & 17.061 & 0.875 & 22.325 & 18.098 & 24.048 & 12.97 & 2.66 & 1.99 & 2.24 & 4.348 & 0.91 & 0.66 & 80.00 \\
MB006 & 17.141 & 0.866 & 21.530 & 17.141 & 0.860 & 21.703 & 18.098 & 26.068 & -68.12 & 3.70 & 3.15 & 3.44 & 4.741 & 0.94 & 1.96 & 121.25 \\
N4& 11.888& 0.747& 24.696& 11.886& 0.747& 25.058& 21.419& 28.393& -35.50& 	2.56& 1.93& 1.87& 4.419& 0.94& 2.37& 89.69\\
N74 &38.886&-0.425&	41.700&	38.885&	-0.425&	41.672&	39.518&	43.669&	17.94&	3.32&2.37&3.56&	2.216&0.82&	0.21& 144.62\\
N75 &38.921& -0.391& 61.649& 38.920& -0.390& 61.385& 60.771& 62.597& 4.20&	4.32& 2.94& 4.60& 5.444& 0.92& 6.29& 13.21\\

\bottomrule
\end{tabular}
\begin{tablenotes}
\item \textbf{Note.}
Column (1): Candidate identifier.
Columns (2)--(4): Galactic longitude, latitude, and LSR velocity of the cavity interior traced by the corresponding weight-clump.
Columns (5)--(6): Centroid of the ellipse fitted to the emission-defined intensity skeleton of the shell.
Column (7): Systemic velocity of the molecular gas associated with the shell.
Columns (8)--(9): Velocity interval within which weight-clump is identified.
Column (10): Position angle of the ellipse fitted to the intensity skeleton, measured counterclockwise from the Galactic longitude direction.
Columns (11)--(12): Semi-major and semi-minor axes of the skeleton ellipse.
Column (13): Shell thickness $T$.
Column (14): Expansion velocity.
Columns (15)--(16): Symmetry score $S_{\rm sym}$ and expansion significance $S_{\rm exp}$.
Column (17): Confidence metric (Conf), defined as the mean bubble-weight value within the corresponding weight-clump. 
\end{tablenotes}
\label{Catalogue_Table_Detection}
\end{minipage}
}
\end{table*}

\end{appendices}

\end{document}